\documentclass[a4paper,11pt,pdftex]{article}
\pdfoutput=1

\usepackage{slashed}
\usepackage{braket}
\usepackage{jheppub}
\usepackage{graphicx}
\usepackage{url}
\usepackage{amssymb,amsmath}
\usepackage{float}
\usepackage{wrapfig}
\usepackage{bm}
\usepackage{CJKutf8}
\usepackage{color}
\usepackage{comment}
\usepackage{xcolor}
\usepackage[normalem]{ulem}
\usepackage{hyperref}
\usepackage{mathrsfs}
\usepackage[version=4]{mhchem}
\usepackage{booktabs}
\usepackage[noabbrev]{cleveref}
\usepackage{rotating}
\hypersetup{
colorlinks=true,
citecolor=blue,
citebordercolor=red,
linkcolor=blue,
urlcolor=blue
}
\catcode`\@=11
\catcode`\@=12

\newcommand{\p}{\partial}

\newcommand{\bp}{\begin{pmatrix}}
\newcommand{\ep}{\end{pmatrix}}

\newcommand{\bs}[1]{\boldsymbol}

\newcommand{\be}{\begin{equation}}
\newcommand{\ee}{\end{equation}}

\newcommand{\ba}{\begin{array}} 
\newcommand{\ea}{\end{array}}

\graphicspath{{./figs/}}
\newbox{\ORCIDicon}
\makeatletter
\gdef\@fpheader{\phantom{prepared for submission to JHEP}}
\makeatother

\makeatletter
\@addtoreset{equation}{section}
\makeatother

\begin{document}
\begin{flushright} 
YGHP-25-01
\end{flushright} 
\title{
Dislocations 
%in dynamical formation of chiral soliton lattices 
and crystallization dynamics of 
chiral soliton lattices 
%edge and screw 
%Dislocations in quantum field theory
}

\author[a,b,c]{Minoru Eto,}%,\,\href{https://orcid.org/0000-0002-2554-1888}{\usebox{\ORCIDicon}}}
\emailAdd{meto@sci.kj.yamagata-u.ac.jp}

\author[d,c,b]{Kentaro Nishimura}
\emailAdd{nishiken.phys@gmail.com}

\author[e,b,c]{Muneto Nitta}%,\,\href{https://orcid.org/0000-0002-3851-9305}{\usebox{\ORCIDicon}}}
\emailAdd{nitta@phys-h.keio.ac.jp}

\affiliation[a]{Department of Physics, Yamagata University, 
Kojirakawa-machi 1-4-12, Yamagata, Yamagata 990-8560, Japan}

\affiliation[b]{Research and Education Center for Natural Sciences, Keio University, 4-1-1 Hiyoshi, Yokohama, Kanagawa 223-8521, Japan}

\affiliation[c]{International Institute for Sustainability with Knotted Chiral Meta Matter(WPI-SKCM$^2$), Hiroshima University, 1-3-2 Kagamiyama, Higashi-Hiroshima, Hiroshima 739-8511, Japan}

\affiliation[d]{Institute of Science and Technology, Niigata University, Niigata 950-2181, Japan}

\affiliation[e]{
Department of Physics, Keio University, 4-1-1 Hiyoshi, Kanagawa 223-8521, Japan
}

\abstract{
Dislocations, as topological defects in crystal lattices, are fundamental to understanding plasticity in materials. Similar periodic structures also arise in continuum field theories, such as chiral soliton lattices (CSLs), which appear in condensed matter systems like chiral magnets and in high-energy contexts such as quantum chromodynamics in strong magnetic field or under rapid rotation.
This work investigates whether dislocations can dynamically form within such emergent CSLs. 
The chiral sine-Gordon model, reduced from the aforementioned examples by certain truncations, is useful to determine the ground state but it cannot describe time evolution,  
lacks dynamical formation or leads to singular dislocations, 
because its equations of motion do not contain a topological term. 
We propose a field-theoretical model including the topological term coupled to external fields resolving these issues by modifying the topological term so it affects the dynamics.
Using numerical simulations, we study the real-time formation of CSLs in two and three spatial dimensions. In 2D, edge dislocations emerge spontaneously, guiding soliton growth and later annihilating to leave a stable CSL. In 3D, both edge and screw dislocations form; the latter exhibits helical structure influenced by the external field. 
We find 
stable double helical screw dislocations looking like a double helix staircase or DNA.
We then demonstrate the formation of helical dislocations and analyze how the external field strength affects CSL density and formation speed. Our results provide a novel theoretical framework for understanding dislocations in solitonic structures, connecting high-energy field theory with materials science phenomena.
}

\maketitle 

%%%%%%%%%%%%%%%%%%%%%%%%%%%%%%%%%%%%%%%%%
\section{Introduction}

Dislocations are linear topological defects in a crystal lattice of atoms, 
that is, 
 crystallographic defect or irregularity within a crystal structure that contains an abrupt change in the arrangement  \cite{Volterra:1907}.
 These topological objects 
 are essential to explain plasticity of materials 
namely, they are the carrier of plastic deformation 
\cite{Orowan:1934,Polanyi:1934,Taylor:1934}.
Since their discovery, 
they have been developed in materials science, 
see textbooks 
\cite{Hull:2001,Anderson:2017}.
In order to describe 
dislocations,
a discrete sine-Gordon model 
was proposed 
\cite{frenkel1939theory,Peierls_1940,
nabarro1947dislocations} 
which is also called  
the Frenkel-Kontorova model \cite{frenkel1939theory, Braun:1998, Braun2004}.\footnote{
It is interesting to point out that 
this was the first time that the sine-Gordon model appears in physics.
}
In two spatial dimensions,
these objects describe 
crystals melting 
to the fluid phase 
in the 
Kosterlitz-Thouless-Halperin-Nelson-Young theory 
\cite{
Kosterlitz:1972,Kosterlitz:1973xp,
PhysRevLett.41.121,
PhysRevB.19.2457,
PhysRevB.19.1855,1520312} 
based on 
the Berezinskii-Kosterlitz-Thouless 
transition
\cite{Berezinsky:1970fr,Berezinsky:1970-2,
Kosterlitz:1972,Kosterlitz:1973xp}.
In three spatial dimensions, these topological defects are 
linearly extended stringy objects, 
and
depending on their directions in a lattice, 
straight dislocations are classified into edge  and screw dislocations. 
Dislocations extended in general direction 
are called mixed dislocations. 
It is known that the shear on crystal proliferates 
dislocation loops via the Franck-Read mechanism  \cite{PhysRev.79.722}. 
Dislocation loops play an important role for crystal melting in smectic liquid crystals  
in three dimensions \cite{Helfrich:1978,PhysRevB.24.363,Moreau_2006,1520312}. 
Dislocation lines can be also knotted or linked in three dimensions 
\cite{PhysRevLett.131.128101,Pieranski13112024,Kamien_2016,severino2024dislocationsfibrationstopologicalstructure}. 
Recently, dislocations together with disclinations have attracted much attentions due to their crucial roles in fractons  
\cite{Gromov:2022cxa}.

Theoretically,  
a continuous translational symmetry ${\mathbb R}^d$
(of the vacuum) in $d$-dimensional space 
is spontaneously broken 
in crystal lattices
down to a discrete translation 
${\mathbb Z}^d$ (in the case of a square lattice)
preserving 
the lattices. 
Roughly speaking, 
a symmetry breaking pattern 
${\mathbb R} \to {\mathbb Z}$ in one spatial direction
results in an order parameter manifold (OPM), 
${\mathbb R} / {\mathbb Z} \simeq S^1$.
Thus, a nontrivial first homotopy group 
$\pi_1(S^1) \simeq {\mathbb Z}$ 
ensures the existence of topological defects 
of codimension two,
similar to vortices, 
which are nothing but dislocations.

On the other hand, 
apart from crystal lattices of atoms, 
periodic configurations or modulations of condensate wave functions or quantum fields 
ubiquitously appear as ground states in various scales in nature 
from superconductors to neutron stars:  
the so-called 
Fulde-Ferrell-Larkin-Ovchinnikov (FFLO) states  
\cite{Fulde:1964zz,Larkin:1964wok}
and 
chiral soliton lattices (CSLs) are typical examples.\footnote{Modulated phases also appear in relativistic scalar field theories with a wrong sign in the kinetic term
\cite{Nitta:2017mgk,Nitta:2017yuf,Gudnason:2018bqb,BjarkeGudnason:2018aij}.}
The former is a periodic condensation in superconductors: 
an  FF state \cite{Fulde:1964zz}
is a monotonic phase rotation,  
an  LO state 
\cite{Larkin:1964wok,Machida:1984zz}
is a periodic array of kinks and anti-kinks, 
 and an FFLO state 
 \cite{Basar:2008im,Basar:2008ki}
 is their combination. 
 These states are considered to appear in 
 various condensed matter systems such as 
 superfluids, 
 ultracold Fermi gasses, 
 and quantum chromodynamics (QCD)
 \cite{Casalbuoni:2003wh,Anglani:2013gfu}.
On the other hand, CSLs are periodic arrays of 
sine-Gordon type solitons 
universally appearing in various condensed matter systems 
such as chiral magnets 
\cite{
togawa2012chiral,togawa2016symmetry,KISHINE20151,
PhysRevB.97.184303,PhysRevB.65.064433,Ross:2020orc,Amari:2023gqv,Amari:2023bmx,Amari:2024jxx} 
with 
the Dzyaloshinskii-Moriya (DM) interaction \cite{Dzyaloshinskii,Moriya:1960zz},
cholesteric (chiral) 
or smectic 
liquid crystals \cite{Chandrasekhar:1992,DEGENNES1968163} 
and QCD. 
CSLs in chiral magnets are expected to have an important nanotechnological application in information processing such as magnetic memory storage devices and magnetic sensors \cite{togawa2016symmetry}.
In QCD, CSLs appear as the ground states 
at finite baryon density 
in strong magnetic fields 
\cite{Son:2007ny,Eto:2012qd,Brauner:2016pko,Chen:2021vou,
Gronli:2022cri,Evans:2022hwr,Evans:2023hms,Qiu:2023guy} 
(or under rapid rotation 
\cite{Huang:2017pqe,Nishimura:2020odq,Eto:2021gyy,Eto:2023rzd})
as a consequence of the chiral magnetic effect 
\cite{Son:2004tq,Son:2007ny} 
(or the chiral vortical effect  
\cite{Vilenkin:1979ui,Vilenkin:1980zv,Son:2009tf}), 
or with thermal fluctuation 
\cite{Brauner:2017uiu,Brauner:2017mui,Brauner:2021sci}
(see also Refs.~\cite{Yamada:2021jhy,Brauner:2019aid,Brauner:2019rjg,Nitta:2024xcu}).
Furthermore, there appear Skyrmions on the soliton surfaces 
in higher-density regions 
of the CSLs, 
where such composite objects were called  
domain-wall Skyrmions,  
as shown in QCD matter 
in strong magnetic fields
\cite{Eto:2023lyo,Eto:2023wul,Amari:2024fbo,Amari:2024mip,Amari:2025twm} 
(or under rapid rotation
\cite{Eto:2023tuu}). 
Such ground states in QCD 
are relevant in the interior of 
neutron stars 
or heavy-ion collisions.

One of the natural questions is whether dislocations are allowed 
in such ``emergent'' solitonic lattice structures of continuum field theories, 
in parallel to crystal lattices of atoms. 
Among such, Weyl semimetals admit 
``dislocations'' 
in charge density waves 
\cite{PhysRevB.87.161107,Roy:2014mia,PhysRevB.94.085102,
PhysRevB.102.115159,Gooth:2019}. 
However, charge density waves 
are just monotonically varying phase shifts, rather than physical soliton lattices, 
and consequently, 
configurations are merely superpositions of vortices 
and monotonically changing phases.
On the other hand, there are also 
examples of dislocations 
in soliton lattices. 
Edge dislocations in a CSL 
were studied in chiral magnets 
in two spatial dimensions  
\cite{Schoenherr:2018,PhysRevB.106.224428}, 
where an end point of the dislocation behaves 
as a meron, that is a half skyrmion. 
Screw dislocations in chiral magnets 
in three spatial dimensions were found in Ref.~\cite{PhysRevLett.128.157204}. 
Dislocations exist also in chiral liquid crystals  
\cite{Smalyukh:2020zin} and 
smectic liquid crystals \cite{PhysRevE.109.L012701,
PhysRevLett.131.128101,Pieranski13112024,Kamien_2016,
severino2024dislocationsfibrationstopologicalstructure,Kamien_2006}. 
Dislocations in FFLO states 
in superconductors carry half magnetic fluxes of those of a singly quantized vortices 
\cite{Agterberg_2008}.
While the aforementioned examples are  dislocations in one dimensional lattices or modulations,  they also can exist in two dimensional lattices:  
an Abrikosov vortex lattice in type-II superconductors allows 
edge \cite{PhysRevLett.67.1926} and screw  \cite{PhysRevB.34.494,PhysRevLett.57.1347,PhysRevB.34.6514,PhysRevB.41.1910,PhysRevB.51.11887} dislocations 
 in two and three dimensions, respectively, 
and 
a skyrmion lattice in a frustrated magnet admits edge dislocations \cite{shimizu2023crystallization}.

In this paper, we propose an efficient model to investigate the dynamics of dislocations in the CSL and use it to study the dynamical formation of CSLs accompanied by dislocations in a soliton lattice.  
Formation of CSLs was studied before in terms of quantum nucleation \cite{Eto:2022lhu,Higaki:2022gnw}. 
In contrast, we focus on dynamical formation of CSLs.
As a field theoretic model admitting soliton lattice as the ground state, 
we consider three models with a topological term coupled to an external field 
that can physically be identified 
with various terms, 
such as a uniform magnetic field 
or global rotation in chiral Lagrangian of QCD or the DM interaction in chiral magnets. 
The first model (the model I) is the chiral sine-Gordon model well studied in the literature. 
It is the simplest model admitting CSL but dislocations are singular objects and formation of CSL cannot be rigorously discussed because the topological term does not contribute to equations of motion (EOM).
In order to make the dislocations regular solitonic objects,
we improve the chiral sine-Gordon model and consider the second model (the model II) that is an axion model with a single complex scalar field, 
coupled to the external field. 
Even in the second model dynamical formation of the CSL is not realized,
so we propose the third model (the model III) by further modifying the topological term and 
overcoming all the problems for the dynamical formation of CSL in the previous models. 
These models contain a mass parameter $m$;
a mass of pseudo-Nambu-Goldstone boson.\footnote{
These models can be obtained as a Ginzburg-Landau low-energy theory 
of a Nambu Jona-Lasino type model of a single fermion of mass $m$ and 
an anomalous coupling.
}
Due to the presence of the topological term, 
the ground state in the massive case $m\neq 0$ is a CSL along the external field \cite{Brauner:2016pko}.\footnote{
In the massless case $m=0$, the situation corresponds to 
the case of charge density waves in Weyl semimetals 
\cite{PhysRevB.87.161107,Roy:2014mia,PhysRevB.94.085102,
PhysRevB.102.115159,Gooth:2019}, 
where the charge density wave is a uniformly varying phase rotation. 
}

The main results of this work are on the dynamical formation of CSL and construction of edge and screw dislocations in the model III. 
To correctly understand the formation of CSL, many factors need to be accounted for.
We clarify the followings in the model III:
a phase diagram for ground state, that is homogeneous vacuum versus a single flat soliton (domain wall), 
dynamics of a finite single soliton ending on dislocations (domain wall suspended by vortices) under the presence of a uniform background field. 
After figuring out these elementary properties,
we investigate dynamical formation of CSL both in two and three dimensions by solving EOM numerically. We succeed in reproducing CSLs both in two and three dimensional simulations, and observe in detail how the CSL is constructed from a soliton-free state.
We find that edge dislocations play a key role in two dimensions.
In the initial stage of evolution, edge and anti-edge dislocations are pair created, then a finite chiral soliton expands along a direction perpendicular to the background field, and finally the (anti-)dislocations annihilate in a pair leaving a CSL. 
In the process of forming an ordered soliton crystal from a random complex state, 
the motion of dislocations both parallel and perpendicular to the external field are important.
Interestingly, we find that these relativistic dynamics are quite similar to non-relativistic dynamics of crystal dislocations in an atomic crystal. 
In three dimensions, another type of dislocation, 
a screw dislocation, also plays an important role.
We show that both edge and screw dislocations appear as boundaries of chiral solitons.
The former extends in a plane perpendicular to the external field,
whereas the latter extends parallel to the external field. 
In general, they are mixed to form mixed dislocations. 
We numerically construct screw dislocations and find that they are helical in the sense that the parity between left-hand and right-hand screws which are correlated to vortex windings is broken by the background external field. 
We also construct a double helical screw dislocation looking like a double helix staircase or DNA. We find that this is stable although two separated screw dislocations are unstable.
The formation of a CSL in three dimensions goes as follows:
In the early stage of the evolution,
small disk-shaped chiral solitons are generated,
and then they expand into the plane perpendicular to the background field,
until a parallel array of chiral solitons occupies the entire space, completing a solid CSL.
We observe a reconnection of the dislocations and collapse of holes on the chiral solitons repeatedly occur during the formation process of CSL.
We also clarify a relation between magnitude of external field and the number density of CSL, the formation speed, and so on.

This paper is organized as follows.
%\textcolor{blue}{
In Sec.~\ref{sec:model} we introduce simple efficient field theoretical models suitable for studying dynamical formation or crystallization of the CSLs.
%Model I is the chiral sine-Gordon model. 
We review the CSLs and explain two major issues in studying the dynamical formation problem in the chiral sine-Gordon model (the model I). We then provide a detailed description of the CSL in the axion model (the model II). 
We show that it partially solves the problems in the chiral sine-Gordon model, but it gives rise to the other issue regarding the stability of solitons. We finally propose the model III as the most appropriate for the purpose of this work. It is the same as the Model II except for the topological term modified in such a way that the presence of external field affects on the EOM.
In Sec.~\ref{sec:2d}, we 
characterize dislocations and vortices as topological defects and
numerically investigate dynamical formation of CSLs in the model III in two spatial dimensions. 
We demonstrate numerical simulations  showing the 2D CSL formation from soliton-free homogeneous state with fluctuations.
Then, we focus on elementary processes related to the edge dislocations in details.
In Sec.~\ref{sec:3D} we numerically study dynamical formation of CSLs in the model III in three spatial dimensions. We first investigate single and double edge and screw dislocations, and then study transitions from soliton-free vacuum with fluctuations to the solid CSLs.
Sec.\ref{sec:summary} is devoted to a summary and discussion.
In Appendix \ref{sec:appendix}, we give an example of the UV theory for the model II. 
In Appendix \ref{sec:atomic_crystal}, we compare dislocations in CSLs and those in atomic crystals.
In Appendix \ref{sec:appendixC}, we give an example of time evolution to form CSL starting from a configuration different from the vacuum.

%%%%
\section{Field theory models for dislocations}
\label{sec:model}

The CSL, a periodic array of the solitons, typically appears in the sine-Gordon model with a scalar potential $\cos \eta$.
We start by pointing out the importance of distinguishing whether a scalar field is compact or noncompact.
Let $\eta$ be a compact scalar field with identification
\be
\eta \sim \eta + 2\pi\,.
\ee
 For the compact scalar field, 
 the following topological number can be defined
\be
Q = \frac{1}{2\pi} \oint_C dl^\mu\p_\mu \eta = n\,,\quad n \in \mathbb{Z}\,,
\ee
where $C$ is a closed pass and $dl^\mu$ is the line element along $C$.
If $n\neq0$ occurs, there must be $n$ points inside $C$ at which $\eta$ is not well-defined.
These singularities are located at topological defects that will play a central role throughout this paper.

In order to study such topological defects, we consider three different field theory models describing the compact scalar field $\eta$ in each subsection below.

\subsection{The model I: The chiral sine-Gordon model}
\label{sec:modelI}

The first model is the usual sine-Gordon model  coupled with a conserved current $S^\mu$ as
\begin{eqnarray}
    {\cal L}_{\rm \eta} = v^2 \p_\mu\eta\p^\mu\eta - 2v^2m^2(1-\cos\eta) +  \kappa S^\mu \p_\mu \eta\,,\qquad
    \p_\mu S^\mu = 0\,.
    \label{eq:Lag_eta}
\end{eqnarray}
This model is called the chiral sine-Gordon model and appears, for instance, in chiral magnets 
\cite{
togawa2012chiral,togawa2016symmetry,KISHINE20151,
PhysRevB.97.184303,PhysRevB.65.064433,Ross:2020orc,Amari:2023gqv} 
and in the chiral Lagrangian in QCD under a magnetic field \cite{Son:2007ny,Eto:2012qd,Brauner:2016pko,Chen:2021vou,
Gronli:2022cri,Evans:2022hwr,Evans:2023hms,Qiu:2023guy}  
or under rotation 
\cite{Huang:2017pqe,Nishimura:2020odq,Eto:2021gyy,Eto:2023rzd}. 
The Greek letters $\mu,\nu,\cdots$ are used for the $3+1$ dimensional spacetime indices,
and our metric has the mostly minus signature $\eta_{\mu\nu} = {\rm diag}(+,-,-,-)$.
We assume that $S^\mu$ is an external conserved current ($\p_\mu S^\mu = 0$).
The last term does not affect the EOM because it can be cast into the total derivative $\kappa S^\mu\p_\mu\eta = \p_\mu(\kappa\eta S^\mu)$.
Hence, the EOM of this model is the same as the conventional sine-Gordon equation
\be
    \p_\mu\p^\mu \eta + m^2 \sin \eta = 0\,.
    \label{eq:sG_Eq}
\ee

The conjugate canonical momentum of $\eta$ is 
\begin{eqnarray}
    \pi_\eta 
    = 2 v^2 \dot \eta + \kappa S^0\,,
\end{eqnarray}
and the Hamiltonian ${\cal H}_\eta = \pi_\eta \dot\eta - {\cal L}_{\eta}$ can be written as 
\be
    {\cal H}_\eta 
    = v^2 \dot\eta^2 + v^2 (\nabla \eta)^2 + 2v^2m^2(1-\cos\eta) - \kappa \vec S \cdot \nabla \eta\,,
\ee
where $S^\mu = (S^0,\vec S) = (S^0,S^1,S^2,S^3)$ and 
$\p_\mu = (\frac{\p}{\p t},\nabla) = (\frac{\p}{\p t},\frac{\p}{\p x^1},\frac{\p}{\p x^2},\frac{\p}{\p x^3})$.
The last term is peculiar to our model.
It will turn out to be crucial when $\eta$ is not homogeneous, $\vec \nabla \eta \neq \vec 0$.

Let us focus on the simplest case with a constant and spatial external current $S^\mu$. Without loss of generality, we can take
\be
    S^0= 0\,,\quad \vec S = (0,0,B)\,,\quad B > 0\,.
    \label{eq:EXC}
\ee
Let us consider the effect of non-zero $B$ in the presence of a static single soliton perpendicular to a unit vector $\vec n$: 
\begin{eqnarray}
    \eta = 4\,\arctan\, \exp(m \xi)\,,
    \quad
    \xi = \vec n \cdot \vec x\,.
    \label{eq:single_soliton}
\end{eqnarray}
This is a solution of the sine-Gordon equation (\ref{eq:sG_Eq}) for arbitrary $\vec n$.
Assuming $m > 0$, its asymptotic behavior reads $\eta \to 0$ as $\xi \to -\infty$ and $\eta \to 2\pi$ as $\xi \to + \infty$.
Hence, the soliton is characterized by the topological winding number $Q=1$.
Using $\p_i \eta = (\p_i\xi) \eta' = n^i\eta'$ with $\eta' = d\eta/d\xi$, the value of the Hamiltonian is evaluated as 
\begin{eqnarray}
    {\cal H}_\eta = {\cal E}_{\rm I} - \kappa B n_3 \eta'\,,
    \label{eq:H_eta_1}
\end{eqnarray}
with
\begin{eqnarray}
    {\cal E}_{\rm I} = v^2 \eta'{}^2 + 2v^2m^2(1-\cos\eta) 
    = 8 v^2 m^2\,{\rm sech}^2 m \xi\,.
\end{eqnarray}
Then, the total tension $\sigma_{\rm I}$ of the soliton (mass per unit area) is given by
\begin{eqnarray}
    \sigma_{\rm I} = \sigma_{\rm I} - 2\pi \kappa B n_3\,,
    \quad \sigma_{\rm I} = \int_{-\infty}^\infty d\xi\, {\cal E}_1 = 16 mv^2\,.
\end{eqnarray}
Here we have used $\int d\xi\, \eta' = 2\pi$.
This is minimized for $\vec n = (0,0,1)$, namely $\xi = x^3$.
Namely, the sine-Gordon soliton with the minimum energy is perpendicular to the background field $\vec S$.
Note that $\sigma_{\rm I}$ is negative when
\be
    B > B_{\rm I} = \frac{\sigma_{\rm I}}{2\pi\kappa} = \frac{8v^2m}{\pi\kappa}\,.
    \label{eq:critical_B}
\ee
Hence, if $2\pi \kappa B$ is larger than $\sigma_{\rm I}$, 
the ground state of the system is not homogeneous $\eta = 2n\pi$ ($n\in\mathbb{Z}$) but inhomogeneous.
Note that the anti-soliton $\eta = -4\,\arctan\, \exp(m x^3)$ [corresponding to $\vec n = (0,0,-1)$] is disfavored because the contribution from the second term in Eq.~(\ref{eq:H_eta_1}) is positive. Hence, the parity symmetry between soliton and anti-soliton is broken. This soliton is called a chiral sine-Gordon soliton, or simply a chiral soliton.

At the ground state, 
we expect more chiral solitons to appear.
We can indeed see this as follows. 
However, multiple sine-Gordon solitons feel repulsion between them,
so there should be some maximal number of solitons in the ground state.
As a static solution to Eq.~(\ref{eq:H_eta_1}),
an array of the sine-Gordon solitons called the chiral soliton lattice (CSL) perpendicular to the $x^3$ axis is known
\be
    \eta(x^3;k) = 2\,{\rm am}\left(\frac{m x^3}{k},k\right) + \pi\,,
    \label{eq:SG_CSL}
\ee
where ${\rm am}(x,k)$ is the Jacobi elliptic amplitude function with $k$ being an elliptic modulus taking value in $0 \le k \le 1$.
Physically, the elliptic modulus $k$ is related to the period $\ell$,
or the distance between adjacent solitons given by
\be
    \ell(k) = \frac{2 k K(k)}{m}\,,
    \label{eq:period}
\ee
where $K(k)$ is the complete elliptic integral of the first kind.\footnote{
Note that $\ell$ goes to infinity at the limit $k\to1$, and $\eta(x^3;k=1)$ coincides with the single soliton solution given in Eq.~(\ref{eq:single_soliton}). In the opposite limit $k\to0$, the period $\ell$ reduces to zero. The solution becomes a linear function $\eta(x^3,k=0) \to \dfrac{2\pi x^3}{\ell}$, although $\ell^{-1}$ diverges in the limit.}
For $\eta$ increases by $2\pi$ for one period $\ell$, there is one soliton ($Q=1$).
The tension of the one soliton among infinite solitons is given by
\be
    \Sigma(k) = \sigma(k) - 2\pi \kappa B\,,
\ee
where
\be
    \sigma(k) = \int_{-\ell(k)/2}^{\ell(k)/2} dx^3\, {\cal H}_\eta
    = mv^2\left\{\frac{16E(k)}{k} + 8 \left(k-\frac{1}{k}\right)K(k)\right\}\,,
\ee
where $E(k)$ is the complete elliptic integral of the second kind. 
Note that in the limit of $k\to1$, 
$\sigma(1)$ coincides with $\sigma_{\rm I}$ because of $E(k)\to 1$
and $\left(k-\frac{1}{k}\right)K(k)\to 0$.
In order to find the $k$ optimizing the average tension $\Sigma/\ell$, we solve
\be
    \frac{d}{dk}\frac{\Sigma(k)}{\ell(k)} = 0
    \quad\Leftrightarrow\quad
    \frac{m E(k)\left( 8mv^2 E(k) - \pi \kappa Bk\right)}{k^3(k^2-1) K(k)^2} = 0\,,
\ee
satisfied by 
\be
    \kappa B = \frac{8v^2mE(k)}{k\pi}\,.
\ee
This determines the elliptic modulus $k$ for given $B$, and we can calculate the period $\ell$ by plugging the optimized $k$ into Eq.~(\ref{eq:period}). Note that this coincides with Eq.~(\ref{eq:critical_B}) in the $k\to 1$ limit. The tension of the single soliton is then calculated as
\be
    \Sigma(k) = 8 mv^2 \left(k-\frac{1}{k}\right)K(k)\,.
\ee
This is negative for $0\le k \le 1$, implying that the CSL is the ground state for the region (\ref{eq:critical_B}).

All explanations so far are clear, but there are at least two subtle points.
The first issue is that the above analysis is based on kinematics only.
So, we cannot answer the natural question of how the CSL with non-zero $Q$ appears from the homogeneous vacuum with $Q=0$ when we gradually increase $B$ from below to above the critical value $B_{\rm I}$.
We can include $t$ dependence to the above analysis, but neither the EOM (\ref{eq:sG_Eq}) nor its solution (\ref{eq:SG_CSL}) depends on the external field $B$.
It means that the dynamics with and without the term $\kappa S^\mu\p_\mu \eta$ are identical,
even though the kinematics are different.
The second issue is related to the infinite world-volume of the CSL.
The sudden appearance/disappearance of such infinitely large objects would not occur in general.
Hence, the static CSL should emerge through a certain physical process involving the growth of some finite object.
In the following section, we will consider two different models to improve the chiral sine-Gordon model to solve these dynamical issues.

\subsection{The model II: the axion model with a topological term}

\subsubsection{Definition of the model}
One simple extension is achieved by introducing a Higgs (amplitude) mode $\rho$ orthogonal to the pseudo-NG mode $\eta$,
Namely, we identify the compact scalar field $\eta$ with a phase of a complex scalar field $\phi$ as
\be
    \phi = \rho e^{i\eta}\,.
\ee
Then, we consider the following Lagrangian for $\phi$
\be
    {\cal L}_{\rm II} = \left|\p_\mu \phi\right|^2 - V
    + \kappa S^\mu \p_\mu \eta\,,
\ee
with the scalar potential
\be
    V =  \frac{\lambda}{4}\left(|\phi|^2 - v^2\right)^2 + v m^2(2v - \phi - \phi^*)\,.
    \label{eq:potential_model_II}
\ee
We give a theoretical background for considering the model II related to the quantum anomaly in Sec.~\ref{sec:appendix}.
The symmetry of ${\cal L}_{\rm II}$ is $\mathbb{Z}_2$ ($\phi \to \phi^*$) for $m\neq0$ which is a subgroup of $O(2)$ for $m=0$.
The scalar potential is minimized at $\phi = v'$ satisfying 
$v^{-3} \phi \left(\phi^2-v^2\right) = 2 m^2/(\lambda v^2)$ and $\phi = \phi^*$. For $\lambda v^2 \gg m^2$, VEV is approximated by
$\phi \simeq v$.\footnote{${\cal L}_{\rm II}$ would be thought of as an UV completion of ${\cal L}_I$ in the previous subsection. Indeed, if $\lambda v^2 \gg m^2$, we can approximately set $\rho = v$, and ${\cal L}_{\rm II}$ reduces to ${\cal L}_I$.}
Note that this is different from the condition $|\phi|=v$ for $m=0$. 

The last term in ${\cal L}_{\rm II}$ is a total derivative as the model I, so it does not affect the EOM:
\be
\p_\mu\p^\mu\phi + \frac{\lambda}{2}\left(|\phi|^2-v^2\right)\phi - vm^2 = 0\,.
\label{eq:EOM_II}
\ee
Assuming $\rho \neq0$,
we can rewrite the EOM in terms of the amplitude $\rho$ and phase $\eta$ as follows:
\begin{eqnarray}
    \p^2 \rho-\rho\p_\mu\eta\p^\mu\eta + \frac{\lambda}{2} \left(\rho^2 -v^2\right)\rho - vm^2 \cos\eta = 0\,,
    \label{eq:EOM_model2_1}\\
    \rho\p^2\eta + 2 \p_\mu\eta\p^\mu \rho + vm^2 \sin \eta = 0\,.
    \label{eq:EOM_model2_2}
\end{eqnarray}
The Higgs mode $\rho$ is frozen at $\rho = v$ when $\lambda v^2 \gg m^2$,
then we are left with only $\eta$ as a dynamical degree of freedom,
and EOM (\ref{eq:EOM_model2_2}) coincides with the sine-Gordon equation (\ref{eq:sG_Eq}).

\subsubsection{Chiral solitons} \label{sec:soliton_model_II}
The model ${\cal L}_{\rm II}$ admits solitonic solutions like the sine-Gordon model ${\cal L}_\eta$.
However, there is an important difference:
they are not topologically stable in ${\cal L}_{\rm II}$ because the target space $\mathbb{C}$ is simply connected,
and the vacuum manifold is a single point $\phi = v'$.
Hence, there are no topological reasons ensuring the stability of the solitons. 
Of course, this does not imply the nonexistence of dynamically stable non-topological solitonic solutions in the model. 
The stability issue is not a topological but a dynamical problem.

Let us construct an infinitely extended flat soliton perpendicular to the $z$ axis.
We impose that $\eta$ winds once when we traverse $z$ from $-\infty$ to $\infty$. 
This is possible when the Higgs mode $\rho$ remains around $\rho \simeq v$.
We expect that this occurs in the parameter region $\lambda v^2 \gg  m^2$ whereas a tachyonic instability in the Higgs mode would arise for $\lambda v^2 \lesssim m^2$.

Let us first assume that there is a meta-stable soliton for $\lambda v^2 \gg m^2$.
Then, its tension reads
\be
\Sigma_{\rm II} = \sigma_{\rm II} - 2\pi Q \kappa B\,,\label{eq:modelII-tension}
\ee
with\footnote{Hereafter, we will use both notations $(x^1,x^2,x^3) = (x,y,z)$.}
\be
\sigma_{\rm II} = \int^\infty_{-\infty}dz\, \left(|\phi'|^2 + V - V_{\rm vac}\right)\,,
\ee
where we have subtracted a constant vacuum energy $V_{\rm vac} = V(v')$,
and the second term in Eq.~(\ref{eq:modelII-tension}) appears due to the external current $S^\mu$ given in Eq.~(\ref{eq:EXC}) with $Q=1$.
Thus, a solitonic configuration becomes energetically favorable when $B$ is larger than the critical value $B$ as
\be
B > B_{\rm II} = \frac{\sigma_{\rm II}}{2\pi \kappa}\,.
\label{eq:critical_model_II}
\ee

In contrast, for $\lambda v^2 \ll m^2$, the flat soliton soon decays into the homogeneous vacuum because the soliton dynamics is governed by EOM (\ref{eq:EOM_II}),
which does not depend on $S^\mu$.
Hence, we have to set $Q = 0$, and $B$ has no contributions to the total tension $\Sigma_{\rm II}$.
But this is an apparent contradiction between kinematics and dynamics: 
From the kinematical viewpoint of lowering the total tension $\Sigma_{\rm II}$ under sufficiently large positive $B$, $Q>0$ should be favored to $Q=0$. 
However, the EOM (\ref{eq:EOM_II}) does not depend on $B$,
so the soliton dynamics is determined in such a way that not $\Sigma_{\rm II}$ but $\sigma_{\rm II}$ is minimized. That is, the dynamics cannot be distinguished by the presence or absence of an external field $B$.
We will solve this puzzle in the next section by providing further improvement in model III.

\begin{figure}[htbp]
    \centering
    \includegraphics[width=0.95\linewidth]{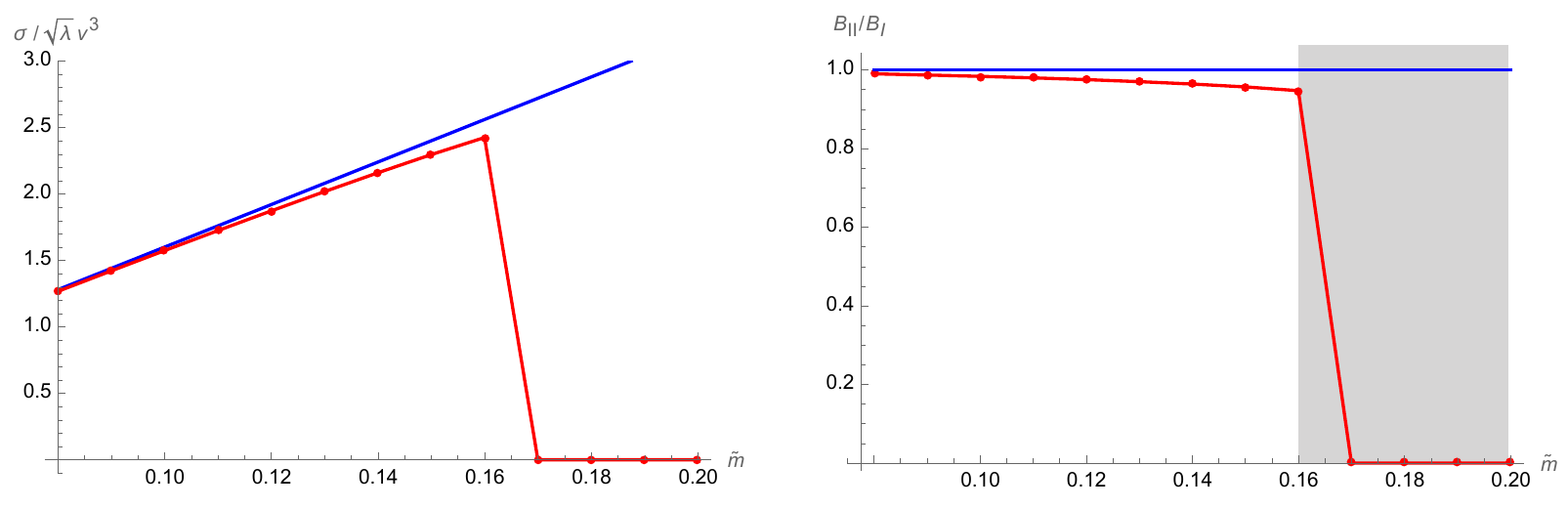}
  \caption{(left) The tensions $\sigma_{\rm I}$ and $\sigma_{\rm II}$ of a single soliton in the model I and II, respectively are plotted as functions of $\tilde m$ by blue and red, respectively. (right) The ratio of the critical $B_{\rm I}$ and $B_{\rm II}$ are plotted as a function of $\tilde m$.}
  \label{fig:tension_criticalB_model_II}
\end{figure}

Let us leave this issue aside for the moment, and explain the features of the solitons in the model II.
The stability of the metastable solitons depends on the unique dimensionless parameter
\be
\tilde m^2 = \frac{m^2}{\lambda v^2}\,.
\ee
In order to numerically solve the EOM,
let us introduce dimensionless variables
\be
\phi = v\tilde \phi\,,\quad z = \frac{\tilde z}{v\sqrt{\lambda}}\,.
\ee
Then, the EOM in Eq.~(\ref{eq:EOM_II}) for the flat soliton can be rewritten as 
\be
- \frac{d^2\tilde\phi}{d\tilde z^2} + \frac{1}{2}
(|\tilde\phi|^2 - 1) \tilde \phi - \tilde m^2 = 0\,,
\label{eq:EOM_II_DL}
\ee
and the tension reads
\begin{eqnarray}
\sigma_{\rm II} = \sqrt{\lambda} v^3
\int^\infty_{-\infty} d\tilde z 
\left[
\left|\frac{d\tilde\phi}{d\tilde z}\right|^2 + \frac{1}{4}\left(|\tilde\phi|^2 - 1\right)^2 + \tilde m^2 \left(2 - \tilde \phi - \tilde \phi^*\right) - \tilde V_{\rm vac}
\right].
\end{eqnarray}
We numerically solve Eq.~(\ref{eq:EOM_II_DL}) for varying $\tilde m$ and find that the solitonic solution only exists in the region
\be
\tilde m < 0.16\,.
\ee 
Then we numerically evaluate the tension $\sigma_{\rm II}/\sqrt \lambda v^3$ in this region, and compare it with $\sigma_{\rm I}/\sqrt \lambda v^3 = 16m/\sqrt \lambda v = 16\tilde m$ of the sine-Gordon soliton in the model I.
We find 
\be
\sigma_{\rm II} < \sigma_{\rm I} = 16 \tilde m\,,
\ee
for $\tilde m < 0.16$ as shown in Fig.~\ref{fig:tension_criticalB_model_II}.
This implies that the critical value of $B_{\rm II}$ in the model II is smaller than $B_{\rm I}$ in the model I: 
\be
B_{\rm II} = \frac{\sigma_{\rm II}}{2\pi \kappa}
= 
\frac{\sigma_{\rm II}}{\sigma_{\rm I}}
B_{\rm I} < B_{\rm I}\,. 
\ee

The reason why the solitons of the models I and II are very similar for $\tilde m \ll 0.16$ is clear:
In that region the Higgs mode is frozen as $|\phi| = v$,
and the dynamical degree of freedom is the phase $\eta$ which obeys the sine-Gordon equation.
The soliton in the model I is free from tachyonic instability for any $m$,
whereas the soliton in the model II is only metastable for $\tilde m < 0.16$ and unstable for $\tilde m > 0.16$.
This is a clear distinction between the models I and II.

\if0
Then the above EOMs reduce to
\begin{eqnarray}
    - \tilde \rho'' +  \tilde \rho \eta'{}^2 + \frac{1}{2}\left(\tilde\rho^2 -1\right)\tilde\rho - \tilde m^2 \cos\eta = 0\,,\\
    - \tilde\rho \eta'' - 2 \eta' \rho' + \tilde m^2 \sin \eta = 0\,,
\end{eqnarray}
where the prime stands for a derivative by $\tilde z$ and we introduced the dimensionless parameter
\be
\tilde m^2 = \frac{m^2}{\lambda v^2}\,.
\ee
Hamiltonian can be expressed by
\be
{\cal H}_\phi = \lambda v^4\left[
\tilde\rho'{}^2 + \tilde\rho^2\eta'{}^2
+ \frac{1}{4}\left(\tilde\rho^2 - 1\right)^2 + 2 \tilde m^2 \left(1 - \tilde \rho \cos\eta\right)
- \tilde B \eta'
\right]\,,
\ee
with
\be
\tilde B = \frac{\kappa B}{\sqrt\lambda v^3}\,.
\ee
The tension of the soliton is given by
\be
M_\phi = \lambda v^4 \left(\tilde T - 2\pi Q \tilde B\right)\,,
\ee
with
\be
\tilde T &=& \int^\infty_{-\infty} d\eta 
\left[
\tilde\rho'{}^2 + \tilde\rho^2\eta'{}^2
+ \frac{1}{4}\left(\tilde\rho^2 - 1\right)^2 + 2 \tilde m^2 \left(1 - \tilde \rho \cos\eta\right)
\right]\,,\\
Q &=& \frac{\eta(\infty) - \eta(-\infty)}{2\pi}\,.
\ee
\fi

\subsubsection{Solitons ending on vortices}
In the region $\tilde m \ll 0.16$ where the Higgs mode is heavy,
the solitons in model II are quite stable against decay into the vacuum by small fluctuations.
Nevertheless, they are not topologically stable due to the presence of vortices to which the solitons can end.
If the explicit $U(1)$ breaking  term does not exist ($m=0$),
model II has topological solitonic vortices (so-called global vortices) as a consequence of spontaneous symmetry breaking of the $U(1)$ symmetry in the vacuum.
Once the explicit breaking term is weakly turned on, the $U(1)$ symmetry is an approximate symmetry.
Then, the vortices are no longer topological and are inevitably accompanied by the solitons.
These phenomena are well-known in axion models,
and from that point of view model II is nothing but the axion model with the domain wall number $N_{\rm DW}=1$.

Here, we show a numerical solution of the flat soliton bridging a pair of parallel vortex and anti-vortex.
The configuration trivially extends along the $x$ axis, so we only consider the $y$-$z$ plane.
Of course, the parameter $\tilde m$ should be below the critical value $0.16$.
To obtain numerical solutions, we make use of the so-called relaxation technique: we modify the EOM by adding a diffusive term with the first order ``time'' ($\tau$) derivative on the right hand side:
\be
- \nabla^2 \phi(x;\tau) + \frac{\lambda}{2}\left(|\phi(x;\tau)|^2-v^2\right)\phi(x;\tau) - vm^2 = - \p_\tau \phi(x;\tau)\,.
\label{eq:relaxation_II}
\ee
Usually, this technique is used to obtain a static configuration corresponding to a local stationary point of the action.
It works as follows: One starts with an initial configuration 
with a correct topology, and numerically solve Eq.~(\ref{eq:relaxation_II}) until the configuration reaches a convergence with $\p_\tau \phi \to 0$.
The convergent configuration is nothing but a static solution of the original EOM belonging to the same topological sector of the initial configuration.
In contrast, here we are interested not in the convergence but in intermediate configurations by pursuing $\tau$-evolution of our initial configuration with the pair of vortex and anti-vortex suspended by the soliton.
Due to the minimization of the system energy,
the distance of the string pair continuously reduces as $\tau$ increases.
We take, as an example, $\tilde m = 0.1$ which is below the critical value $0.16$.
The size of our numerical box is $[-\tilde L,\tilde L]^2 = [-100,100]^2$ with the 2-dimensional lattice $1000^2$,
and we prepare the initial configuration with the vortex separation $\tilde d=50$.
Namely, the vortices are located at $(\tilde y,\tilde z) = (\pm 25,0)$,
and the soliton stretched in between them is on the $\tilde y$ axis in the beginning.
Here, $\tilde y$ ($\tilde z$) stands for the dimensionless coordinate normalized by $v\sqrt\lambda$.
\begin{figure}[htbp]
    \centering
    \includegraphics[width=0.98\linewidth]{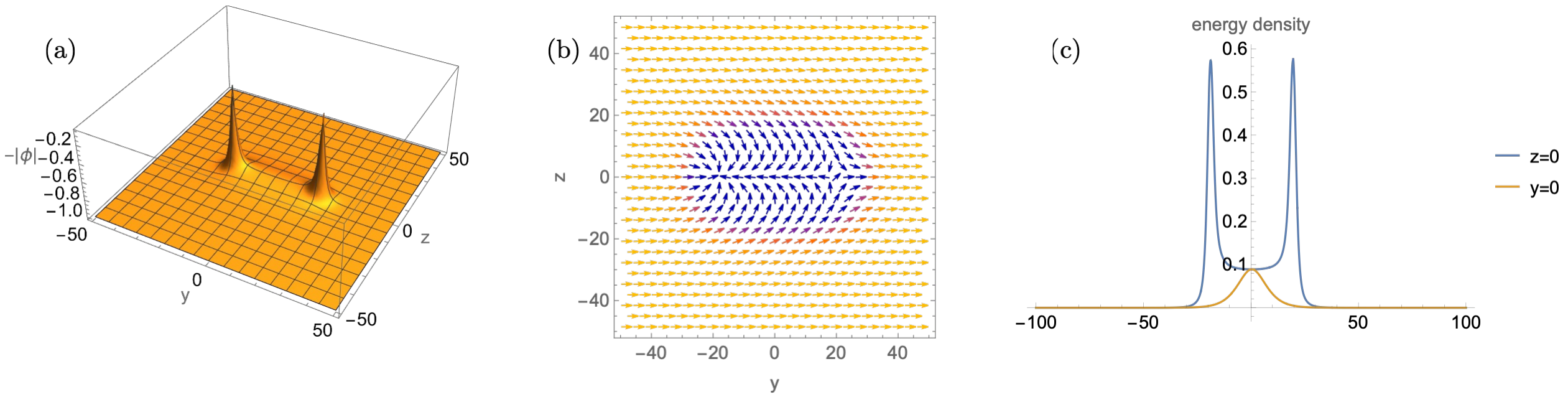}
  \caption{A snapshot (at an appropriate $\tau$) of the vortex pair suspended by the soliton. The panel (a) shows the amplitude ($-|\phi|$), (b) shows the phase of $\phi$ by the vector plot, and (c) shows the energy densities (without contribution from $\kappa S^\mu\p_\mu\eta$ term) at the slices $z=0$ and $y=0$.}
  \label{fig:string_wall_model2}
\end{figure}
Fig.~\ref{fig:string_wall_model2} shows a snapshot of the numerical configuration at the moment when the vortex separation $\tilde d = 40.12$ (the vortex centers are defined as the positions of $\phi = 0$).
The amplitude and phase of $\phi$ shown in (a) and (b) of Fig.~\ref{fig:string_wall_model2},
respectively clearly show the two vortices at the edges of the soliton.
The soliton between the vortices is also clearly seen in panel (c),
showing two cross sections of the energy densities at $\tilde z= 0$ (blue) and $\tilde y = 0$ (yellow);
the low bank between two blue spikes and the yellow small bump.

There is one important comment which we would like to make here. 
The vortices in model II are global vortices whose tensions are logarithmically divergent if the explicit $U(1)$ breaking term is turned off by setting $m=0$.
Once $m$ is turned on, the global vortices cannot exist alone because the solitons are inevitably accompanied. The nonzero $m$ transforms the logarithmically divergent energy of the axially symmetric global vortex into the linearly divergent tension of the soliton attached to the vortices \cite{Eto:2022lhu,Eto:2023gfn}. We should emphasize the fact that the vortex tension no longer logarithmically diverges for $m\neq 0$, contrary to the seemingly common misunderstanding that it logarithmically diverges like the usual global vortices \cite{Vilenkin:2000jqa}. 
Qualitatively, this can be explained by the absence of the massless Nambu-Goldstone mode in the bulk that is lifted by the explicit $U(1)$ breaking term.
It can also be explained quantitatively as follows.
We measure the separation $\tilde d$ of the two vortices and the tension $T_{\rm II}$ at various $\tau$, resulting in the $T_{\rm II}$ as a function of $\tilde d$
\begin{eqnarray}
T_{\rm II}^{(\tilde L)}(\tilde d) =  v^2
\int^{\tilde L}_{-\tilde L} d\tilde y 
\int^{\tilde L}_{-\tilde L} d\tilde z 
\left[
\left|\frac{\p\tilde\phi}{\p\tilde y}\right|^2 + \left|\frac{\p\tilde\phi}{\p\tilde z}\right|^2 + \frac{1}{4}\left(|\tilde\phi|^2 - 1\right)^2 + \tilde m^2 \left(2 - \tilde \phi - \tilde \phi^*\right) - \tilde V_{\rm vac}
\right]\nonumber\\
\end{eqnarray}
\begin{figure}[htbp]
    \centering
    \includegraphics[width=0.65\linewidth]{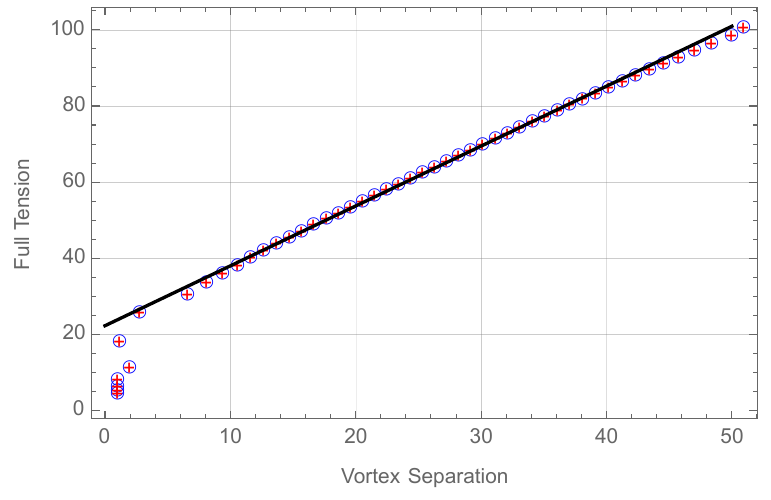}
  \caption{The tension $T_{\rm II}$ (in the unit of $v^2$) as the function of vortex separation $\tilde d$.
  We start with $\tilde d = 50$ and pursue the relaxation $\tau$-evolution to obtain data for smaller $\tilde d$. The blue circles and the red crosses $T_{\rm II}^{(50)}$ and $T_{\rm II}^{(100)}$, respectively. The black solid line corresponds to the linear function $v^{-2}T_{\rm II}(\tilde d) = 1.57 \tilde d + 22.3$.}
  \label{fig:dvsT_model2}
\end{figure}
The result is summarized in Fig.~\ref{fig:dvsT_model2} for $\tilde m = 0.1$.
For the numerical evaluation of $T_{\rm II}^{(\tilde L)}$, we choose two different box sizes with $\tilde L = 50$ and $100$,
corresponding to the blue circles and red crosses in Fig.~\ref{fig:dvsT_model2}.
They coincide very well, implying that the energy density is well localized at each $\tau$, and there is no logarithmic divergence of the axially symmetric global strings.
Furthermore, $T_{\rm II}$ is well approximated by the linear function
\be
T_{\rm II}(d)
= \sigma_{\rm II} d + 2\mu_{\rm II}\,,\quad \sigma_{\rm II} = 1.57 \sqrt{\lambda}v^3\,,\quad
\mu_{\rm II} = 11.2 v^2\,,
\ee
which corresponds to the black solid line in Fig.~\ref{fig:dvsT_model2}.
Physically, the coefficient $\sigma_{\rm II}$ should be interpreted as the tension of the soliton. Indeed, the soliton tension $\sigma_{\rm II}$ for $\tilde m = 0.1$ is given in Fig.~\ref{fig:tension_criticalB_model_II}, and agrees well with $1.57 \sqrt{\lambda}v^3$.
Then, the remaining constant term $2 \mu_{\rm II}$ should be interpreted as the sum of the two vortex tensions because it does not depend on $d$ for the region where the vortex separation is large enough,
while it shrinks for small $d$ region where the vortices start to annihilate in pairs. 
Thus, we conclude that the vortex tension is not logarithmically divergent but a finite constant, and its numerical value is about $\mu_{\rm II} = 11.2 v^2$ for $\tilde m = 0.1$.

Finally, let us take the topological term $\kappa S^\mu \p_\mu \eta$ into account, and figure out the condition that the soliton-vortex composite is energetically favored than the vacuum. Firstly, note that the phase $\eta$ is not well defined at the center of vortex because the amplitude vanishes as $\phi = 0$ there. It is difficult to numerically deal with such singularities. However, it is easy to handle them analytically: $\eta$ transits from 0 to $2\pi$ only when we go across the soliton. Namely, we have
\be
\int dy dz ~
\kappa S^\mu \p_\mu \eta 
= \int^{d/2}_{-d/2} dy \int^\infty_{-\infty} dz ~
\kappa B \p_z \eta 
= 2\pi \kappa B d .
\ee
Thus the tension of the whole system reads
\be
E_{\rm II} = T_{\rm II} - 2\pi \kappa B d = (\sigma_{\rm II}-2\pi \kappa B)d + 2\mu_{\rm II}\,.
\ee
Therefore the soliton of the finite length $d$ is energetically favored than the homogeneous vacuum if $B$ satisfies the following condition
\be
B > B_{\rm II}(d) \equiv \frac{\sigma_{\rm II}}{2\pi \kappa} + \frac{\mu_{\rm II}}{\pi\kappa d}\,.
\ee
Of course, this conclusion is valid only for the assumption that $\tilde m < 0.16$ holds (the condition for the soliton to stably exist).
When $d = \infty$, we have 
$B_{\rm II}(d \to \infty) = \sigma_{\rm II}/2\pi\kappa$ which
is the previously found condition (\ref{eq:critical_model_II})
for the infinitely large flat soliton.

Before closing this subsection, let us make a comment on our question at the end of the previous subsection \ref{sec:modelI}. The model II partially answers to it: 
the chiral soliton in the model II can end on the global vortex. Namely, the chiral soliton is not necessary to be infinitely large but it can be a finite object if it is surrounded by a closed vortex string. This is  fine, but it is valid only for the parameter region $\tilde m < 0.16$. More seriously, the dynamics in the model II is insensitive to the surface term $\kappa S^\mu\p_\mu \eta$ as in the model I. Therefore, the model II is still insufficient to reveal the dynamical formation of the CSL.

\subsection{The model III: the axion model with a modified topological term}

\subsubsection{Definition of model}
We finally come to the model III that is free from all the problematic points of the model II.
The model III is obtained as a minor change of the model II. The difference is only the surface term as
\be
{\cal L}_{\rm III} = \left|\p_\mu \phi\right|^2 - V
+ \frac{\kappa}{v^2} j_\mu S^\mu\,,
\label{eq:Lag_phi}
\ee
where the scalar potential is the same as $V$ in Eq.~(\ref{eq:potential_model_II}).
The current $j_\mu$ is given by
\be
j_\mu = \frac{-i}{2}\left(\phi^*\p_\mu\phi - \phi\p_\mu\phi^*\right)
= \rho^2 \p_\mu \eta \,,
\ee
with $\phi = \rho e^{i\eta}$. 
The third term in this Lagrangian  is {\it not} topological anymore.
The EOM reads
\be
\p_\mu\p^\mu\phi  + \frac{\p V}{\p \phi^*} + \frac{i\kappa}{v^2}(\p_\mu\phi) S^\mu = 0\,.
\label{eq:EOM_phi}
\ee
This can be reexpressed in terms of $\rho$ and $\eta$ as
\begin{eqnarray}
    \p^2 \rho - \rho \p_\mu\eta\p^\mu\eta + \frac{\p V}{\p\rho} - \frac{\kappa}{v^2}\rho S^\mu\p_\mu\eta = 0\,,\label{eq:EOM_phi_2}\\
    \rho \p^2\eta + 2 \p_\mu\eta\p^\mu \rho + \frac{1}{2\rho}\frac{\p V}{\p \eta} + \frac{\kappa}{v^2}S^\mu\p_\mu \rho =0\,,
    \label{eq:EOM_phi_3}
\end{eqnarray}
where we have used 
$\frac{\p V}{\p\phi^*} = \frac{\p\rho}{\p\phi^*}\frac{\p V}{\p \rho} + \frac{\p\eta}{\p\phi^*}\frac{\p V}{\p\eta} = e^{i\eta}\left(\frac{\p V}{\p\rho}+\frac{i}{2\rho}\frac{\p V}{\p\eta}\right)$.
One should note that the EOM of the model III depends on $\kappa$ since the last term is not topological. This is the crucial difference from the model II.

When $m = 0$, the Lagrangian has the global $U(1)$ symmetry $\phi \to e^{i\theta}\phi$. The corresponding Noether current is
\begin{eqnarray}
    J^\mu = \frac{-i}{2}\left[
    \phi^*\left(\p^\mu\phi + \frac{i\kappa}{2v^2}S^\mu\phi\right)
    - \phi \left(\p^\mu\phi^* - \frac{i\kappa}{2v^2}S^\mu\phi^*\right)
    \right] = j^\mu + \frac{\kappa}{2v^2}|\phi|^2S^\mu\,.
\end{eqnarray}
One can verify $\p^\mu J_\mu = 0$ for $m=0$ by using EOM (\ref{eq:EOM_phi}).\footnote{If the $U(1)$ symmetry is exact, the $U(1)$ symmetry can be local symmetry and $S^\mu$ can be regarded as the corresponding gauge field. But this interpretation does not work for $m \neq 0$.}
For the generic case that the potential $V$ includes the explicit $U(1)$ breaking term, the conservation equation is modified as
\be
\p^\mu J_\mu = - i \phi^* \frac{\p V_2}{\p\phi^*} + i \phi \frac{\p V_2}{\p\phi}\,,
\ee
where we have decomposed $V$ into the $U(1)$ symmetric part $V_1$ and the $U(1)$ asymmetric part $V_2$ as $V = V_1 + V_2$. For $V$ given in Eq.~(\ref{eq:potential_model_II}), we have $V_1 = \frac{\lambda}{4}(|\phi|^2-v^2)^2$ and $V_2 = vm^2(2v-\phi-\phi^*)$.

Let us derive the Hamiltonian for the model III. The conjugate momentum is given by
\be
\pi_\phi = \dot\phi^* - \frac{i\kappa}{2v^2}\phi^* S^0\,,\quad \pi_{\phi^*} = \dot\phi + \frac{i\kappa}{2v^2}\phi S^0\,.
\ee
Then, the Hamiltonian is obtained as 
\begin{eqnarray}
{\cal H}_{\rm III} &=& \pi_\phi \dot\phi + \pi_{\phi^*} \dot\phi^* - {\cal L}_\phi \nonumber\\
&=& 2|\dot\phi|^2 
+ \frac{\kappa}{v^2}\left\{-\frac{i}{2}\left(\phi^*\dot\phi - \phi\dot\phi^*\right)\right\}S^0
- {\cal L}_\phi \nonumber\\
&=& |\dot\phi|^2 + |\nabla \phi|^2 + V - \frac{\kappa}{v^2}\vec j \cdot \vec S\,,
\end{eqnarray}
with $\vec j = (j_1,j_2,j_3)$ and $\vec S = (S^1,S^2,S^3)$.
As before, we are interested in the uniform and constant background current $S^i = (0,0,B)$. Then, the Hamiltonian is reduced to 
\be
{\cal H}_{\rm III} = 
|\dot\phi|^2 + |\nabla \phi|^2 + V - \frac{\kappa B \rho^2}{v^2}\p_3\eta\,.
\ee
The last term negatively contributes to the energy for a configuration with $\p_3 \eta > 0$.

For later convenience, let us give expressions in the dimensionless unit
($\tilde\phi = \phi/v$, $\tilde x^\mu = v\sqrt{\lambda} x^\mu$, $\tilde \p_\mu = \frac{\p_\mu}{v\sqrt\lambda}$):
\be
{\cal L}_{\rm III} = v^4\lambda \left[\left|\tilde\p_\mu \tilde\phi\right|^2 - \frac{1}{4}\left(|\tilde\phi|^2-1\right)^2 - \tilde m^2 \left(2-\tilde\phi - \tilde\phi^*\right)
+ \kappa \tilde j_\mu \tilde S^\mu \right]\,,
\ee
with
\be
j_\mu = v^3\sqrt\lambda\, \tilde j_\mu\,,\quad
\tilde j_\mu = \frac{-i}{2}\left(\tilde\phi^*\tilde\p_\mu\tilde\phi - \tilde\phi\tilde\p_\mu\tilde\phi^*\right) \,,\quad
S_\mu = v^3\sqrt\lambda \tilde S_\mu\,,\quad
\tilde m^2 = \frac{m^2}{v^2\lambda}\,.
\ee
The EOM can be rewritten 
in the dimensionless unit as
\be
\tilde\p^2\tilde\phi + \frac{1}{2}\left(|\tilde\phi|^2-1\right)\tilde\phi - \tilde m^2 + i \kappa \tilde B \tilde\p_z \tilde\phi = 0\,,
\label{eq:DL_EOM_III}
\ee
where we have defined $\tilde S^\mu = (0, 0,0,\tilde B)$ with $\tilde B = (v^3\sqrt{\lambda})^{-1}B$.

\subsubsection{Chiral solitons}
\label{sec:chiral_soliton_modelIII}

Let us 
study chiral solitons in the model III.
We assume $\phi$ to depend only on $z$, and numerically solve EOM (\ref{eq:EOM_phi}).
It is similar to EOM (\ref{eq:EOM_II}) in the model II but there is a crucial difference: it includes the term proportional to $j^\mu\p_\mu\phi$. In the models I and II the surface term does not affect the EOMs, so we first solved the $B$-independent EOMs and then we included the surface term when we calculated the full tension $\Sigma_{\rm I,II}$ of solitons. This means that we only needed to solve the EOMs once for various $B$. On the other hand, in the model III we need to solve Eq.~(\ref{eq:EOM_phi}) for each value of $B$ because the EOM is manifestly dependent on $B$.

In order to compare with the previous models, let us divide the soliton tension as follows:
\be
\Sigma_{\rm III} = \sigma_{\rm III} - 2\pi \kappa B {\cal Q}\,,
\ee
with the partial tension
\begin{eqnarray}
\sigma_{\rm III} = \sqrt{\lambda} v^3
\int^\infty_{-\infty} d\tilde z 
\left[
\left|\frac{d\tilde\phi}{d\tilde z}\right|^2 + \frac{1}{4}\left(|\tilde\phi|^2 - 1\right)^2 + \tilde m^2 \left(2 - \tilde \phi - \tilde \phi^*\right)
- \tilde V_{\rm vac}
\right]\,,
\end{eqnarray}
and we have introduced the quasi winding number
\be
{\cal Q} = 
\frac{1}{2\pi v^2} \int dz\, \rho^2 \frac{d\eta}{dz}
= \frac{1}{2\pi} \int d \tilde z\, \tilde \rho^2 \frac{d\eta}{d\tilde z}\,,\qquad
\left(\rho \equiv v \tilde \rho\right)\,.
\ee
Note that ${\cal Q}$ is not integer in general because the amplitude $\rho$ is not a constant but nontrivially depends on $z$.
Intuitively, the gradient term $(d\phi/dz)^2$ makes the soliton fat, the potential makes it thin, and the $\rho^2(d\eta/dz)$ term favors a large $\rho$ and $d\eta/dz > 0$.
Although the soliton is not topological, the last term allows the soliton to be in a true ground state of the model III not only kinematically but also dynamically, as we will show below.

\begin{figure}[htbp]
    \centering
    \includegraphics[width=0.65\linewidth]{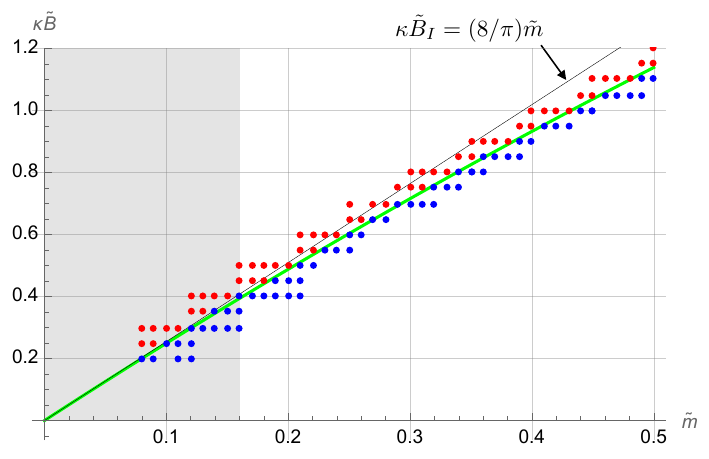}
  \caption{The phase diagram in the $\tilde m$-$\kappa\tilde B$ plane. The blue dots correspond to the homogeneous vacuum whereas the red dots correspond to the solitonic ground state. The black line corresponds to the phase boundary of the model I given in Eq.~(\ref{eq:critical_B}). The green curve is the phase boundary that can be fit by $\kappa \tilde B_{\rm III} = - a \tilde m^2 + \frac{8}{\pi}\tilde m$ with $a = 0.54$. The gray color region stands for the region where the solitonic solution exists in the model II.}
  \label{fig:phase_modeIII}
\end{figure}

By using the relaxation method explained above, we numerically solve the dimensionless EOM (\ref{eq:DL_EOM_III}) and evaluate the full tension $\Sigma_{\rm III}$ for each $\tilde m$ and $\kappa \tilde B$. We survey the ground state by examining the sign of $\Sigma_{\rm III}$: If $\Sigma_{\rm III}$ is positive, the ground state is the homogeneous configuration $\phi = v'$. On the other hand, the ground state is the soliton when $\Sigma_{\rm III}$ is negative. The result is summarized in Fig.~\ref{fig:phase_modeIII}. 
The green curve denotes the phase boundary separating 
the homogeneous states denoted by the blue dots and 
the solitonic ground states denoted by the red dots. 
As a reference, we also denote  
the phase boundary in Eq.~(\ref{eq:critical_B}) 
of the model I by the black solid line. 
We find that the phase boundary of the model III 
is below that of the model I.
The phase boundary can be well fitted by
$\kappa \tilde B_{\rm III} = - a \tilde m^2 + \frac{8}{\pi}\tilde m$
with $a = 0.54$, and this in the physical unit reads
\be
B_{\rm III} = B_{\rm I} - a \frac{m^2v}{\kappa \sqrt\lambda}\,,
\label{eq:criB_III}
\ee
implying $B_{\rm III} < B_{\rm I}$.
Therefore, the solitonic state becomes more favorable in model III compared to model I for a common parameter combination of $v$, $m$ and $\kappa$ in Lagrangians (\ref{eq:Lag_eta}) and (\ref{eq:Lag_phi}). 

Furthermore, the model III is preferable to the model II in the sense that the phase diagram extends beyond the limit ($\tilde m = 0.16$) of the model II.
  As we seen in Sec.~\ref{sec:soliton_model_II}, 
  in the model II 
the solitonic solutions exist only for $\tilde m < 0.16$ \
indicated by the gray color; 
in the model II, the solitons do not exist for $\tilde m > 0.16$ regardless of $B$. 
In contrast, there is no such limitation in the model III, and the solitons do exist for any $\tilde m$ with sufficiently large $\kappa B$ as shown in Fig.~\ref{fig:phase_modeIII}.
In this respect, the model III is a better extension of the model I compared to the model II.

\subsubsection{Solitons ending on vortices}
\label{sec:modelIII_single_finite_soliton}

Let us next study a soliton with finite length suspended by a vortex and an anti-vortex.
Particularly we focus on the real time dynamics of such finite-size soliton reflecting the term $(\kappa/v^2) j_\mu S^\mu$ in the Lagrangian (\ref{eq:Lag_phi}), that is the unique property of the model III.
The process we are going to investigate below is the following. 
Firstly, we prepare a finite-size soliton as an initial configuration ($t=0$; $t$ is physical time which should be distinguished from the relaxation time $\tau$). It is perpendicular to the $z$ axis since we take $\tilde S^\mu = (0,0,0,\tilde B)$. Then we numerically solve EOM (\ref{eq:DL_EOM_III}) with it and zero initial velocity $\p_0\tilde\phi\big|_{t=0}=0$. We adopt the periodic boundary condition for both the $y$ and $z$ directions. The numerical box is $[-100,100]^2$ with $10^3\times 10^3$ lattice points. Namely, the spatial lattice size is $0.2$ and the temporal lattice size is $0.03$. We take $\tilde m = 0.1$ and choose the initial configuration with vortex separation $\tilde d = 31$ throughout the simulations in this subsection. On the other hand, we will vary $\kappa \tilde B$. From Eq.~(\ref{eq:criB_III}) the critical value for the infinitely large flat soliton is $\kappa \tilde B_{\rm III} \simeq 0.25$ for $\tilde m = 0.1$. Thus we naively expect that the finite-size soliton grows up for $\kappa \tilde B > 0.25$ whereas it shrinks for $\kappa \tilde B < 0.25$.

\begin{figure}[htbp]
    \centering
    \includegraphics[width=0.99\linewidth]{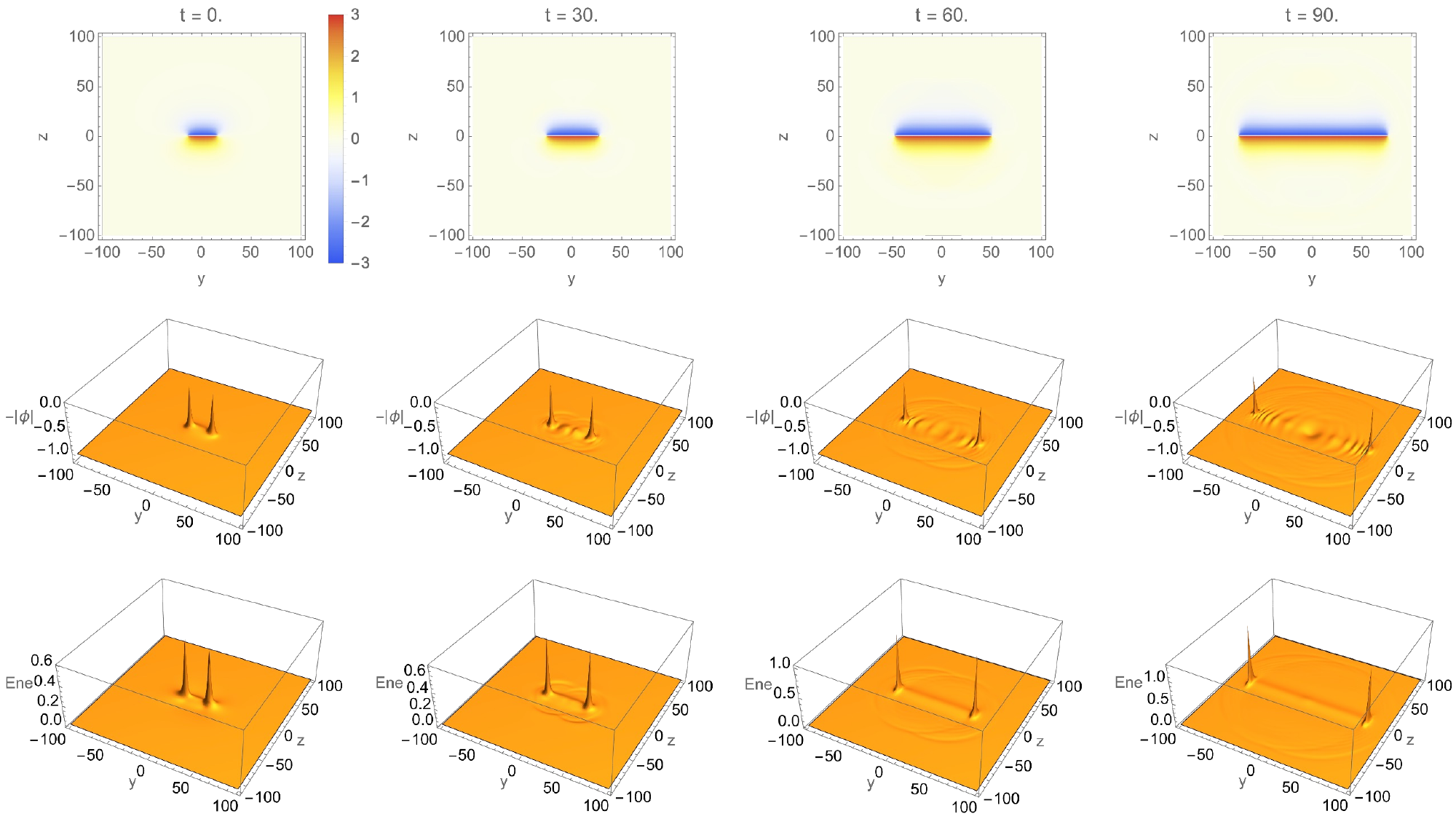}
  \caption{Dynamical evolution of the single soliton in the model III with $\tilde m = 0.1$ and $\kappa \tilde B=0.3$. The figures at the top row show $\arg\tilde \phi$, those in the middle row show $-|\tilde\phi|$, and those in the bottom row show the energy densities. The figures in the each column have the same $t = \{0,30,60,90\}$.}
  \label{fig:single_soliton_model_III_S0p3}
\end{figure}

Let us first consider the case with $\kappa \tilde B$ above $0.25$. 
We take $\kappa \tilde B = 0.3$ as an illustration.
Fig.~\ref{fig:single_soliton_model_III_S0p3} shows snap shots of $\arg \tilde\phi$, $-|\tilde\phi|$, and energy density at $\tilde t = 0$, 30, 60, and 90. The numerical simulation clearly shows that the soliton size (the separation between the vortex and anti-vortex) grows larger with time.  

\begin{figure}[htbp]
    \centering
    \includegraphics[width=0.4\linewidth]{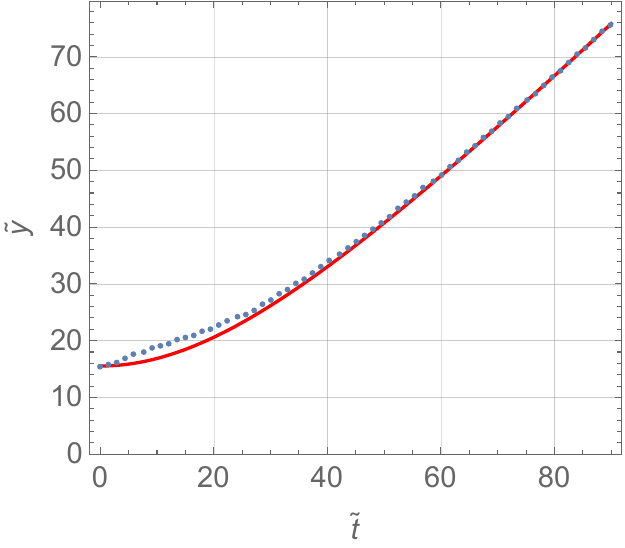}
  \caption{The position of the vortex at the edge of soliton as the function of $t$ in the model III with $\tilde m = 0.1$ and $\kappa \tilde B=0.3$. The dots are the numerical data, and the red curve is the motion of a relativistic particle with the constant acceleration $\alpha=0.027$.}
  \label{fig:acceleration_modelIII}
\end{figure}

We have several observations. The soliton has the negative tension under the presence of non-zero $B$. This means that the vortices at the edges feel a constant repulsive force (opposite to the confinement) by the linear soliton. Thus, the vortices are accelerated outward by the constant negative tension of the soliton. Indeed, we can see that the vortices initially have round shape, but over time they become flatter due to the Lorentz contraction. Furthermore, the motion of the vortex can be approximated quite well by the relativistic particle accelerated by a constant force along the $y$ axis:
\be
y_\pm(t) = \pm \frac{1}{\alpha}\sqrt{1+(\alpha t)^2} + y_\pm(0)\,,
\label{eq:relativistic_particle}
\ee
where $\alpha$ corresponds to the constant acceleration for an observer on the vortex. In order to confirm if this observation is correct or not, we plot the position of the vortex in Fig.~\ref{fig:acceleration_modelIII} and compare it with Eq.~(\ref{eq:relativistic_particle}). They agree quite well by taking $\alpha=0.027$.

\begin{figure}[htbp]
    \centering
    \includegraphics[width=0.99\linewidth]{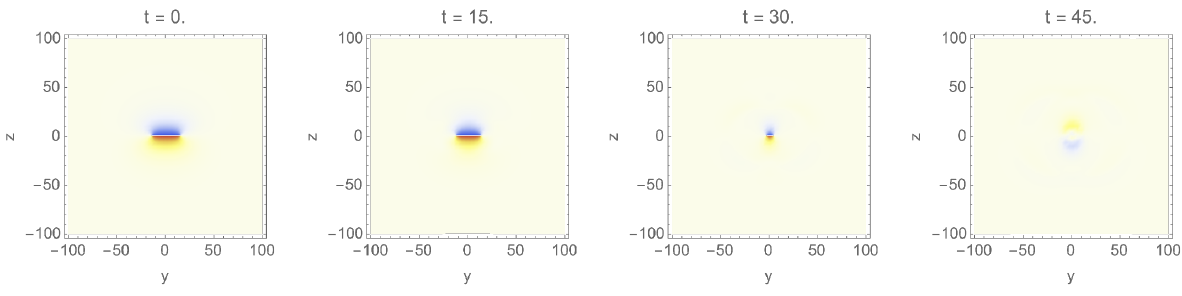}
  \caption{Dynamical evolution of the single soliton in the model III with $\tilde m = 0.1$ and $\kappa \tilde B=0.2$. The figures show $\arg\tilde \phi$ for $t = \{0,15,30,45\}$. The pair of vortex and anti-vortex annihilates.}
  \label{fig:single_soliton_model_III_S0p2}
\end{figure}

We next consider $\kappa \tilde B$ which is smaller than the critical value $0.25$. As an example, we choose $\kappa \tilde B = 0.2$. Since the soliton tension is positive, the vortex and anti-vortex at the edges of the soliton attract each other by a constant confinement force. They eventually annihilate as shown in Fig.~\ref{fig:single_soliton_model_III_S0p2}.

\begin{figure}[htbp]
    \centering
    \includegraphics[width=0.99\linewidth]{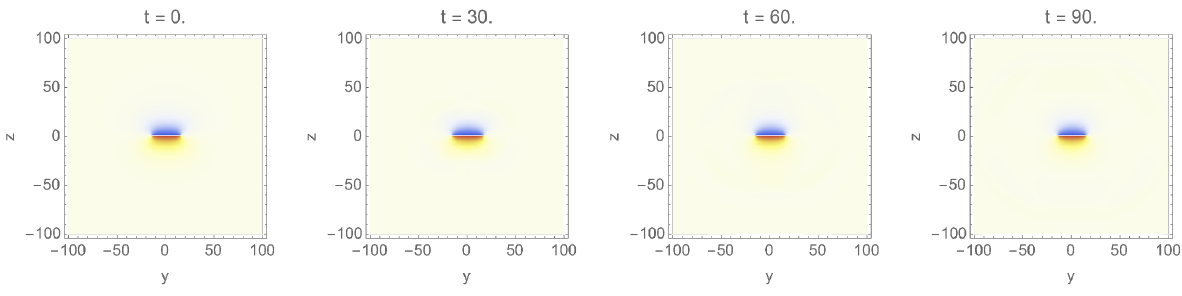}
  \caption{Dynamical evolution of the single soliton in the model III with $\tilde m = 0.1$ and $\kappa \tilde B=0.256$. The figures show $\arg\tilde \phi$ for $t = \{0,30,60,90\}$. The pair of vortex and anti-vortex almost does not move.}
  \label{fig:single_soliton_model_III_S0p256}
\end{figure}

We also examine the dynamics near the critical point. We find that the soliton almost does not move from the initial configuration when $\kappa\tilde B = 0.256$ as shown in Fig.~\ref{fig:single_soliton_model_III_S0p256} (we observed the soliton shrinks for $\kappa \tilde B = 0.255$ whereas it grows larger for $\kappa \tilde B = 0.257$). Thus the critical value found through the dynamical simulation is $\kappa \tilde B \simeq 0.256$ for $\tilde m = 0.1$. It is a good agreement with $0.25$ estimated previously.

\section{2D dynamical formation of chiral soliton lattices with dislocations} 
\label{sec:2d}

In this section, 
we characterize edge dislocations as topological defects or vortices, and numerically investigate the dynamical generation of CSL 
with dislocations
in the model III in two spatial dimensions. We consider a strong external current which is sufficiently larger than the critical value, and pursue the real time evolution of a randomly fluctuating initial state.

%%%%%%%%%%%%%%%%%%%%%%%
\subsection{Symmetry and topology of edge dislocations
and vortices}

Here let us discuss the spontaneous symmetry breaking in the CSL background. 
We should discuss it separately for the massless $m=0$ and massive $m\neq 0$ cases. 
Let us denote the translational symmetry in the lattice direction ($z$-direction) by ${\mathbb{R}_T}$ and ignoring the other directions and rotational symmetries since they are irrelevant. 
We also denote the phase shift of the scalar field $\phi = |\phi|e^{i\eta}$ by $U(1)_{\phi}$. The action of $U(1)_\phi$ is $\eta \to \eta + \theta$ with $\theta \in [0,2\pi)$. We use the universal covering group $\mathbb{R}_\phi$ whose action is defined by $\eta \to \eta + \alpha$ with $\alpha \in \mathbb{R}$. Having its discrete subgroup $\mathbb{Z}_\phi$: $\eta \to \eta + 2n\pi$ with $n \in \mathbb{Z}$, 
$U(1)_\phi $ %$= \mathbb{R}_\phi/\mathbb{Z}_\phi$ 
can be expressed as $U(1)_\phi \simeq \mathbb{R}_\phi/\mathbb{Z}_\phi$. We denote the action of $\mathbb{R}_T$ as $z \to z + \beta/c$ with $c$ being a nonzero constant given below and $\beta \in \mathbb{R}$.

In the massless case $m=0$, the relevant symmetry $G$ of the Lagrangian is 
\be
G_{m=0} = \mathbb{R}_T \times U(1)_\phi 
\simeq \mathbb{R}_T \times \frac{ \mathbb{R}_\phi}{\mathbb{Z}_\phi}
\simeq \frac{\mathbb{R}_{T+\phi} \times \mathbb{R}_{T-\phi}}{\mathbb{Z}_\phi}\,,
\ee
where $\mathbb{R}_{T+\phi}$ is a simultaneous transformation of $\mathbb{R}_T$ and $\mathbb{R}_\phi$ defined by $z \to z + \alpha/c$ and $\eta \to \eta - \alpha$ with $\alpha \in \mathbb{R}$, whereas $\mathbb{R}_{T-\phi}$ is a simultaneous  opposite transformation defined by $z \to z + \beta/c$ and $\eta \to \eta + \beta$ with $\beta \in \mathbb{R}$. 

In the massive case 
$m \neq 0$, $U(1)_\phi$ is explicitly broken by the scalar potential. Thus, the relevant symmetry $G$ of the Lagrangian is
\be
G_{m \neq 0} = \mathbb{R}_T\,.
\ee

Let us next clarify how the symmetry $G$ of the Lagrangian is spontaneously broken to a subgroup $H$ in the ground state for
the background field $B \ge B_{\rm c}$.
In the massless case $m=0$, the critical value is $B_{\rm c} = 0$. 
The EOM (\ref{eq:EOM_phi}) is analytically solved by
\be
\phi = v e^{i\eta}\,,\quad \eta = c z + \eta_0\,,\quad c\equiv \frac{\kappa B}{v^2}\,,
\label{eq:CSL_m=0}
\ee
with a constant $\eta_0$, and this is an inhomogeneous ground state for $B > 0$.
The translational symmetry ${\mathbb{R}_{T}}$ and $U(1)_\phi$ are broken while the simultaneous transformation $\mathbb{R}_{T+\phi}$ remains unbroken: 
$\eta = c z + \eta_0 \to c(z+\alpha/c) + \eta_0 - \alpha = c z + \eta_0$.
Therefore, the unbroken subgroup of $G_{m=0}$ is 
\begin{eqnarray}
   H_{m=0} = \frac{\mathbb{R}_{T + \phi} \times \mathbb{Z}_{T-\phi}
   }{\mathbb{Z}_\phi }
   (\subset G_{m=0}).
\end{eqnarray}
Therefore,
the order parameter manifold (OPM) 
parametrized by Nambu-Goldstone modes is 
\begin{eqnarray}
   m=0:\quad 
    M_{m=0} =  \frac{G_{m=0}}{H_{m=0}} \simeq 
    \left(
    \frac{\mathbb{R}_{T+\phi} \times \mathbb{R}_{T-\phi}}{\mathbb{Z}_\phi}
    \right) /
    \left(
    \frac{\mathbb{R}_{T+\phi} \times \mathbb{Z}_{T-\phi}}{\mathbb{Z}_\phi}
    \right)
    \simeq \frac{\mathbb{R}_{T-\phi}}{\mathbb{Z}_{T-\phi}}
 \simeq 
    S^1.
    \label{SSB:m=0}
\end{eqnarray}

In the massive case $m\neq 0$, the inhomogeneous ground state for $B > B_{\rm c}$ is
the CSL solution of the lattice constant $\ell$
\be
\phi = \rho(z) e^{i\eta(z)}\,,\quad \rho(z + \ell) = \rho(z)\,,\quad \eta(z +  \ell) = \eta(z) + 2\pi \,.
\ee
The translational symmetry $\mathbb{R}_T$ is  spontaneously broken as $\mathbb{R}_T \to 1$.
However, the discrete shift symmetry $\mathbb{Z}_{T+\phi}$: $z \to  z + n\ell$ and $\eta \to \eta - 2\pi n$ with $n\in \mathbb{Z}$ keeps $\rho$ and $\eta$ unchanged due to the periodicity of the ground state.
Namely, the unbroken subgroup is 
\begin{eqnarray}
  H_{m\neq0} 
= 
\frac{\mathbb{Z}_{T+\phi} \times \mathbb{Z}_{T-\phi}}{\mathbb{Z}_\phi}
\simeq \frac{\mathbb{Z}_{T} \times \mathbb{Z}_{\phi}}{\mathbb{Z}_\phi}
\simeq \mathbb{Z}_T 
(\subset G_{m\neq0})\,,
\end{eqnarray}
where $\mathbb{Z}_{T-\phi}$ is another discrete symmetry defined by
$z \to  z + m\ell$ and $\eta \to \eta + 2\pi m$ with $m\in \mathbb{Z}$.
Then, 
the OPM becomes 
\begin{eqnarray}
  m\neq 0: \quad  
  M_{m\neq 0} 
  =  \frac{G_{m\neq 0}}{H_{m\neq 0}} \simeq 
  \frac{{\mathbb{R}_T}}{{\mathbb{Z}_{T}}}\simeq S^1.
  \label{SSB:m=/=0}
\end{eqnarray}

For both the massless and massive cases, 
the OPM is $S^1$ characterized by
a nontrivial fundamental group 
\begin{eqnarray}
    \pi_1 (M_{m=0}) \simeq
     \pi_1 (M_{m\neq 0}) \simeq
    {\mathbb Z}.
    \label{eq:homotopy}
\end{eqnarray}
This admits vortices for $m=0$ or dislocations for $m\neq 0$ in the inhomogeneous ground state as we will explain below separately.

In the massless case $m=0$, 
a vortex  exists alone as a topological object.
It is characterized by the winding number associated with $\pi_1(\mathbb{R}_{T-\phi}/\mathbb{Z}_{T-\phi}) \simeq \mathbb{Z}$.
Let $C$ be a small circle encircling the vortex. 
Then, we count the winding number as
\be
\oint_C d\vec\ell\cdot \nabla \arg\phi = \pm 1\,\label{eq:winding}
\ee
where $d\vec\ell$ is the line element along $C$, and $+1$ is for the vortex while $-1$ is for the antivortex.
Note that here we only look at the phase $\eta = \arg \phi$ but the OPM is $S^1 \simeq \mathbb{R}_{T-\phi}/\mathbb{Z}_{T-\phi}$. This mismatch does not matter because the linear $z$ dependence of the background in Eq.~(\ref{eq:CSL_m=0}) can be always absorbed by the field redefinition $\tilde \phi = e^{-icz} \phi$. For the new field $\tilde\phi$, $c$ is interpreted as a mass, and Eq.~(\ref{eq:CSL_m=0}) becomes a homogeneous solution $\arg\tilde\phi = \eta_0$. Hence, the OPM and the vortices in the backgrounds $\eta = \eta_0$ and $\eta = cz + \eta_0$ are physically equivalent.

\begin{figure}[htbp]
    \centering
    \includegraphics[width=0.99\linewidth]{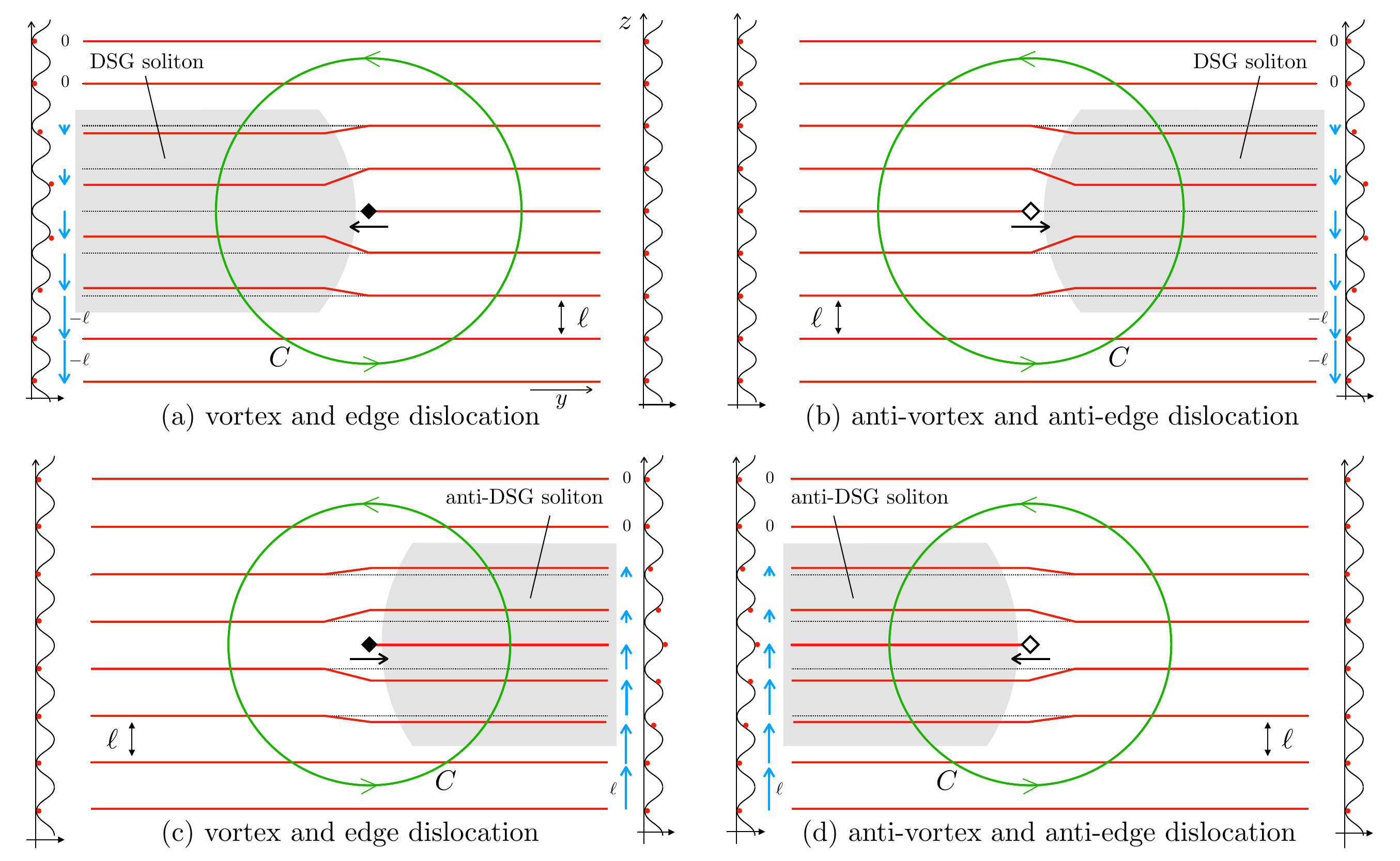}
  \caption{
  The red curves show the center of solitons identified by $\eta = (2n+1)\pi$. The black (white) diamond in (a) and (c) [(b) and (d)] stands for the (anti-)vortex which in the CSL is the (anti-)edge dislocation. 
  The effective potential induced in the lattice is plot on the left and right side of each subfigure, where the red points indicate the chiral solitons 
   and the blue arrows represent the shift $\delta z$ of the solitons from the bottoms.
  The thick gray shaded region in (a) and (b) [(c) and (d)] stands for the (anti-)discrete sine-Gordon (DSG) soliton induced by a mismatch between the soliton positions and the bottoms of the effective potential. The DSG soliton
  pulls the edge dislocation toward the direction indicated arrow below the diamond. Note that the scalar field configurations
  (a) and (c) [(b) and (d)] are the same, however, the strength of the external field are different. 
  The lattice spacing of the CSL is the ground state at the right and left parts of configurations  (a) and (c) [(d) and (b)], respectively, 
  and consequently the directions of the (anti-)DSG solitons are opposite to each other.
  }
  \label{fig:new_pi1_dislocation}
\end{figure}

In the massive case $m\neq 0$, however, physics is different.
First, the topological charge in Eq.~(\ref{eq:winding}) 
responsible for 
$\pi_1 (M_{m=0})  \simeq
    {\mathbb Z}$ in 
    Eq.~(\ref{eq:homotopy})
is not conserved under the presence of the explicit $U(1)$ symmetry breaking term. 
The vortex does not exist alone but is inevitably attached by a soliton, as the same as the case of axion strings with the domain wall number one.
The fact that the vortex and soliton cannot be separated implies that neither vortices nor solitons are truly topological objects, but they are only approximately topological.
Eq.~(\ref{eq:winding}) simultaneously counts the number of solitons crossing the contour $C$. This can be schematically shown in Fig.~\ref{fig:new_pi1_dislocation}
. For example, in the panel (a), the green contour $C$ encircles a single vortex (black diamond) and crosses seven solitons upwards and six solitons downwards. 
In general, the number $N_{\rm v}$ of vortex and the numbers 
$N_{\text{s}}$ ($N_{\text{as}}$) 
of (anti-)soliton are related by
\be
N_{\rm v} = N_{\text{s}} - N_{\text{as}}\,.
\ee

\begin{figure}[htbp]
    \centering
    \includegraphics[width=0.8\linewidth]{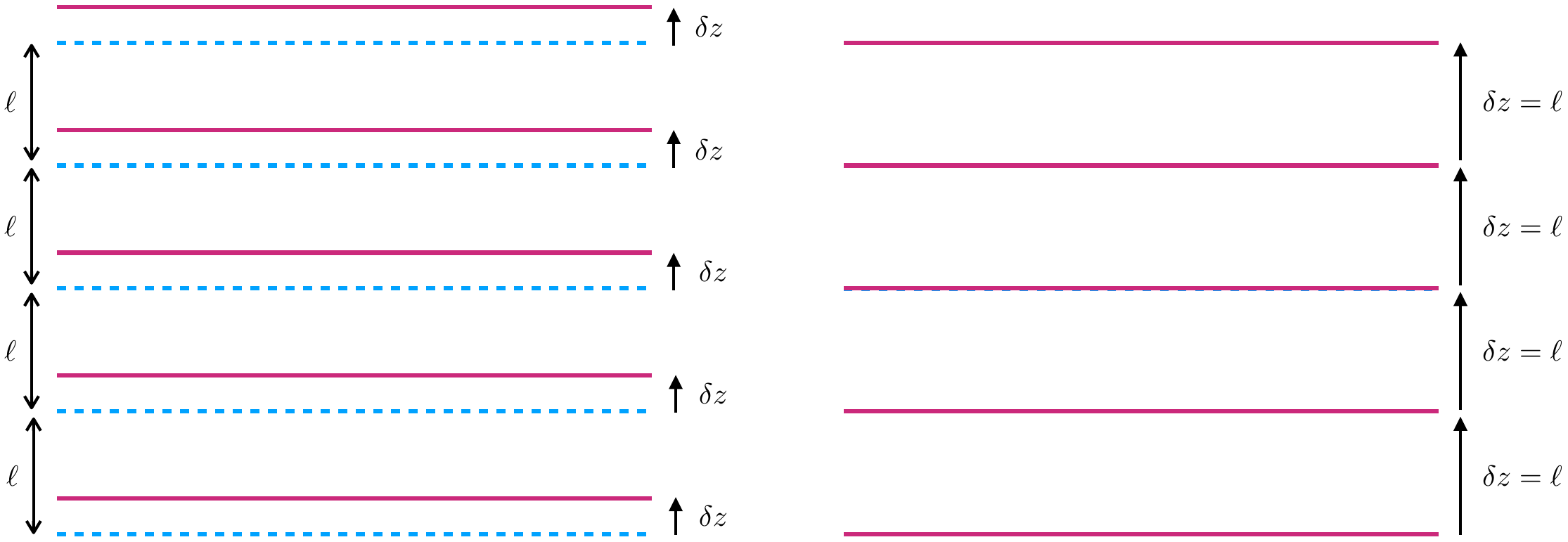}
  \caption{The translation of CSL with the period $\ell$ by $\delta z$ (left) and $\delta z = k\ell$ with $k \in \mathbb{Z}$ (right).}
  \label{fig:CSL_S1}
\end{figure}

All of the above is valid for both the homogeneous vacuum for $B < B_{\rm c}$ and the CSL background for $B > B_{\rm c}$.
However, the vortex-soliton composite in the homogeneous vacuum and that in the CSL background should be  distinguished by means of the fundamental group of the pure translational symmetry breakings 
instead of the full symmetry breakings discussed above: The former is trivial ($\mathbb{R}_T \to 1$) but the latter is nontrivial due to $\pi_1(\mathbb{R}_T/\mathbb{Z}_T) \simeq \mathbb{Z}$ as in Eq.~(\ref{eq:homotopy}).
We call the topological defect characterized by $\pi_1(\mathbb{R}_T/\mathbb{Z}_T)$ the edge dislocation in this paper.
Let $\delta z$ be a coordinate of OPM $M_{m \neq 0} \simeq S^1$. Namely, it is a constant shift $\delta z$ with the identification $\delta z \sim \delta z + \ell$ as schematically explained in Fig.~\ref{fig:CSL_S1}.
Now we are ready to clarify topological nature of the edge dislocations.\footnote{ 
Identification of dislocations in atomic crystals as topological defects dates back to 1990 \cite{YISHI1990689}. }
In the CSL background without the edge dislocations the constant shift $\delta z$ is everywhere zero. In contrast, when the edge dislocation exists,
$\delta z$ is a nontrivial function of $y$ and $z$ as $\delta z(y,z)$. Hence, $\delta z$ on a closed curve $C$ in the $yz$ plane gives a map from the loop $C$ to the OPM $M_{m \neq 0} \simeq S^1$. 
The map is characterized by the fundamental group
\be
\pi_1\left(M_{m \neq 0}\right) \simeq \mathbb{Z}\,.
\label{eq:pi1_massive}
\ee 
To be more concretely, let us consider a configuration shown in Fig.~\ref{fig:new_pi1_dislocation}(a) where the black diamond represents the vortex and the red curves correspond to the soliton centers [$\eta = (2n+1) \pi$].
We assume that the solitons on the right side of the vortex are placed in the correct positions, namely $\delta z = 0$,
for the ground state in the CSL 
with a particular strength  of the external field.
On the other hand, those on the left side which are closed to the vortex are shifted along $z$ axis, 
namely $\delta z \neq 0$,
because one soliton is missing. The blue arrow on the left side of the panel (a) represents the shift $\delta z$ of the soliton beside it.\footnote{The longest vector $-\ell\hat e_y$ is called the Burgers vector of an edge dislocation that describes the magnitude and direction of the lattice distortion caused by the dislocation.} Let us pursue $\delta z$ on the loop $C$. Clearly, $\delta z \neq 0$ appears only when $C$ passes through the gray shaded region, where $\delta z$ varies continuously from $0$ to $-\ell$ from top to bottom.
Due to the identification $\delta z \sim \delta z - \ell$, the solitons below the gray  region having $\delta z = - \ell$ are regarded as they are not shifted. Therefore, 
when we travel  along $C$ clockwise once, 
we  go around $M_{m\neq 0} \simeq S^1$ counterclockwise once.  
 We count the corresponding winding number as $+1$, and it corresponds to the number of the edge dislocation
\be
N_{\rm DS} = 1\,.
\label{eq:N_DS=1}
\ee

Note that the edge dislocation is accompanied with a lattice soliton called a 
discrete sine-Gordon (DSG) soliton \cite{frenkel1939theory,Peierls_1940,nabarro1947dislocations,Braun:1998, Braun2004}. 
In the panel (a) it is depicted by the gray shaded region which includes several chiral solitons and ends on the dislocation. The DSG soliton costs energy. This can be intuitively explained as follows. 
First, let us consider the effective   
periodic potential 
of the period $\ell$ along the $z$ axis, which is   
effectively induced on the lattice structure by the soliton-soliton interaction \cite{Peierls_1940,
nabarro1947dislocations}. 
We denote it on the left and right of each panel in Fig.~\ref{fig:new_pi1_dislocation}.
There, the red dots represent the positions of the chiral solitons,
and all the solitons sit on the bottoms of the effective potential in the CSL ground state. 
The displacement of the solitons  from the ground state can be described by the DSG or Frenkel-Kontorova model 
\cite{frenkel1939theory, Braun:1998, Braun2004}.
Now, let us go back to the configuration (a). 
One can find that the solitons on the right side of the vortex is in the CSL ground state.
On the other hand, 
the chiral solitons on the left side must be deviated from the CSL background
 because the number of soliton is less than that of the CSL ground state on the right side. 
 Therefore, the chiral solitons in the gray shaded region have to cross the effective potential barrier once due to the missing soliton.
This excited state in the CSL background 
is nothing but the DSG soliton.
Therefore, the DSG soliton costs energy due to the effective potential and it pulls the edge dislocation to the left in the panel (a).

The configuration in the panel (b) is obtained by the parity transformation $y\to -y$ of (a), so that the white diamond represents the anti-vortex and also the dislocation is replaced by the anti-dislocation because the associated edge dislocation number is $N_{\rm DS} = -1$. Note that two more different configurations exist as shown in the panels (c) and (d). The scalar field configuration in (c) is the same with that in (a).
However, the strength of the external field is different from that of (a) so that the left side of the vortex becomes the CSL background;
the solitons on the left side are placed in the correct positions with $\delta z = 0$ at the bottoms of the effective potential.
In this case, those on the right hand are shifted because an additional soliton is inserted compared with the CSL background.
Again, $\delta z$ is non-zero only in the gray shaded region, which continuously changes from $\delta z = \ell$ to $\delta z = 0$ for one clockwise cycle along $C$. This gives $N_{\rm DS} = +1$, so that this configuration has the edge dislocation as the same as (a). However, the anti-DSG soliton extends to the opposite direction.

The same relation holds between the configurations (b) and (d). 
Although their scalar configurations are the same, the strengths of the external fields are different.
In (b) the left side is the CSL background and the DSG soliton is attached to the anti-vortex, 
while in (d)  the right side is the CSL background and the anti-DSG soliton is attached to the anti-vortex.

We have explained that the configurations (b) and (d) are obtained by the parity transformation $y \to -y$ of the configurations (a) and (c), respectively. These are energetically degenerated. One can also take another parity transformation $z \to -z$ of the configurations (a), (b), (c), and (d). This exchanges a vortex and an anti-vortex, and also an edge dislocation and an anti-edge dislocation, whereas the DSG and anti-DSG remain unchanged.
However, all these are stable only in the external field with $B<0$ because the CSL is transformed to the anti-CSL in this transformation.

One can understand that in the massless case $m=0$, no DSG soliton is attached to the dislocation and they are merely vortices without DSG solitons.

\subsection{Solutions of edge dislocations in two dimensions}
\begin{figure}[htbp]
    \centering
    \includegraphics[width=0.99\linewidth]{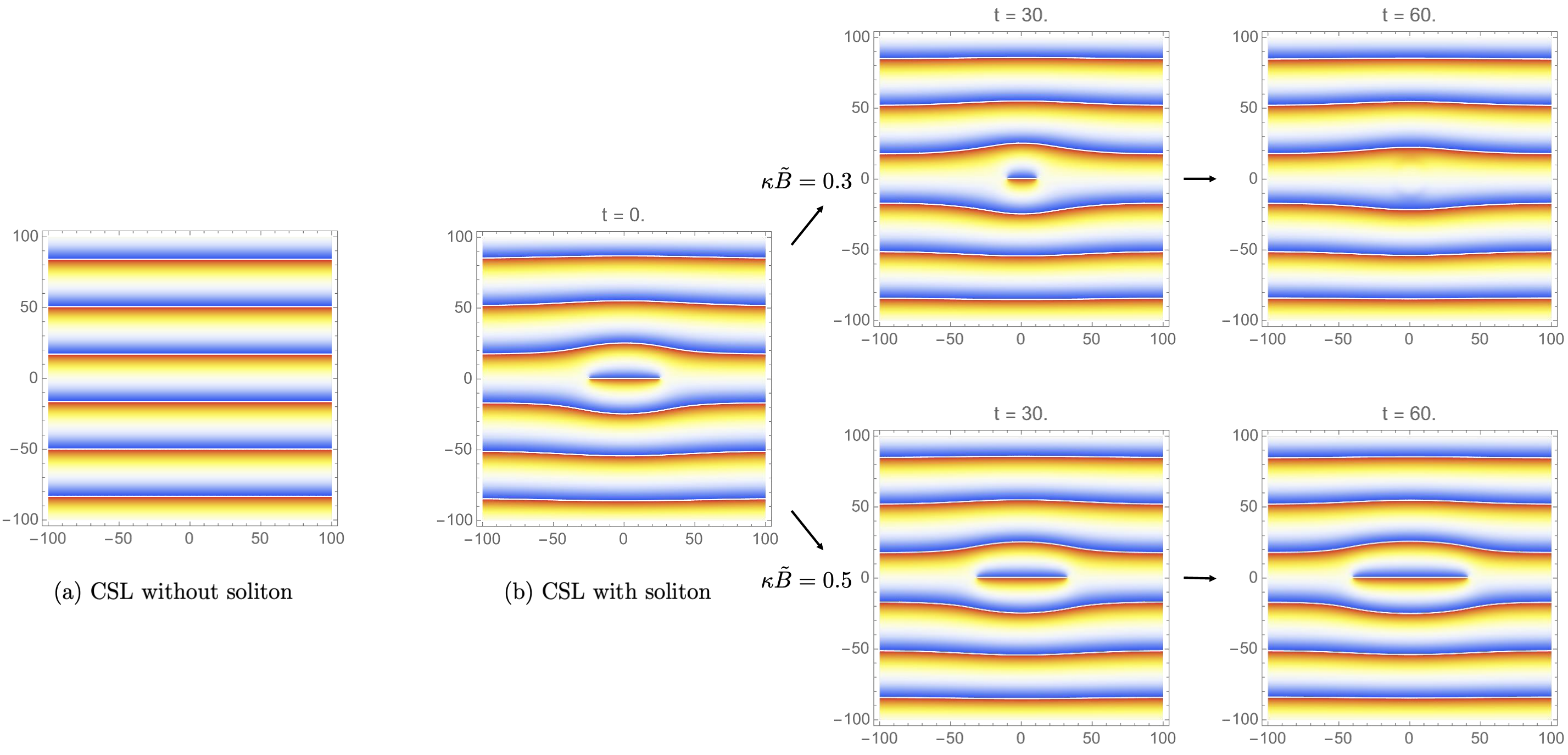}
  \caption{(a) The CSL without finite solitons. (b) The CSL with a finite soliton, and its dynamical evolution. The parameters are $\tilde m = 0.1$ and $\kappa \tilde B=0.3$ and $0.5$. The figures show $\arg\tilde \phi$. }
  \label{fig:soliton_in_CSL}
\end{figure}

We studied dynamics of the single finite soliton in homogeneous vacuum in Sec.~\ref{sec:modelIII_single_finite_soliton}. Below in Sec.~\ref{sec:2d_CSL_formation}, we observe that the generation and evolution of the single finite soliton occur 
in the process of CSL formation.
In this subsection, we numerically construct edge dislocations in 
the CSL ground state in the model III.
For this purpose, we construct a pair of dislocation and anti-dislocation, resulting in a single finite soliton stretching this pair.
Let us consider the model III with $\tilde m = 0.1$, and prepare the CSL that has 6 solitons aligned along the $z$ axis within $\tilde z \in [-100,100]$, 
as shown in Fig.~\ref{fig:soliton_in_CSL}(a).  
Then, we insert a finite soliton of the size $\tilde d = 50$ between the third and fourth solitons of CSL as shown in Fig.~\ref{fig:soliton_in_CSL}(b). We take this as the initial configuration and numerically solve the EOM (\ref{eq:DL_EOM_III}) for the two cases with $\kappa \tilde B = 0.3$ and $0.5$. Note that both the values are larger than the critical value $\kappa \tilde B = 0.256$ with the same $\tilde m = 0.1$. Nevertheless, the finite soliton in the CSL background shrinks and annihilates for the case of $\kappa \tilde B = 0.3$. This indicates that the CSL background relatively diminishes the generation of solitons in comparison with the homogeneous vacuum. For the larger case with $\kappa \tilde B = 0.5$ the finite soliton grows up like the previous case of the vacuum background. In general, the critical value $\kappa \tilde B$ for generating new soliton depends on the background configuration.

%%%%%%%%%%%%%%%%%%%%%%

\subsection{Dynamical formation of chiral soliton lattices in two dimensions}
\label{sec:2d_CSL_formation}

\begin{figure}[htbp]
    \centering
    \includegraphics[width=0.99\linewidth]{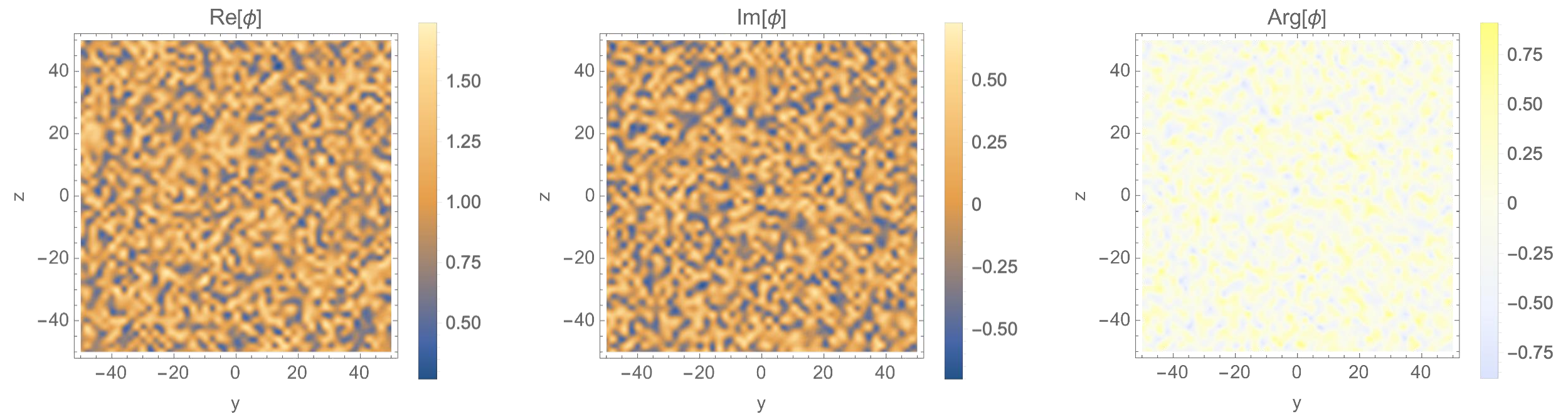}
  \caption{The initial configuration $\tilde \phi$ for the numerical simulations in this subsection.}
  \label{fig:random_ini_2d}
\end{figure}

In this subsection, we study crystallization dynamics of CSL with dislocations in 
two spatial dimensions $(y,z)$ ignoring the $x$ direction. We first prepare a random distribution of $\phi$ as an initial configuration for our numerical study. We set our initial configuration as $\phi = v' + \delta$ with $\delta\sim {\cal O}(v'/2)$, see Fig.~\ref{fig:random_ini_2d} for the concrete example.
We numerically solve the EOM (\ref{eq:DL_EOM_III}) of the model III. We adopt a periodic boundary condition for both the $y$ and $z$ directions. To suppress random noises that persist during the simulation, we modify the equation by adding a small diffusion term on the right-hand side as
\be
\tilde\p^2\tilde\phi + \frac{1}{2}\left(|\tilde\phi|^2-1\right)\tilde\phi - \tilde m^2 + i \kappa \tilde B \tilde\p_z \tilde\phi = \epsilon \tilde\p_t\tilde\phi\,.
\label{eq:DL_EOM_III_modify}
\ee
In the following simulations,
we will fix $\tilde m = 0.1$ and $\epsilon = 0.05$,
and take two different values of $\kappa \tilde B = 1$ and $2$ which are sufficiently large compared to the critical value $0.25$.
The size of a numerical box is $[-50,50]^2$ with $500^2$ lattice points.
Namely, the spatial lattice size is $0.2$ and the temporal lattice size is $0.03$.

\begin{figure}[htbp]
    \centering
    \includegraphics[width=0.99\linewidth]{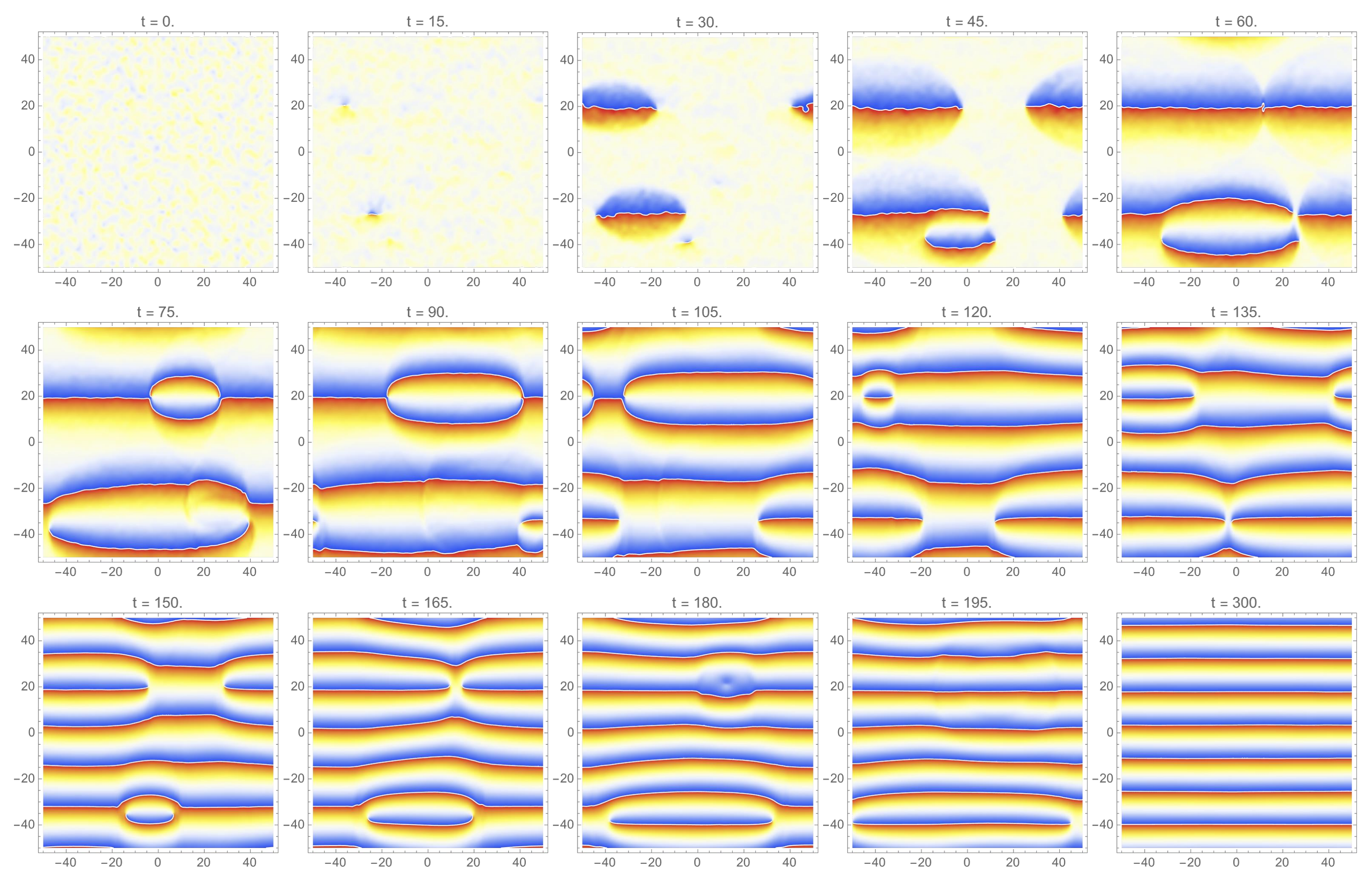}
  \caption{Dynamical evolution of the single soliton in the model III with $\tilde m = 0.1$ and $\kappa \tilde B=1.0$. The figures show $\arg\tilde \phi$ at several $t$. The configuration at $t=0$ is identical to that given in Fig.~\ref{fig:random_ini_2d}. The horizontal axis is $\tilde y$ and the vertical one is $\tilde z$.}
  \label{fig:2d_random_L50_S1p0_alpha0p05}
\end{figure}

\begin{figure}[htbp]
    \centering
    \includegraphics[width=0.99\linewidth]{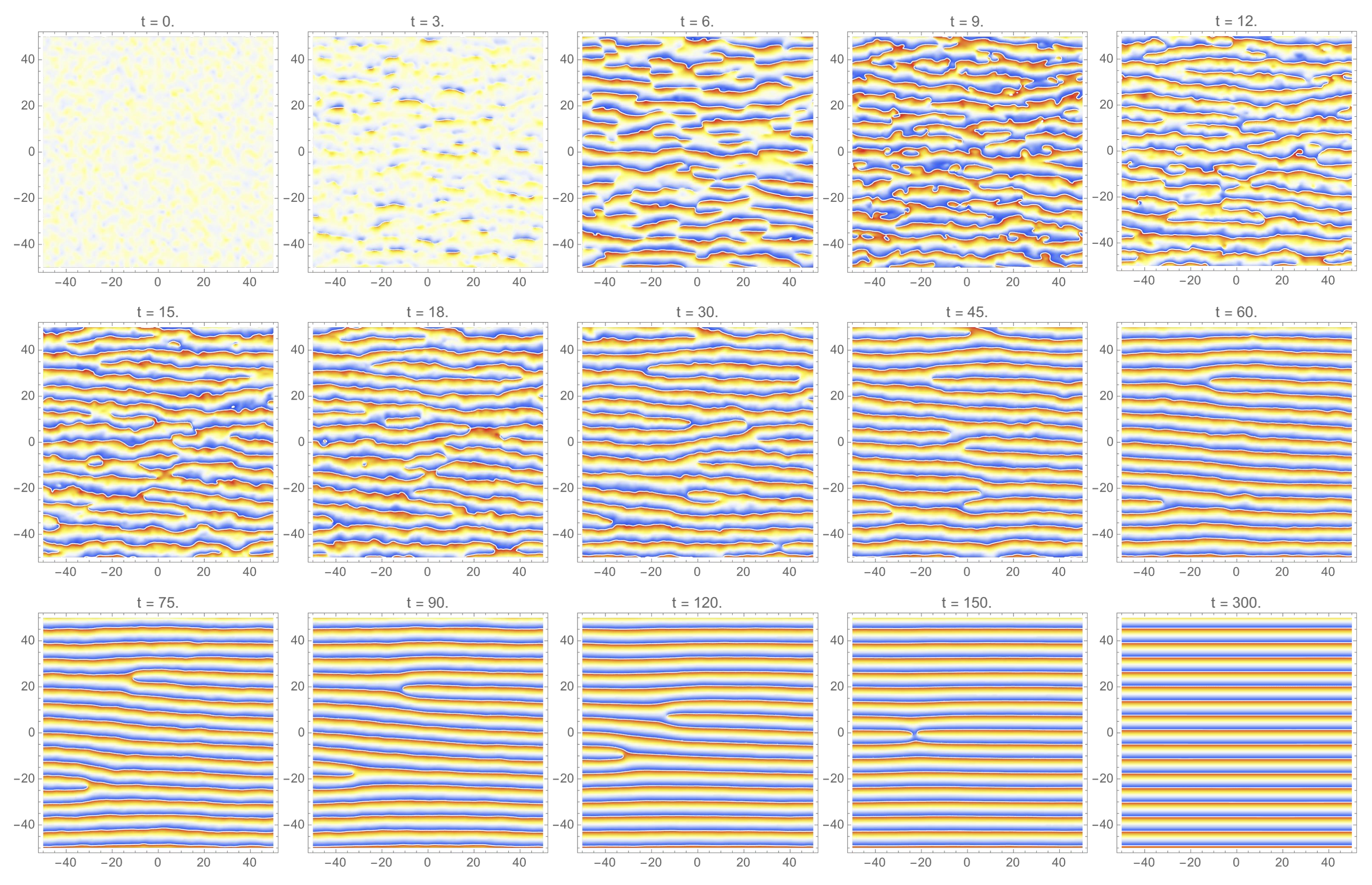}
  \caption{Dynamical evolution of the single soliton in the model III with $\tilde m = 0.1$ and $\kappa \tilde B=2.0$. The figures show $\arg\tilde \phi$. }
  \label{fig:2d_random_L50_S2p0_alpha0p05}
\end{figure}

We show the results of the numerical simulations for $\kappa\tilde B = 1$ and $2$ in Figs.~\ref{fig:2d_random_L50_S1p0_alpha0p05} and \ref{fig:2d_random_L50_S2p0_alpha0p05}, respectively.
Starting from the same initial configuration at $t=0$ for both the simulations (see Fig.~\ref{fig:random_ini_2d}), we obtain different final CSL states.
The final CSL for $\kappa \tilde B = 1$ has 7 solitons in $2\tilde L = 100$ whereas the one for $\kappa\tilde B = 2$ has 16 solitons.

There is a general caveat to our numerical simulations: We observed that the numerical simulation is sensitive to various factors such as the random initial configuration, the box size $\tilde L$, the artificial diffusion coefficient $\epsilon$, the boundary condition, and so on. Therefore, we should not extract a typical phenomenon from only one simulation.
Instead, we should focus on phenomena commonly observed in various simulations. 
Let us list them in order: 1) In the early stages of the simulation, several small vortex-antivortex pairs appear as seeds of solitons, and they grow along the $y$-axis. The larger the external field $\kappa\tilde B$, the more seeds will emerge. In the middle stages of simulation, a left-moving vortex at the left end of a soliton and a right-moving antivortex at the right end of another soliton at the same $z$-height on the right collide and annihilate, causing the two solitons to merge. See for example the third soliton from the top of the panels at $t=165$ and $t=180$ of Fig.~\ref{fig:2d_random_L50_S1p0_alpha0p05}. The expanding motion of a soliton
can stimulate the emergence of another seed around it. See the panels at $t=30$ and $t=45$ of Fig.~\ref{fig:2d_random_L50_S1p0_alpha0p05} where one can observe the small third seed appears near the right edge of the large soliton in the left-bottom region. The annihilation of a vortex and an antivortex can generate another soliton, increasing the number of solitons. This occurs, for example, on the first soliton from the top in the panels $t=45$, $60$, and $75$ of Fig.~\ref{fig:2d_random_L50_S1p0_alpha0p05}. 
At the final stage of the simulation, the system settles to a horizontal CSL. The larger the external magnetic field, the greater the final number of solitons and the shorter the time to reach equilibrium.

\subsection{Dyamics of edge dislocations}

\subsubsection{The motion perpendicular to the external field}

\begin{figure}[htbp]
    \centering
    \includegraphics[width=0.8\linewidth]{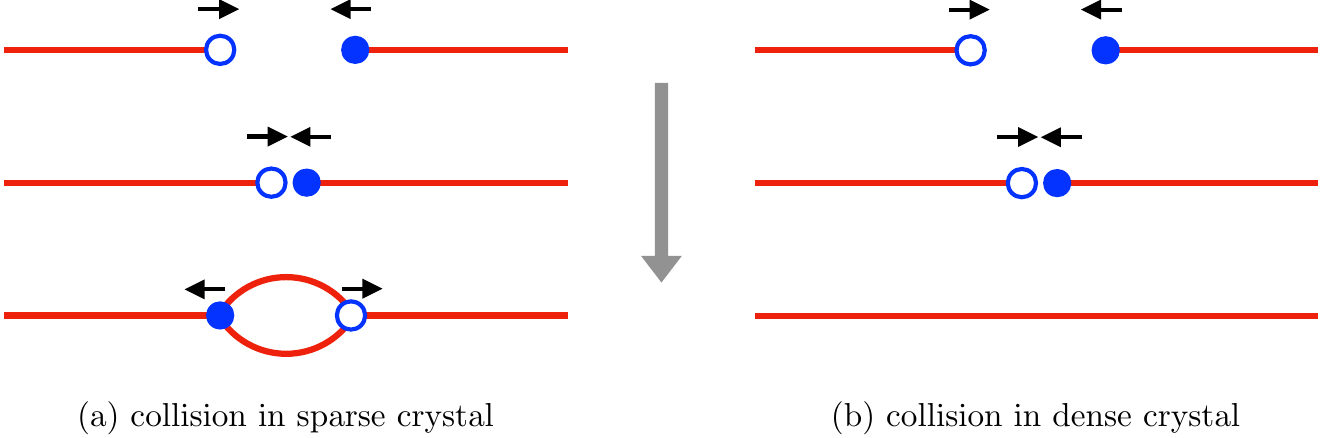}
  \caption{Collisions of dislocation and anti-dislocation. (a) The dislocations pass through leaving two solitons between them. (b) The dislocations annihilate leaving one soliton.}
  \label{fig:collision_edge_dislocation}
\end{figure}

The most basic motion of an edge dislocation is a straight line motion perpendicular to the external field as we have seen in Figs.~\ref{fig:2d_random_L50_S1p0_alpha0p05} and \ref{fig:soliton_in_CSL}.
Edge and anti-edge dislocations move in opposite directions, so if they are at the same $z$ level, they will collide head-on.
What happens after the collision depends on the environment.
Let us take another look at Fig.~\ref{fig:2d_random_L50_S1p0_alpha0p05} again.
Because the soliton crystal is sparse in the initial stage, dislocations move relatively fast, see the evolution of the finite soliton at the top of the panels with $t=45$, $60$, $75$ of Fig.~\ref{fig:2d_random_L50_S1p0_alpha0p05}. The dislocation passes through, leaving two solitons between them. We illustrate this phenomenon in Fig.~\ref{fig:collision_edge_dislocation}(a). This is one mechanism by which solitons multiply to form dense crystal.
A different outcome can be seen in the collision in the final stage.  Since the soliton crystal is dense, the dislocations move relatively slow and they annihilate leaving one soliton, see the panels with $t=150$, $165$, $180$ of Fig.~\ref{fig:2d_random_L50_S1p0_alpha0p05}.
Fig.~\ref{fig:collision_edge_dislocation}(b) illustrates the pair annihilation. This process finalizes formation of the ideal soliton crystal.

\begin{figure}[htbp]
    \centering
    \includegraphics[width=0.8\linewidth]{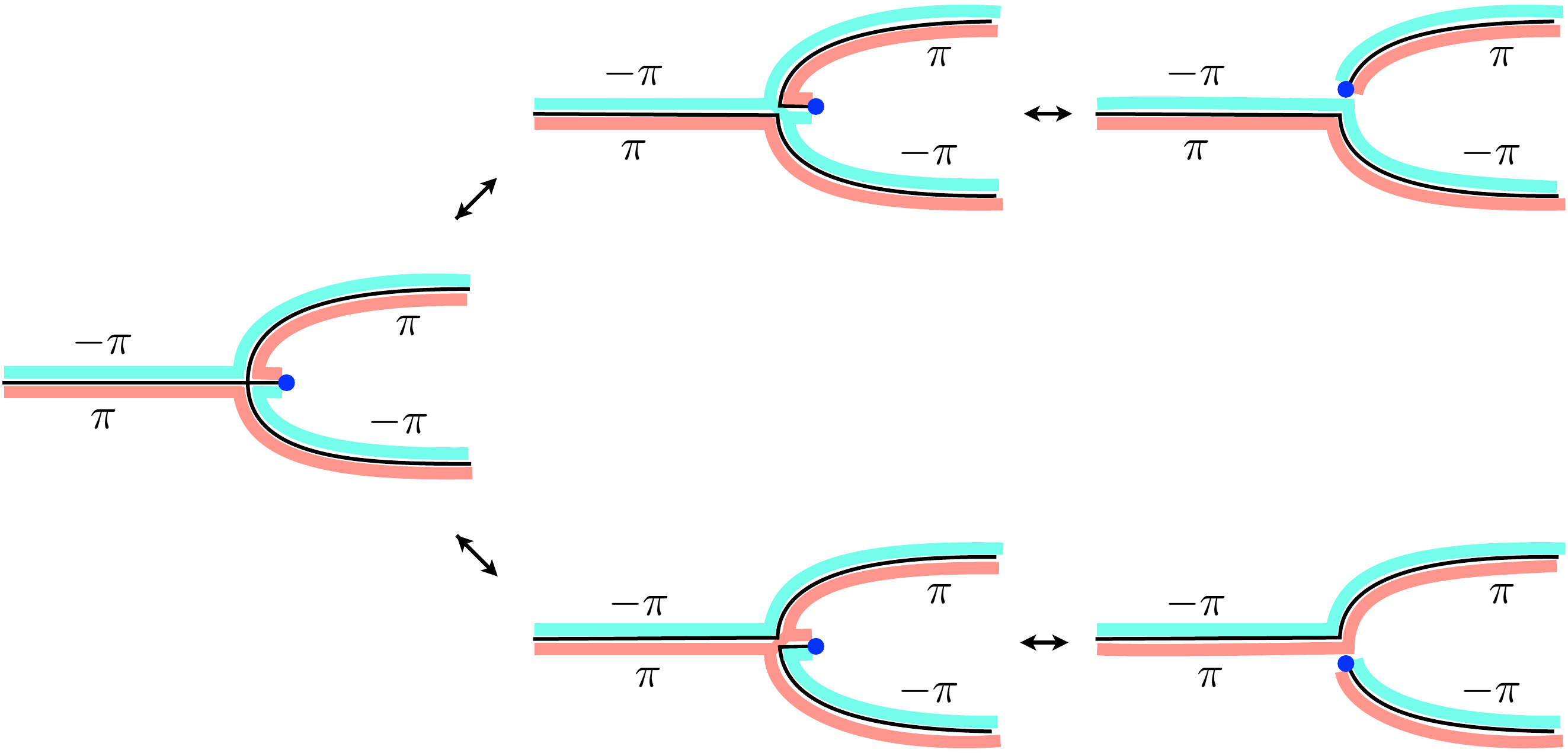}
  \caption{The Y-junction: The composite of solitons with and without an edge dislocation. The black solid curve stands for the solitons that are boundaries between $\theta = \pi$ (pink) and $-\pi$ (cyan). The transition can be understood as pair annihilation of the soliton and anti-soliton.
  }
  \label{fig:Y_junction}
\end{figure}

Next, let us describe the Y-junction appearing in Fig.~\ref{fig:collision_edge_dislocation}(a) in more detail.
Note that only one soliton can attach to each  edge dislocation. The Y-junction seems to violate this rule since it terminates three solitons. %But this is not true. 
However, we observe that an approximate Y-junction is possible: by closely looking at the Y-junction, we observed that it consists of one long soliton and one dislocation as illustrated in the top right-most and the bottom right-most figures of Fig.~\ref{fig:Y_junction}. These two configurations can continuously transit to each other % through the intermediate states 
through a reconnection or 
a pair annihilation of soliton and anti-soliton as shown in Fig.~\ref{fig:Y_junction}. 

\subsubsection{The motion parallel to the external field}

\begin{figure}[htbp]
    \centering
    \includegraphics[width=0.99\linewidth]{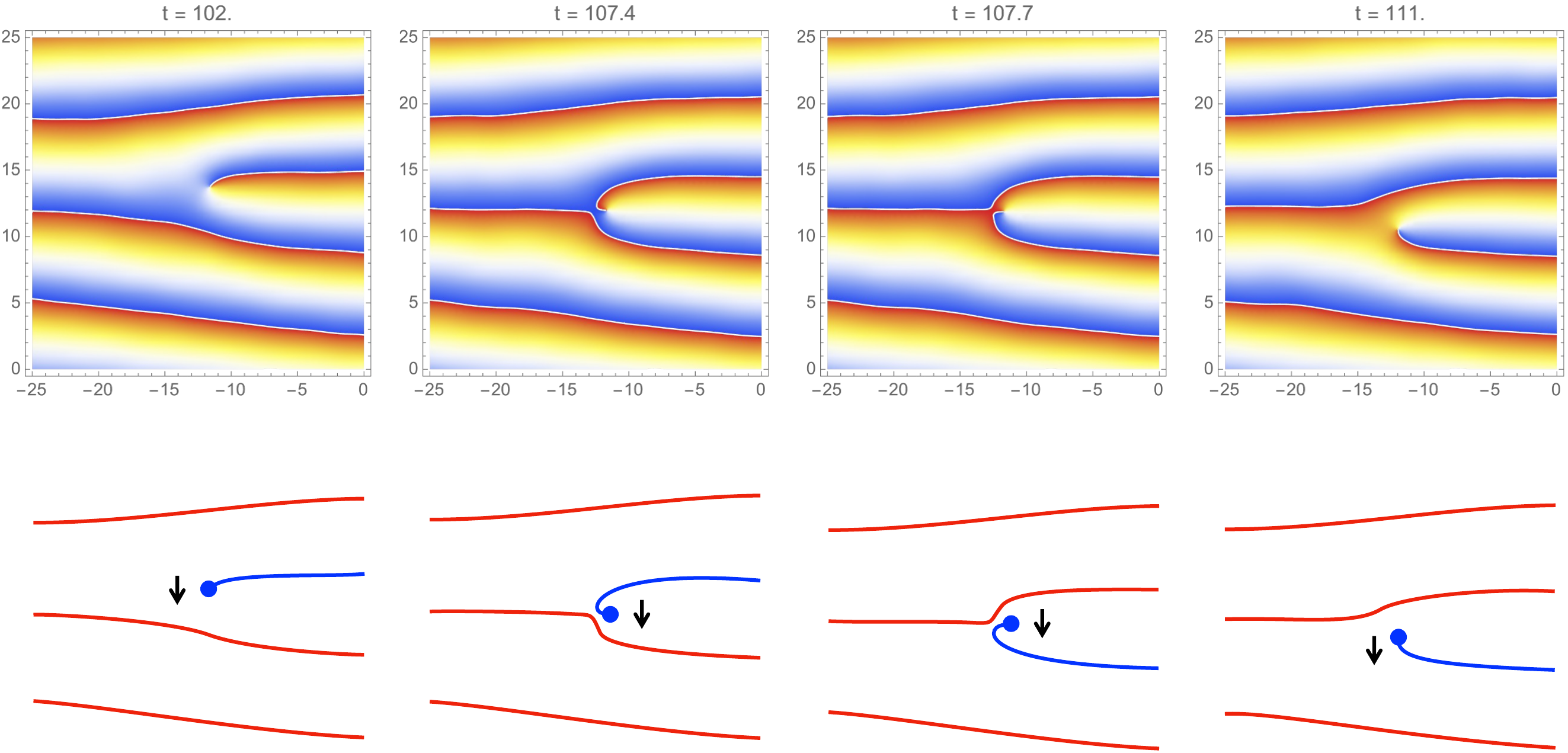}
  \caption{
  The movement of an edge dislocation in the direction of the CSL.
  The first line is an enlarged portion of Fig.~\ref{fig:2d_random_L50_S2p0_alpha0p05} at $t=102$, $107.4$, $107.7$ and $111$. The second line shows the illustration of how the dislocation moves parallel to the external field.}
  \label{fig:hopping_edge_dislocation}
\end{figure}

We also observed the edge dislocation moving parallel to the external field (orthogonal to the soliton lattice). See, for example, the panels with $t=60$, $75$, $90$, and $120$ of Fig.~\ref{fig:2d_random_L50_S2p0_alpha0p05}.
The edge dislocation hops downward and the anti-edge dislocation hops upward from one soliton to another. Fig.~\ref{fig:hopping_edge_dislocation} is an enlarged portion of Fig.~\ref{fig:2d_random_L50_S2p0_alpha0p05} to show the dislocation hopping. As illustrated in Fig.~\ref{fig:hopping_edge_dislocation}, 
a Y-junction shown in Fig.~\ref{fig:Y_junction} is formed during the hopping. The edge and anti-edge dislocations come to the same layer after the hopping several times, and they finally annihilate to form a pure crystal. This motion is very similar to the motion of crystal dislocations within a crystal as illustrated in Fig.~\ref{fig:hopping_edge_dislocation_atom} in Appendix \ref{sec:atomic_crystal}.
During this hopping, 
the dislocation will feel 
the so-called Peierls-Nabarro potential 
\cite{Peierls_1940,
nabarro1947dislocations}.
%%%%%%%%

%%%%%%%%%%%%%%%%%
\section{3D dynamical formation of chiral soliton lattices with dislocations}
\label{sec:3D}

In this section, we study dislocations in CSL in the model III in three spatial dimensions. 
In this dimensionality, 
there are two kinds of dislocations: one is an edge dislocation as the same with two spatial  dimensions, and the other is a screw dislocation. Both are stringy defects terminating solitons. 
The former is a string extending parallel to the solitons (as a linear extension of dislocations in two dimensions), whereas the latter is one extending orthogonal to the solitons. In general, mixtures of these two also exist.

We numerically solve Eq.~(\ref{eq:DL_EOM_III}) and investigate dislocations in three spatial dimensions. We fix the size of a numerical box as $[-20,20]^3$ and the lattice points as $200^3$. Namely, the spatial lattice size is $0.2$, and the temporal lattice size is $0.01$. Furthermore, we take $\tilde m = 0.5$ for all the simulations in this section. The critical value for one soliton in the vacuum can be read from Fig.~\ref{fig:phase_modeIII}  to be $\kappa\tilde B = 1.1$. The critical value for the CSL background should be larger than this, so that we mostly consider the case with $\kappa \tilde B > 1.1$ in what follows.

\subsection{Edge dislocations in three dimensions}
\label{sec:edge_3d}

\begin{figure}[htbp]
    \centering
    \includegraphics[width=0.3\linewidth]{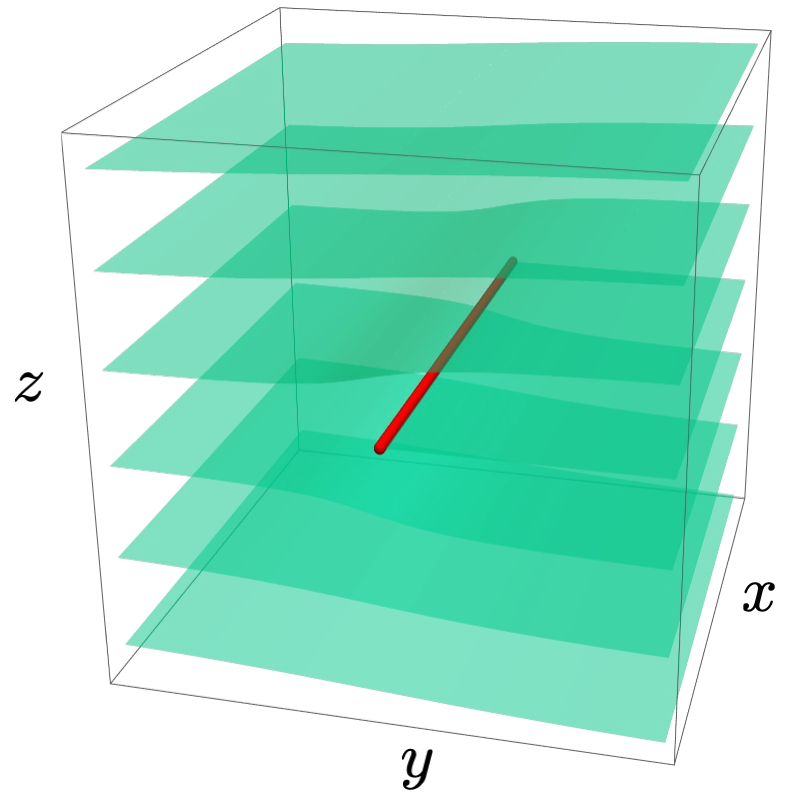}
  \caption{A straight edge dislocation (red) in the CSL (green) in three dimensions.}
  \label{fig:3d_edge_dislocation}
\end{figure}

Let us first investigate edge dislocations 
in three dimensions. We have already studied those in two dimensions.
They are defects of codimensions two, 
and are point-like and  string-like defects in two and three dimensions, respectively. 
Therefore, 
straight edge dislocations in three dimensions can 
be obtained from those in two dimensions linearly extended into the third direction as shown in Fig.~\ref{fig:3d_edge_dislocation}.
We here study truly three dimensional defects that cannot be obtained from trivial embedding of two dimensional dislocations.   

\begin{figure}[htbp]
    \centering
    \includegraphics[width=0.99\linewidth]{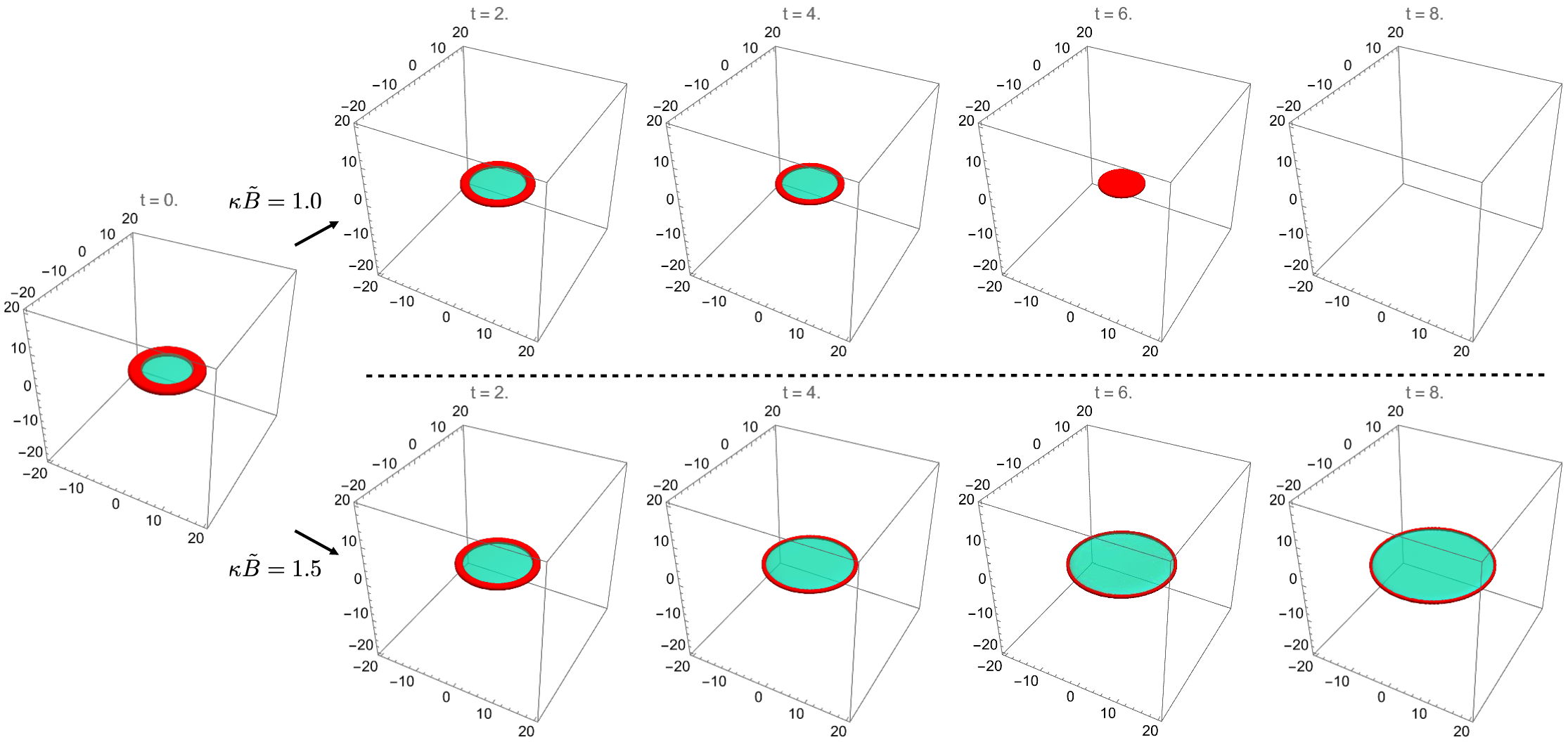}
  \caption{Real time evolutions of a single ring-shaped edge dislocation in vacuum with $\kappa \tilde B = 1.0$ (top) and with $\kappa \tilde B = 1.5$ (bottom). The red surface has $|\tilde \phi|=0.5$ and the cyan surface has $|\arg \tilde \phi| = 0.8 \pi$. The periodic boundary conditions are imposed for $x$, $y$, and $z$ directions.}
  \label{fig:single_ring}
\end{figure}

In Sec.~\ref{sec:modelIII_single_finite_soliton} we first took the single finite soliton ending on strings. The three dimensional analogue that we consider here is a finite soliton of a disk shape surrounded by a closed string on its edge. 
We prepare a disk soliton in the vacuum and use it as an initial configuration of the real time evolution with the zero initial velocity. 
The results are shown in Fig.~\ref{fig:single_ring}.
We simulated two cases with $\kappa\tilde B = 1.0$ and $1.5$. The former is below the critical coupling while the latter is above it. Therefore, one would expect that the initial disk soliton shrinks in the former case and expands in the latter case. 
The results agree very well with the expectation as shown in Fig.~\ref{fig:single_ring}.

\begin{figure}[htbp]
    \centering
    \includegraphics[width=0.9\linewidth]{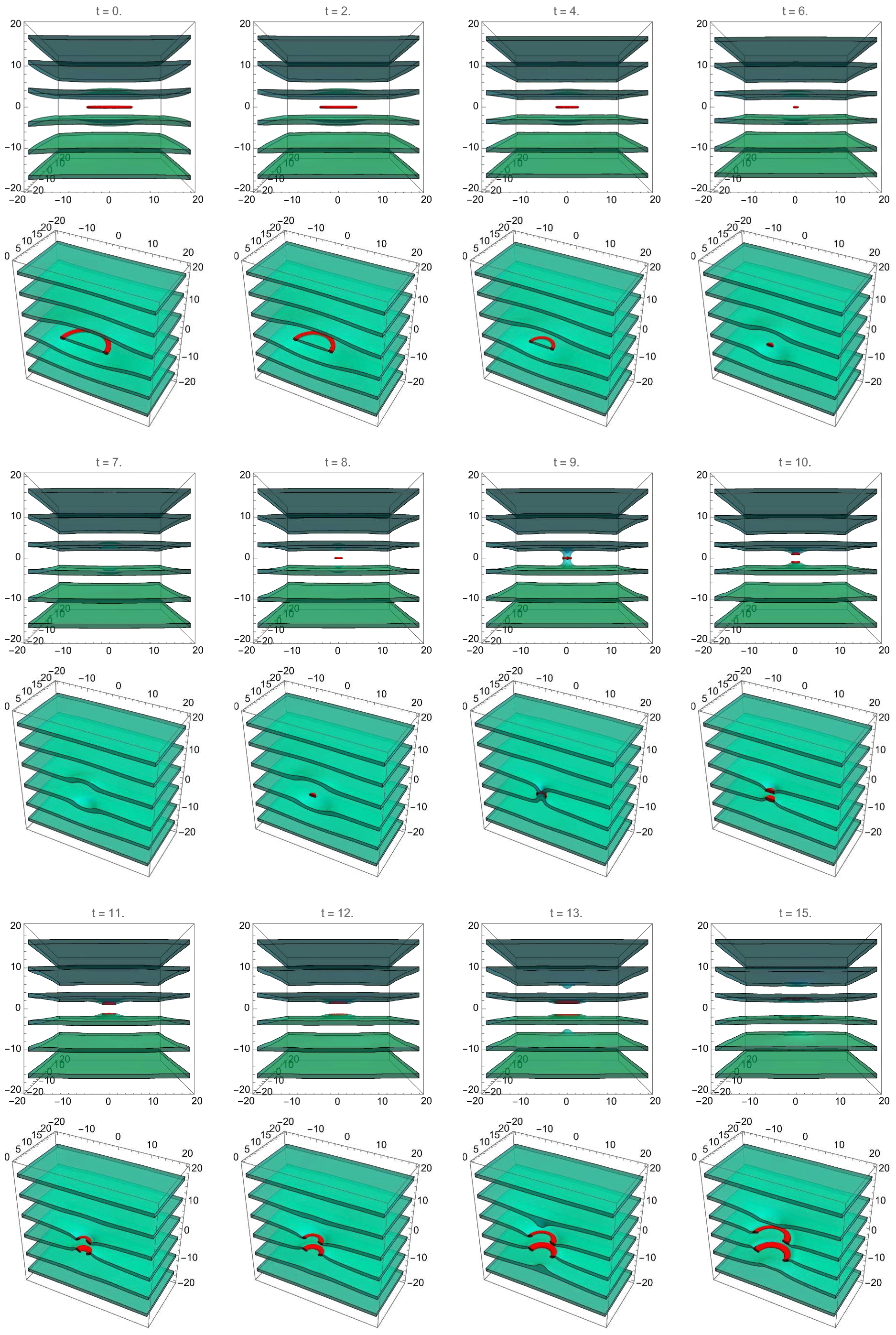}
  \caption{Real time evolutions of a single ring-shaped edge dislocation in CSL with $\kappa \tilde B = 1.5$. The periodic boundary condition is imposed in the $x$, $y$, and $z$ directions.}
  \label{fig:single_ring_inCSL}
\end{figure}

Next, we put the disk soliton in the CSL background and study the time evolution in the two cases with $\kappa \tilde B = 1.5$ and $\kappa \tilde B = 2.5$.
The simulation result in the former case is shown in Fig.~\ref{fig:single_ring_inCSL}.
The initial configuration is shown in the left-top corner of Fig.~\ref{fig:single_ring_inCSL}.
We observe that the disk soliton shrinks and disappears soon even though $\kappa \tilde B = 1.5$ is larger than the critical value for the vacuum.
This happens because the background is not the vacuum but CSL.
An interesting phenomenon occurs after the disk collapses.
A small bubble nucleates from the remnants of the collapse,
and it grows into two string loops that are the boundaries of two solitons just above and below the original disk soliton.
This results in expanding holes in both solitons,
see the panels $t=8$ to $t=15$ of Fig.~\ref{fig:single_ring_inCSL}.
This is a process reducing several solitons from CSL.

\begin{figure}[htbp]
    \centering
    \includegraphics[width=0.9\linewidth]{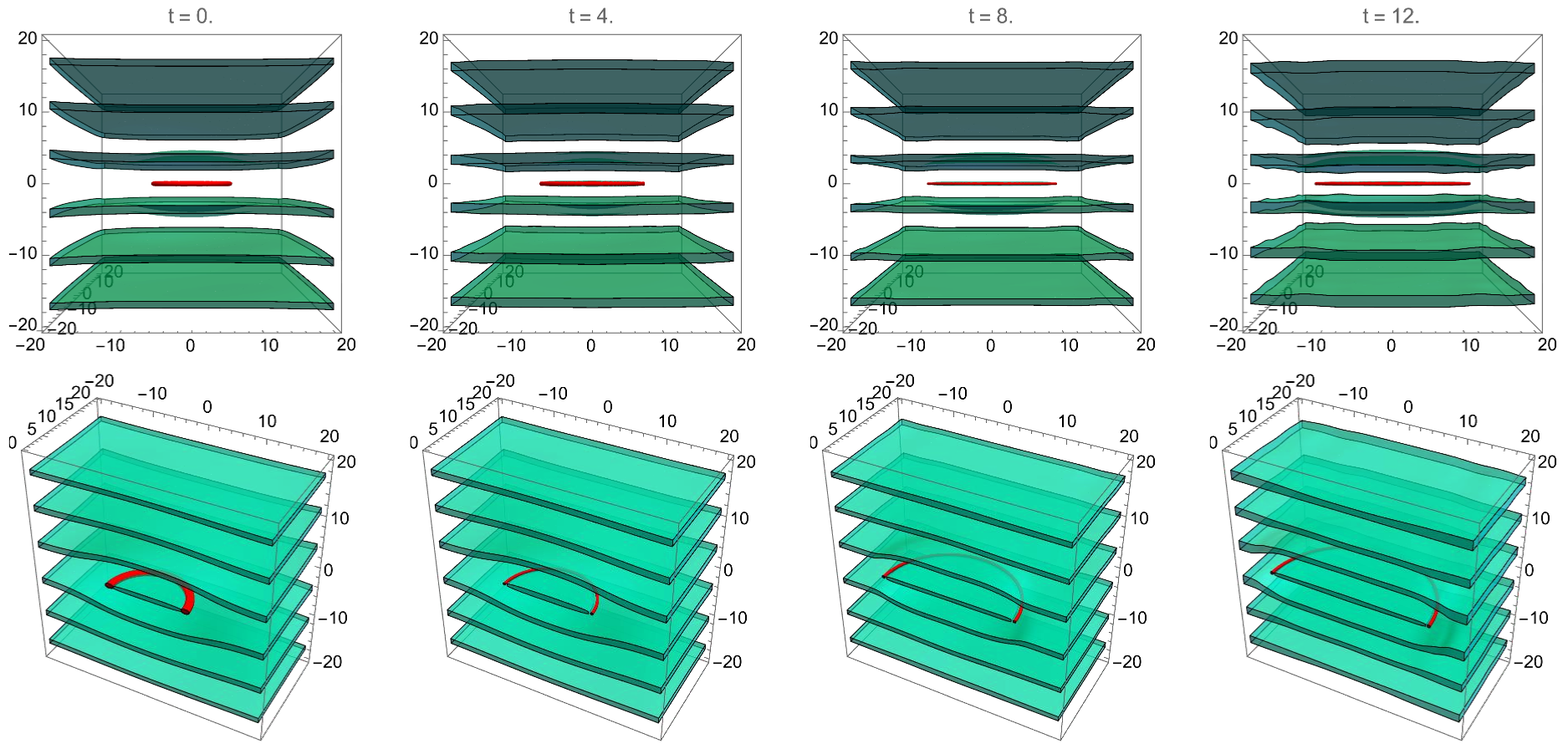}
  \caption{Real time evolutions of a single ring-shaped edge dislocation in CSL with $\kappa \tilde B = 2.5$. The periodic boundary condition is imposed in the $x$, $y$, and $z$ directions.}
  \label{fig:single_ring_inCSL_2}
\end{figure}

The other simulation with $\kappa\tilde B = 2.5$ is shown in Fig.~\ref{fig:single_ring_inCSL_2}. The initial configuration is the same as that for Fig.~\ref{fig:single_ring_inCSL}. Since the external current $\kappa \tilde B = 2.5$ is sufficiently large in this case, the disk soliton grows up to increase the number of solitons of CSL.

\subsection{Topology of screw dislocations and vortices}\label{sec:screw}

\begin{figure}[htbp]
    \centering
    \includegraphics[width=0.8\linewidth]{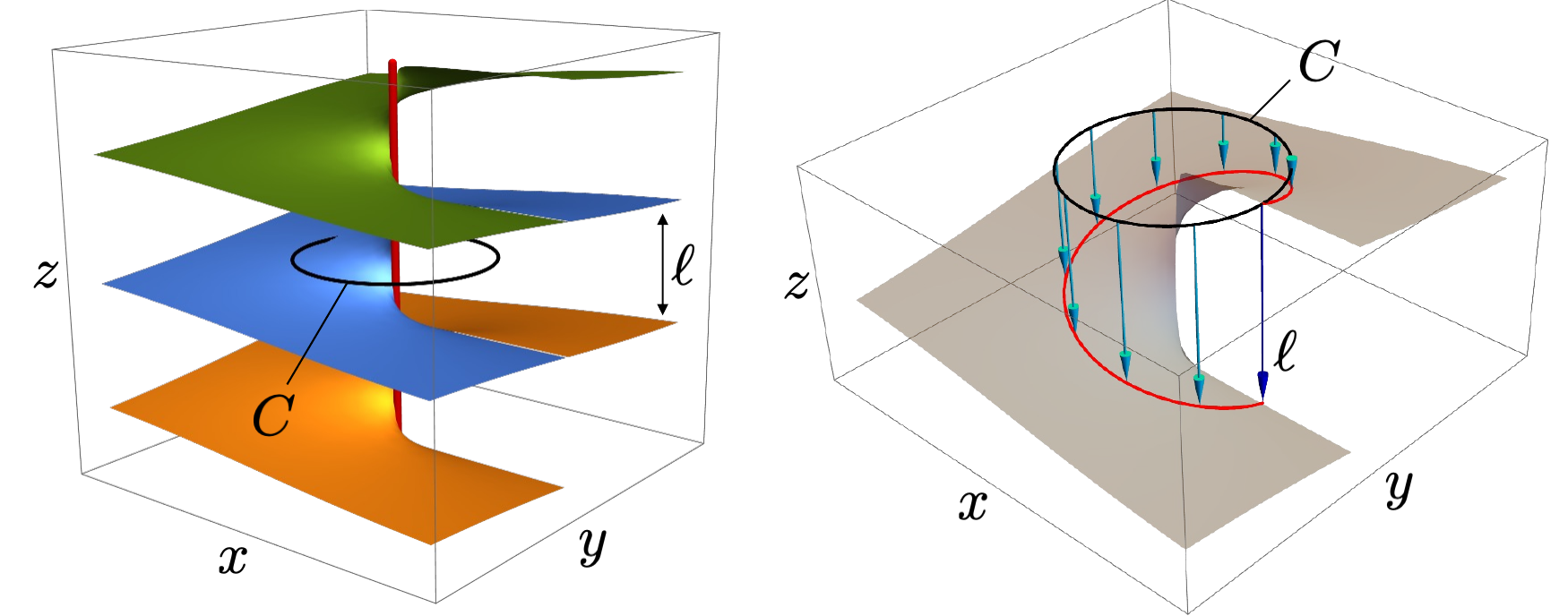}
  \caption{The screw dislocation with $N_{\rm DS} = 1$ and the left-handed helicity in the CSL with the period $\ell$. The surfaces are determined by $\arg\phi = 2n+1$. The red thick line in the left panel stands for the vortex, and the arrows in the right panel show displacements $\delta z$ on the loop $C$.}
  \label{fig:screw_homotopy}
\end{figure}

Next, let us move to screw dislocations which exist only in three dimensions. 
A screw dislocation in a crystal is a line defect with a spiral stacking of crystal planes around a central line, namely the screw dislocation.
The screw dislocations in the CSL have very similar spiral structures.
Since here we are interested in a topological aspect of the screw dislocation, we focus on the phase $\eta = \arg\phi$ only.
The screw dislocation is essentially the vortex perpendicular to the CSL.
Suppose the array of parallel solitons in the CSL stacked along the $z$-axis with period $\ell$. The phase $\eta = \arg \phi$ of the CSL background can be roughly approximated by a linear function $\eta = \frac{2\pi}{\ell}z$ up to a constant. Now putting the vortex along the $z$-axis can be realized by
the following function 
in the cylindrical coordinates 
$(r,\theta, z)$: 
\be
\arg \phi(x,y,z) = \theta(x,y) + \frac{2\pi}{\ell}z\,,
\label{eq:ini_screw}
\ee
with $\theta(x,y) = \arg (x+iy)$. A constant $\arg\phi$ surface is shown in Fig.~\ref{fig:screw_homotopy}.
%%%%%%%%%%%%%%%%
At a constant $z$ slice, this gives a vortex characterized by a map from $S^1$ in the $xy$-plane to the phase of $\phi$ with the winding number one. At a fixed point $(x, y)$ in the $xy$-plane, this can be regarded as 
a single chiral soliton, 
characterized by a map from another $S^1$ which is the line segment of the length $\ell$ whose endpoints are identified to the phase of $\phi$ with the winding number one.
Note that, however, these winding number makes sense only approximately under the presence
of the explicit $U(1)$ symmetry breaking term with $m\neq 0$. The genuine topological winding number is the fundamental group given by Eq.~(\ref{eq:pi1_massive}) and the corresponding topological soliton is the screw dislocation. Let us consider a closed loop $C$ surrounding the vortex as is shown in Fig.~\ref{fig:screw_homotopy}.
Since we have $\frac{2\pi}{\ell}z = - \theta + \text{const.}$, 
when we travel along $C$ clockwise once with $\theta = 0$ to 
$\theta = 2\pi$, we go
around $M_{m\neq0} \simeq S^1$ counterclockwise once $\delta z = 0 \to \delta z =- \ell$ as shown in Fig.~\ref{fig:screw_homotopy}(right).\footnote{The displacement vector $-\ell \hat e_z$ corresponds to the Burgers vector of a screw dislocation.} 
We count the corresponding winding number as
$+1$ in the same manner as before, and it corresponds to the number of the screw dislocation as Eq.~(\ref{eq:N_DS=1}).

Let us mention several differences between edge and screw dislocations: 1) The edge dislocation\footnote{The phase structure of the edge dislocation in the CSL is $\arg\phi = \arg(y+iz) + \frac{2\pi}{\ell}z$ while the one for the screw dislocation is $\arg\phi = \arg(x+iy) + \frac{2\pi}{\ell}z$.} is parallel to the CSL whereas the screw dislocation is perpendicular to the CSL. 2) The edge dislocation is accompanied by the DSG soliton but the screw dislocation is not because all the solitons around it sit at the bottom of the effective potential. Therefore the screw dislocation can be static. 3) The screw dislocations have a helicity.
The helicity of the screw dislocation generated by Eq.~(\ref{eq:ini_screw}) is left-handed since a constant $\arg \phi$ gives a Riemann surface $z = - \frac{\ell}{2\pi}\theta + {\rm const.}$.

\begin{figure}[htbp]
    \centering
    \includegraphics[width=0.8\linewidth]{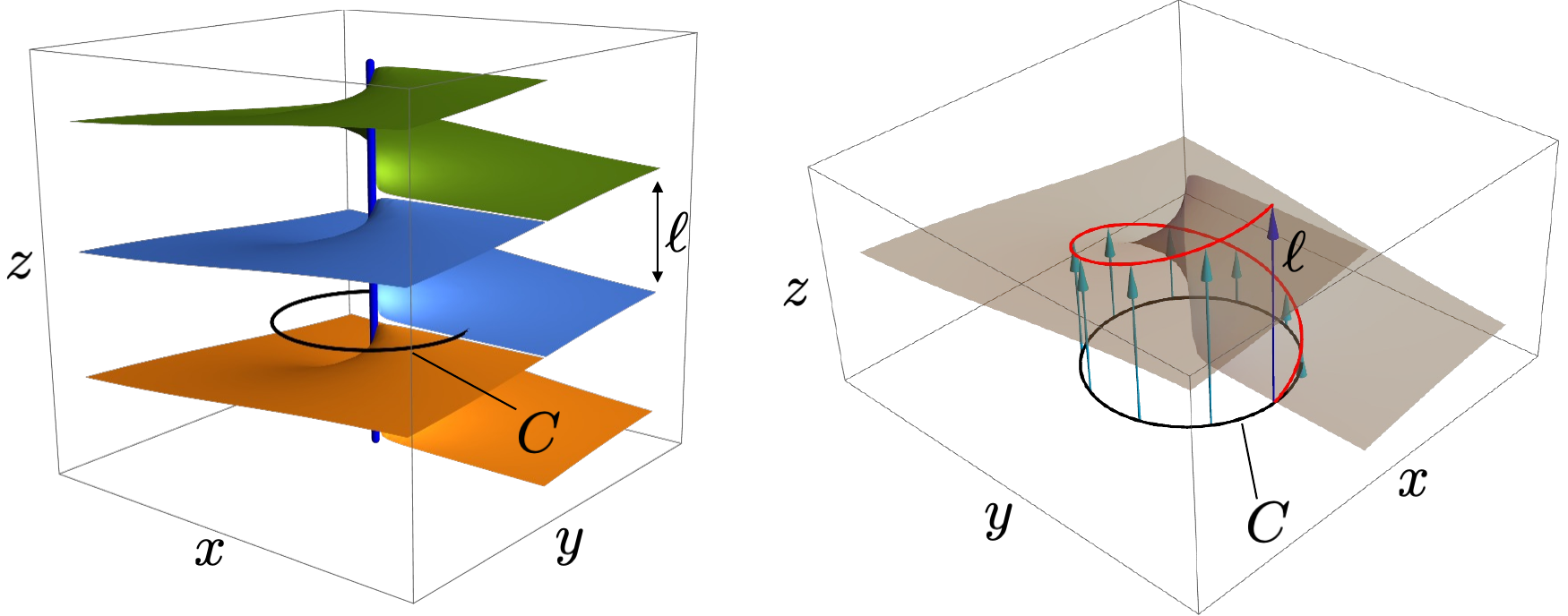}
  \caption{The anti-screw dislocation with $N_{\rm DS} = -1$ and the right-handed helicity in the CSL with the period $\ell$. The surfaces are determined by $\arg\phi = 2n+1$. The blue thick line in the left panel stands for the anti-vortex, and the arrows in the right panel show displacements $\delta z$ on the loop $C$.}
  \label{fig:screw2_homotopy}
\end{figure}

Next, we consider a mirror image (the $y$ parity: $(x,y,z) \to (x,-y,z)$) of Eq.~(\ref{eq:ini_screw}). Since the $y$ parity flips the vortex winding at a constant $z$ plane as $\theta = \arg(x+iy) \to \arg(x-iy) = - \theta$, we have 
\be
\arg \phi = -\theta + \frac{2\pi}{\ell}z\,.
\label{eq:ini_screw_right}
\ee
This is the right-handed screw dislocation because the constant $\arg \phi$ surface is $z = \frac{\ell}{2\pi}\theta + {\rm const.}$. 
The constant $\arg \phi$ surface for Eq.~(\ref{eq:ini_screw_right}) is shown in Fig.~\ref{fig:screw2_homotopy}.
%%%%%%%%%%%%%%%%
The $x$ parity $(x,y,z) \to (-x,y,z)$ is essentially the same as the $y$ parity because they are related by the rotation along the $z$ axis.
Let us count the winding number of the screw dislocation. Since we have $\frac{2\pi}{\ell}z = \theta + \text{const.}$, when we go around $C$ clockwise once with $\theta \to \theta + 2\pi$, the displacement changes as $\delta z = 0 \to \ell$.
Namely, we go around $M_{m\neq0}\simeq S^1$ clockwise once, giving the winding number $N_{\rm DS} = -1$.
Hence, Eq.~(\ref{eq:ini_screw_right}) corresponds to the anti-screw dislocation.

In contrast, the $z$ parity $(x,y,z) \to (x,y,-z)$ should be distinguished from the $x$ or $y$ parity, because it is broken by the background external current $\vec S = (0,0,B)$. Indeed, the $z$ parity exchanges the left-handed and right-handed screws with keeping the dislocation winding number $N_{\rm DS}$ (and also vortex winding number) on the $xy$ plane. This can be easily understood by writing down explicitly the phase
\be
\arg \phi = \pm \theta - \frac{2\pi}{\ell}z\,,
\ee
where $+$ ($-$) is for the $z$ parity of Eq.~(\ref{eq:ini_screw}) [(\ref{eq:ini_screw_right})]. 
Instead, it makes $\p_z\arg \phi > 0$ to be $\p_z\arg \phi < 0$, so that the mirror images by the $z$ parity are energetically disfavored in the presence of the background current with $B > 0$. 
In other words, these configurations live in an anti-CSL that cannot be the ground state 
and is unstable for $B > 0$.
Instead they are stable for $B < 0$. 
Hence, the helicity and the dislocation windings are locked in the CSL and anti-CSL. 
While the screw dislocation with $N_{\rm DS} = 1$ (the vortex winding number $1$) which has the left-handed helicity and the anti-screw dislocation with $N_{\rm DS}=-1$ (the vortex winding number $-1$) which has the right-handed helicity {are energetically favored under the positive external field $B>0$, the screw dislocations with opposite combinations of the helicity and the dislocation charge are energetically favored for $B < 0$.
These are summarized in Fig.~\ref{fig:screw_chirality}. 
\begin{figure}[htbp]
    \centering
    \includegraphics[width=0.99\linewidth]{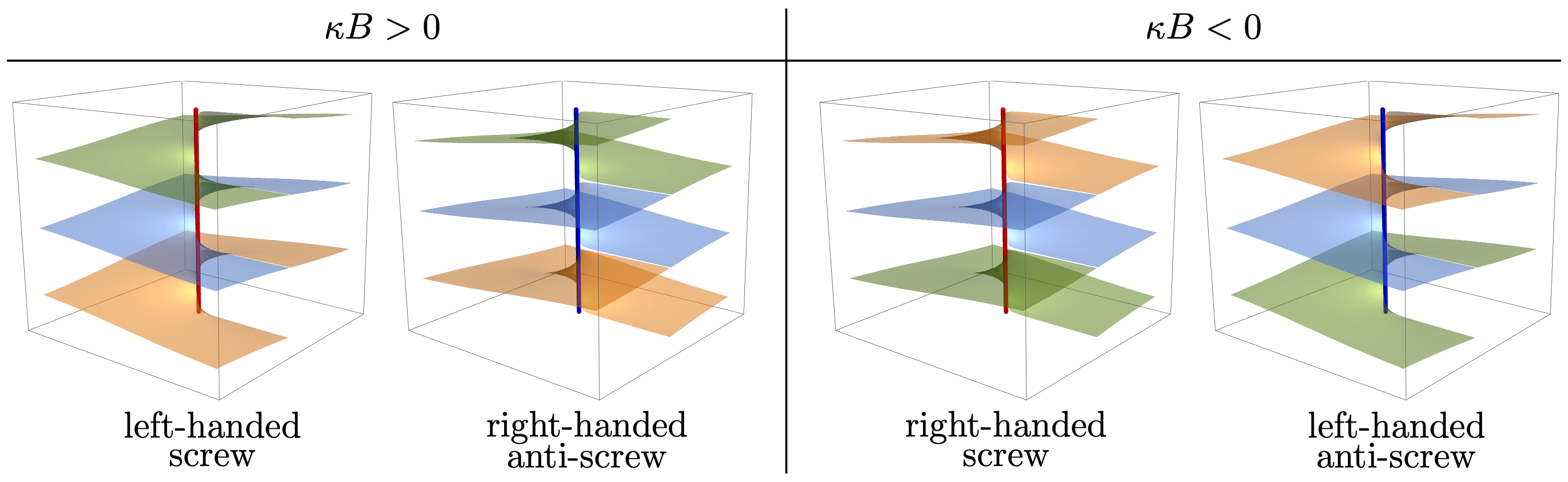}
  \caption{The left-handed screw and the right-handed anti-screw are stable for $\kappa B >0$, whereas their $z$-parity images, the right-handed screw and the left-handed anti-screw are stable  for $\kappa B <0$. The orange surface is $\arg\phi = -\pi$, the blue one has $\arg\phi = \pi$, and the green one has $\arg\phi = 3\pi$.}
  \label{fig:screw_chirality}
\end{figure}

%%%%%%%%%%%%%%%%%
\subsection{Solutions of screw dislocations
in three dimensions}

%%%%%%%%%%%%%%%%%%%%%%
\subsubsection{Single screw dislocations}

Let us explain how (meta)stable screw dislocations are obtained in the CSL. We first need to prepare a suitable initial configuration that has a spiral structure. The concrete initial configuration we used is shown in the top row of Fig.~\ref{fig:screw}
whose phase is essentially same as the one given in Eq.~(\ref{eq:ini_screw}).\footnote{
One can construct a better ansatz by 
replacing the second term 
with a soliton lattice in 
Eq.~(\ref{eq:SG_CSL}). 
In purpose for the initial configuration for the relaxation, the linear dependence in Eq.~(\ref{eq:SG_CSL}) is enough.
}
We take this as the initial configuration and solve EOM (\ref{eq:DL_EOM_III_modify}) with a large diffusion coefficient $\epsilon = 1$. The boundary condition along the $z$-axis is periodic whereas we impose the Neumann  boundary conditions in the $x$ and $y$ directions.
\begin{figure}[htbp]
    \centering
    \includegraphics[width=0.99\linewidth]{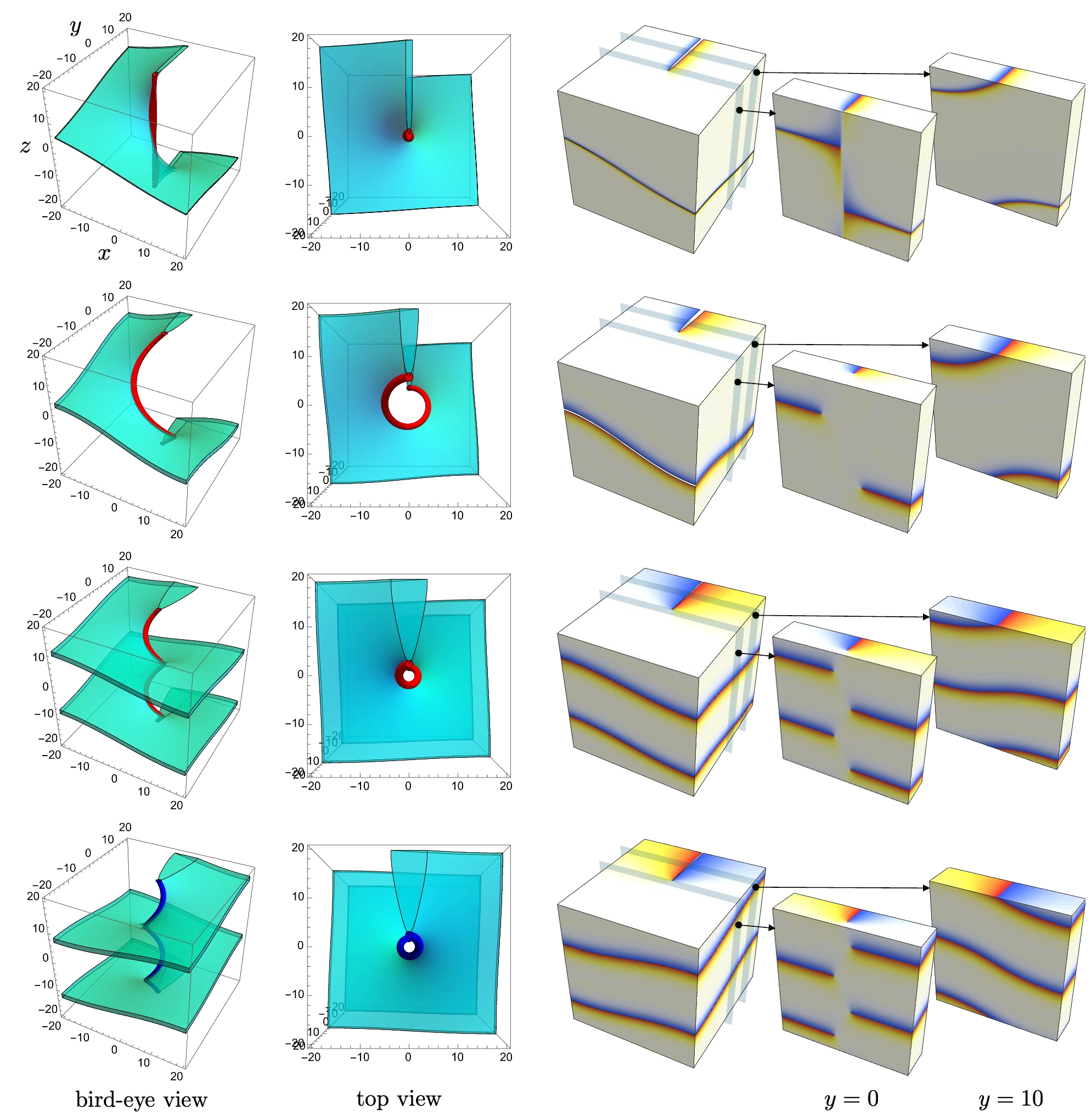}
  \caption{Simple screw dislocations for $\tilde m = 0.5$ and $\kappa \tilde B = 1.5$. The first row shows the hand-made configuration for the initial state of the relaxation method. The second row is the configuration at the convergence evolved from that in the first row. The third and fourth rows show the left-handed and right-handed screws, obtained by the relaxation method, of a half period of that in the second row. We impose the periodic boundary condition in the $z$ direction while the Neumann boundary condition for $x$ and $y$ directions. The green surfaces have $|\arg \tilde\phi| = 0.8\pi$, and the red (vortex) and blue (anti-vortex) surfaces correspond to $|\tilde\phi|=0.5$.}
  \label{fig:screw}
\end{figure}
We numerically integrated Eq.~(\ref{eq:DL_EOM_III_modify}), and observed that it quickly reached a convergence due to the large diffusion term with $\epsilon = 1$. The result is shown in the second row from the top of Fig.~\ref{fig:screw}. The screw dislocation is initially straight, but the tension of the soliton caused it to bend into a spiral shape. If the background field $\kappa \tilde B$ is smaller than the critical value, the soliton continuously pulls the string toward the boundaries of the $xy$ plane and eventually the soliton disappears. However, we take $\kappa \tilde B = 1.5$ large enough so that the spiral soliton remains as a stable configuration.
We also show a screw dislocation with a half period  in the third row from the top of Fig.~\ref{fig:screw} which is obtained from the initial configuration similar to 
Eq.~(\ref{eq:ini_screw}) with $2\pi/\ell$ replaced by $4\pi/\ell$.
The right-most column of Fig.~\ref{fig:screw} shows the phase of $\tilde \phi$. One can clearly see from the slices at $y=0$ that 
the solitons on one side of the screw dislocation slide by half a period from the solitons on the other side.

The right-handed screw in the forth row from the top of Fig.~\ref{fig:screw} is a mirror image (under the $x$ parity: $(x,y,z) \to (-x,y,z)$) of the right-handed one in the third row. Note that the $x$ parity also flips the vortex winding at a constant $z$ plane as can be seen in the right-most column of Fig.~\ref{fig:screw}. 
The corresponding initial configuration is given in Eq.~(\ref{eq:ini_screw_right}).
This is the right-handed anti-screw dislocation with $N_{\rm DS} = -1$ as explained above.

%%%%%%%%%%%%%%%%%%%%%%%
\subsubsection{Multiple screw dislocations}

Multiple screw dislocations can be constructed as well. We show left-handed multiple screw dislocations in Fig.~\ref{fig:coaxial_multiple_spiral}. 
\begin{figure}[htbp]
    \centering
    \includegraphics[width=0.99\linewidth]{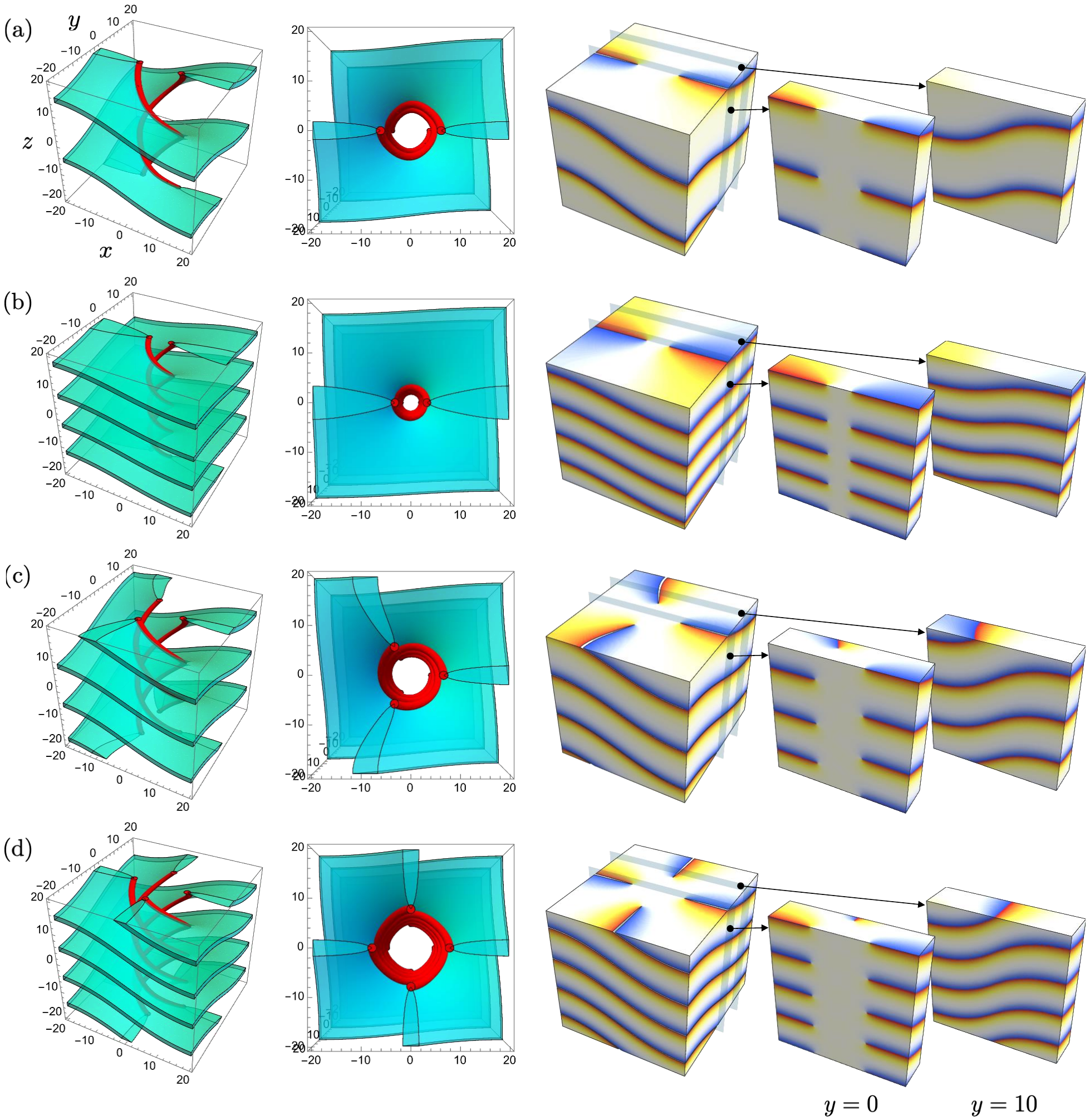}
  \caption{(a) Double spiral (b) Double spiral (2 periods) (c) Triple spiral (d) Quadruple spiral: The left-handed multiple spiral screw dislocations obtained by the relaxation method for $\tilde m = 0.5$ and $\kappa\tilde B = 1.5$. We impose the periodic boundary condition in the $z$ direction while the Neumann boundary condition for $x$ and $y$ directions.}
  \label{fig:coaxial_multiple_spiral}
\end{figure}
The top row shows one period of a double screw dislocation 
with the common center of spiral.
This has the same structure as a {\it double helix staircase}\footnote{
Da Vinci designed a double helix staircase at the Ch$\hat{\rm a}$teau de Chambord, 
a castle located in the Loire Valley in France.
} or DNA.
It consists of two soliton sheets that are not connected, and the helical strings at the end of soliton sheets have the common center along the $z$ axis. The second row from the top shows two periods of the double spiral with a double frequency that we present for comparison with the latter configurations. The third and forth rows from the top show the triple and quadruple spiral screw dislocations, respectively. The helical strings have the common center, and the soliton sheets are not connected. Although the double spiral of two periods (in the second row) and the quadruple spiral of one period (in the fourth row) look  very similar when one observes them from a distance, one should distinguish them because the numbers of the strings (or the disconnected solitons) are different (two for the former and four for the latter).

%%%%%%%%%%%%%%%%%%%%%%%%%%%%%%%%
\begin{figure}[htbp]
    \centering
    \includegraphics[width=0.75\linewidth]{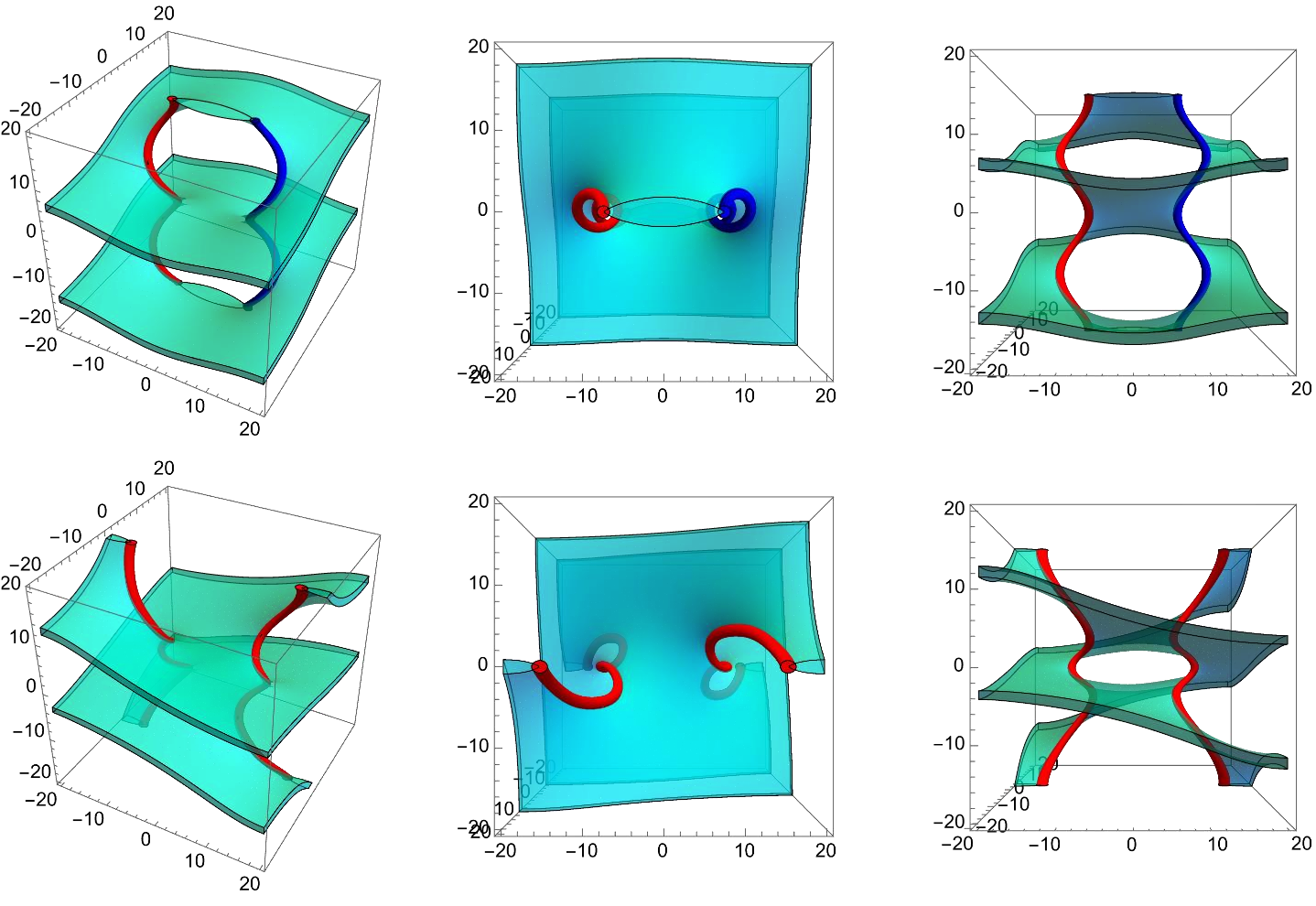}
  \caption{The non-coaxial two screw dislocations. The first row shows the composite of the left- and right-handed screw dislocations. The second row shows %the soliton lattice with 
  two left-handed non-coaxial screw dislocations.}
  \label{fig:screw_antiscrew}
\end{figure}
%%%%%%%%%%%%%%%%%%%%%%%%%%%%
Superposition of screw dislocations with separated centers of spirals is also possible.
We show two examples in Fig.~\ref{fig:screw_antiscrew}. 
The configuration at the first row consists of left- and right-handed screw dislocations, 
while one in the second row 
consists of the two left-handed screw dislocations. 
One can find an apparent difference between  screw dislocations 
with the common and separated centers 
in the top views in Figs.~\ref{fig:coaxial_multiple_spiral} and \ref{fig:screw_antiscrew}. 
The configurations with common center have a hole piercing from top to bottom at the center while the separated one does not have such holes. 
More interesting topological difference is that 
all the soliton sheets
attached to the separated screw dislocations 
are connected with consisting of one soliton, 
while those with common center consist of the two separated solitons sheets. 
Therefore, 
one can continuously move from one point to the other in the former, but one has to jump from one soliton to another in the latter.

As mentioned, 
the configuration at the first row 
in Fig.~\ref{fig:screw_antiscrew}
consists of left- and right-handed screw dislocations. 
Since the left-handed and right-handed screw dislocations have the opposite dislocation  charges (vortex windings), 
they eventually annihilate each other 
to reach a pure crystalline soliton lattice without dislocations.

The separated dislocations 
in the second row of Fig.~\ref{fig:screw_antiscrew} consist of the two left-handed screw dislocations. 
This is topologically distinguishable 
with 
the double screw dislocations with the common center
in the top row of Fig.~\ref{fig:coaxial_multiple_spiral} 
in the sense that the linking number for the former is unity per one period while that for the latter is zero. 
However, a reconnection 
between two dislocations is allowed 
so that the double screw dislocations 
with the common center 
in the top row of Fig.~\ref{fig:coaxial_multiple_spiral} and the
and separated screw dislocations 
in the second row of Fig.~\ref{fig:screw_antiscrew} 
can be continuously deformed each other through a reconnection. 
 Nevertheless,  
there is a significant difference between their stability.
Given the general fact that two global vortices repel each other, we expect that the separated screws dislocations further separate and become eventually destabilized.
We confirm it  
in the relaxation process 
[Eq.~(\ref{eq:DL_EOM_III_modify}) with the diffusion coefficient $\epsilon = 1$].
Fig.~\ref{fig:edge_screw} shows that 
it can be continuously deformed to the pure CSL.\footnote{
We have observed the intermediate states appearing in a transition from the double screw dislocation to the pure CSL include a mixture of the edge and screw dislocations. Although the transition in Fig.~\ref{fig:edge_screw} is not real time dynamics, we will see that the edge and screw dislocations are elementary even in formation of the CSL by the real time dynamics in the next section.}
\begin{figure}[htbp]
    \centering
    \includegraphics[width=0.99\linewidth]{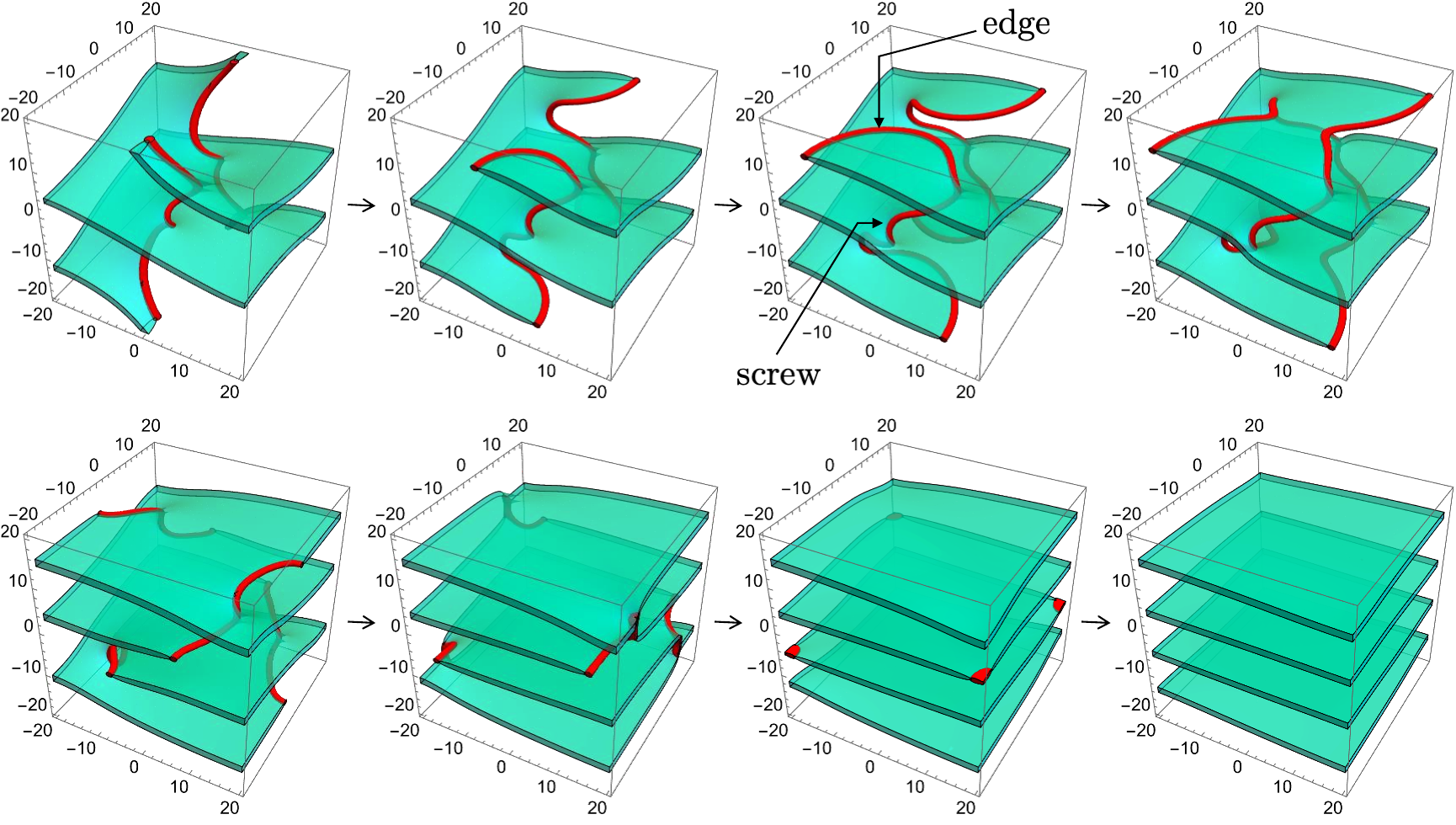}
  \caption{The relaxation process from the non-coaxial double screw dislocation to the pure CSL. The initial state (the left top figure) is identical to the left-most panel at the second row of Fig.~\ref{fig:screw_antiscrew}. The initial configuration containing screw dislocations continuously deforms, edge dislocations appear at an intermediate state, then the dislocations disappear, and finally a pure CSL is reached. We impose the periodic boundary condition in the $z$ direction and the Neumann boundary conditions in the $x$ and $y$ directions.}
  \label{fig:edge_screw}
\end{figure}
On the other hand, the  double screw dislocation, the double helix, studied in the last subsection is metastable 
although it can be continuously 
deformed to the separated one through a reconnection. 
This stability can be explained as follows. Since the spiral screws are intertwined with each other, if one tries to separate one part at a constant $z$-plane, the other parts at a different constant $z$-plane cannot avoid to move closer together. 
Thus, they are locked and are stable.

%%%%%%%%%%%%%%%%%%%%%%%
\subsection{Dynamical formation of chiral soliton lattices in three dimensions}
\label{sec:CSL_formation_3d}

Finally, we investigate dynamical formation of the CSL in three dimensions. We have already studied the similar problem in two dimensions in Sec.~\ref{sec:2d_CSL_formation} where we found that the edge dislocation is the essential intermediate state appearing/disappearing during evolution from vacuum with small fluctuations to the pure CSL. In three dimensions, in addition to the edge dislocation, the screw dislocations appear as an intermediate state in transitions between two different configurations. 
In this section, we will focus only on the real time evolution, so we will set $\epsilon = 0$ in Eq.~(\ref{eq:DL_EOM_III_modify}) throughout this section.

We investigate dynamical formation of the CSL from fluctuations around the vacuum in three spatial dimensions. In order to get better resolution, we slightly increase the number of lattice for all the simulations in this subsection. The size of numerical box $[-20,20]^3$ is the same as before and the lattice points are $250^3$. Namely, the spatial lattice size is $0.16$, and the temporal lattice size is $0.008$. We adopt periodic boundary conditions in all the $x$, $y$, and $z$ directions.

We first prepare an initial configuration with random fluctuations around the vacuum
as $\tilde \phi = v' + \delta$ with $v' \simeq 1.19149$ for $\tilde m = 0.5$ and $|\delta| \sim {\cal O}(v'/2)$. The concrete initial configurations are shown in Fig.~\ref{fig:seed8_initial} from which one can see neither solitons nor dislocations initially exist. We numerically solve Eq.~(\ref{eq:DL_EOM_III}) with this as the initial configuration and zero initial velocity.
\begin{figure}[htbp]
    \centering
    \includegraphics[width=0.99\linewidth]{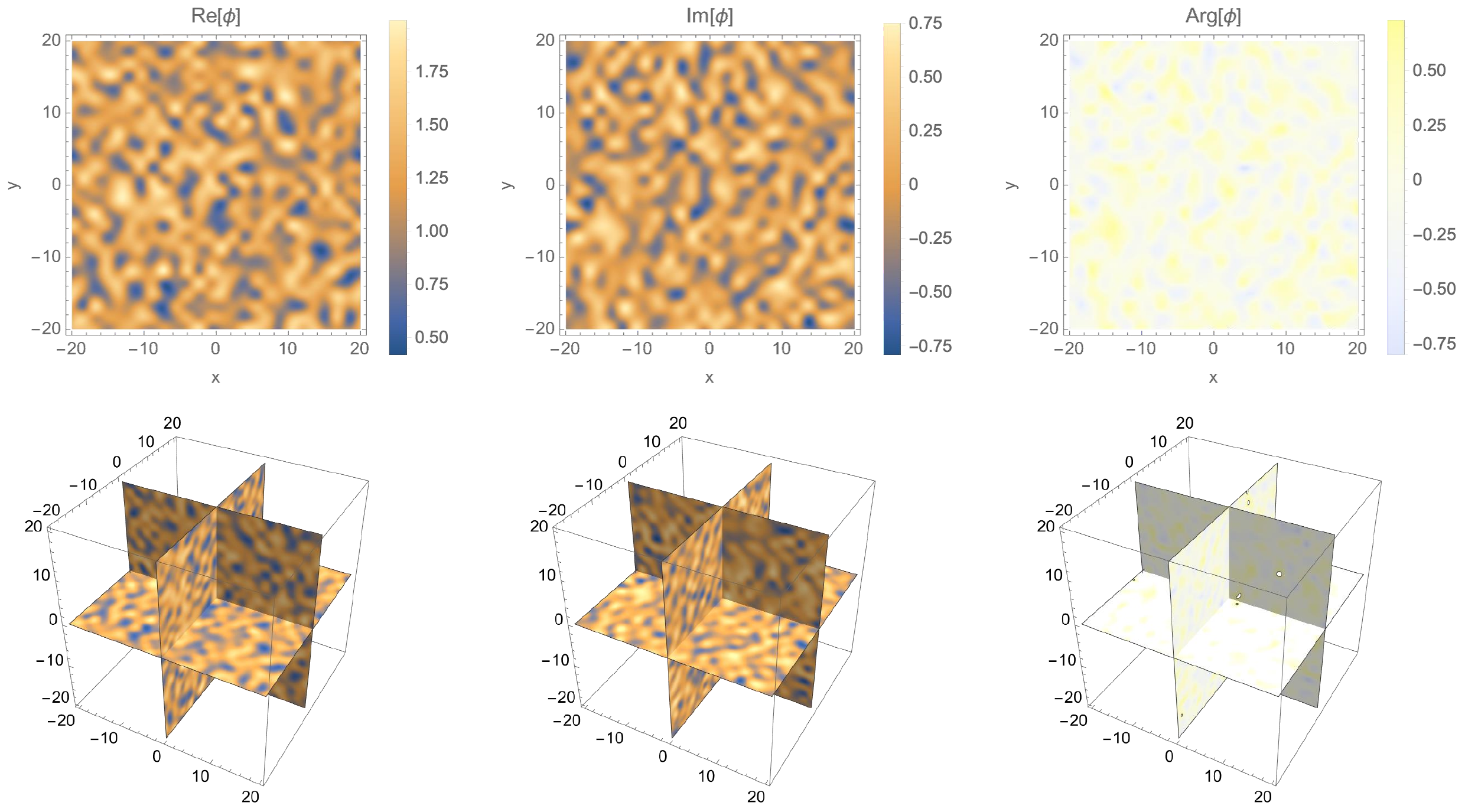}
  \caption{The initial configuration $\tilde \phi$ for the numerical simulations in this subsection. The bottom row shows ${\rm Re}(\tilde\phi)$, ${\rm Im}(\tilde\phi)$, and $\arg\tilde\phi$, respectively. The top row shows the slices at $z=0$.}
  \label{fig:seed8_initial}
\end{figure}

We vary $\kappa \tilde B$ whereas $\tilde m = 0.5$ is fixed, and the initial configuration and the boundary conditions are unchanged. We first show the simulation result for $\kappa \tilde B = 1.5$ in Fig.~\ref{fig:seed8_s1p5}. 
\begin{figure}[htbp]
    \centering
    \includegraphics[width=0.99\linewidth]{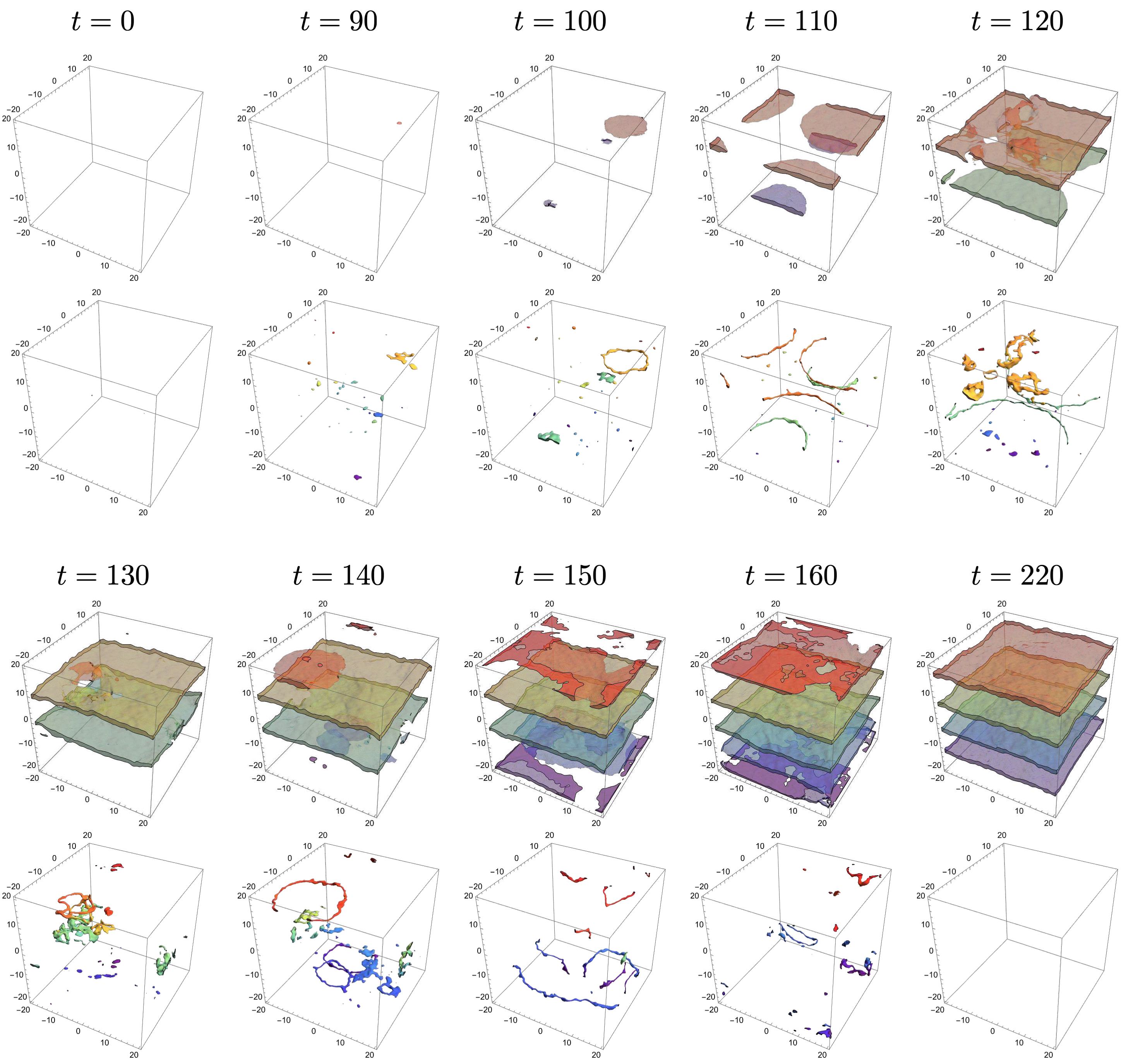}
  \caption{The real time dynamics of the CSL formation from the vacuum with random fluctuations for $\kappa \tilde B = 1.5$. The initial configuration is given in Eq.~\ref{fig:seed8_initial}. The first and the third rows show surfaces of $|\arg\tilde\phi|=0.8\pi$, and the second and fourth rows correspond to surfaces of $|\tilde \phi|=0.3$. The periodic boundary conditions are imposed for $x$, $y$, and $z$ axes.}
  \label{fig:seed8_s1p5}
\end{figure}
At first, it appears to oscillate randomly around the vacuum for a fairly long time (from $t=0$ to $t\simeq 90$). The amplitude can accidentally be amplified locally somewhere, forming tiny seeds of solitons. Small seeds will shrink and disappear, but large enough seeds grow into flat solitons.
The first large seed emerges at $t=90$ and the second one appears slightly later at $t=100$. Each seed expands horizontally to form a finite, closed edge dislocation with a disk soliton, and it eventually collides with itself due to the periodic boundary conditions. The impact of self collision stirs up the surrounding fields, and it further generates the third and fourth large seeds. They also grow up to form other flat solitons, but the dynamical process is not so simple because the environment is mixed up already. We observe that edge and screw dislocations emerge in the intermediate states, they repeatedly reconnect, and finally all the dislocations disappear and the CSL with four flat solitons is completed.

\begin{sidewaysfigure}[h]
%\begin{figure}[htbp]
    \centering
    \includegraphics[width=0.99\linewidth]{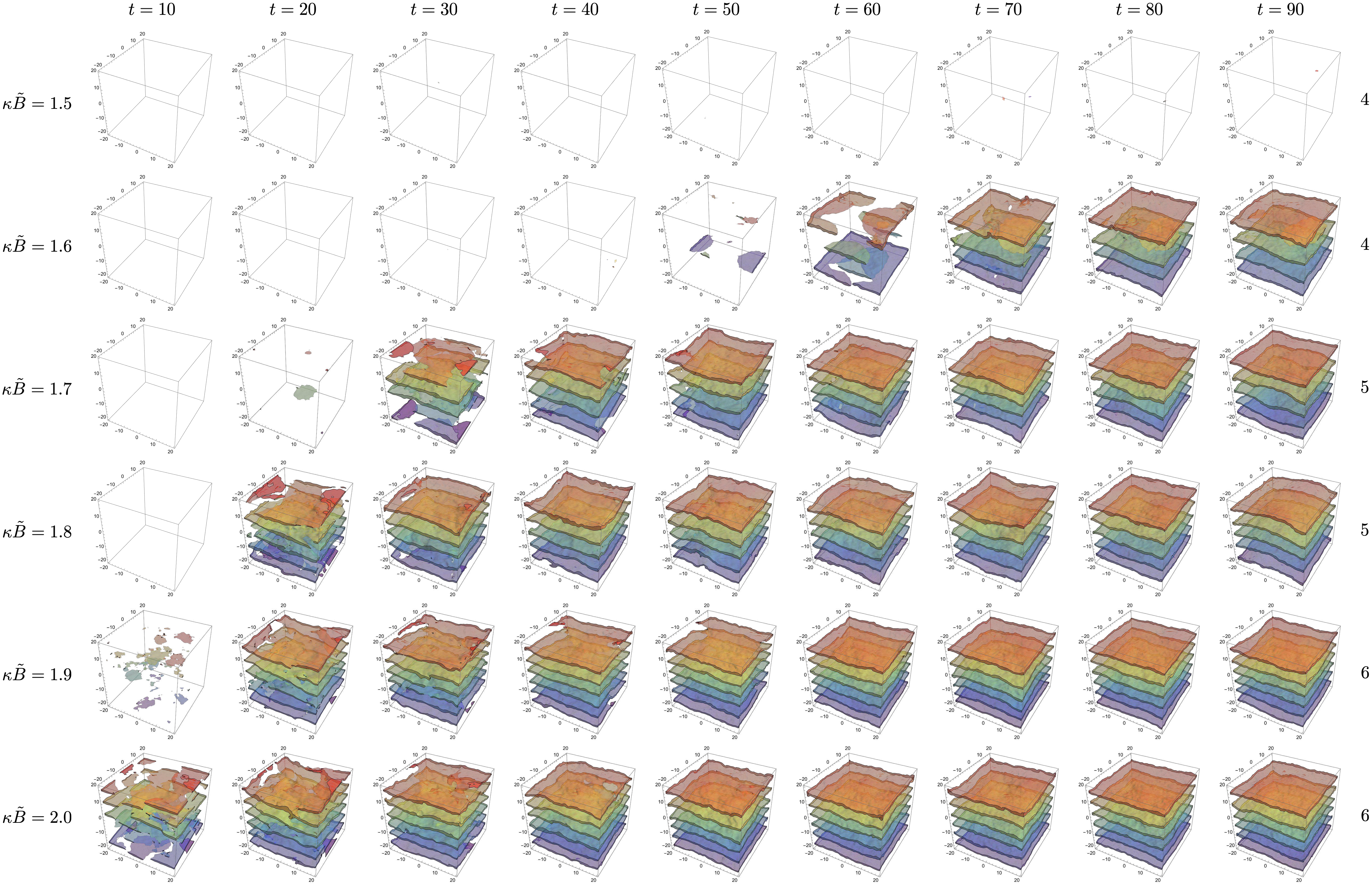}
  \caption{The real time dynamics of the CSL formation from the same initial state as the one given in the top-left panel of Fig.~\ref{fig:seed8_s1p5}. We show the surfaces of $|\arg\tilde\phi| = 0.8 \pi$. From top to bottom, we vary $\kappa \tilde B$ from $1.5$ to $2.0$. The number on the right-most column stands for the number of solitons in the final CSL. The periodic boundary conditions are imposed for $x$, $y$, and $z$ axes.}
  \label{fig:seed8_soliton_various_S_longtime}
%\end{figure}
\end{sidewaysfigure}
We also investigate evolutions from the same initial configuration with different $\kappa \tilde B$. The formation of the CSL is qualitatively the same for different $\kappa \tilde B$. We see that as $\kappa \tilde B$ increases, three things happen: 1) many large seeds appear all over the place at the same time (very short instance), 2) they appear earlier in the simulation time, and 3) the number of solitons in the final CSL becomes larger. We show snap shots for $t = \{10,\,20,\,\cdots, 90\}$ of the real time evolution with $\kappa \tilde B = \{1.5,\,1.6,\,\cdots,2.0\}$ in Figs.~\ref{fig:seed8_soliton_various_S_longtime} and \ref{fig:seed8_dislocation_various_S_longtime}.
Fig.~\ref{fig:seed8_soliton_various_S_longtime} shows the surfaces of $|\arg\tilde\phi| = 0.8 \pi$ and one can easily see how the solitons are generated and evolve into the pure CSL. For small $\kappa \tilde B$, the CSL formation begins late and only a few initial seeds are produced. On the other hand, for large $\kappa \tilde B$, the many initial seeds are quickly produced, and they soon transit to the CSL. The number of solitons in the final states read $4$, $5$, and $6$ for $\kappa\tilde B = 1.5$ and $1.6$, $1.7$ and $1.8$, and $1.9$ and $2.0$, respectively. 
Fig.~\ref{fig:seed8_dislocation_various_S_longtime} shows the surfaces of $|\tilde \phi| = 0.3$ of the same simulation by which one can clearly observe the dislocations, and how they are produced, evolve, and disappear. When $\kappa \tilde B$ is large, many small dislocations are present at the beginning of the CSL formation. As they collide repeatedly, they recombine and gradually connect with each other. Small dislocations shrink and disappear, but large ones grow into larger dislocations.  All the dislocations are closed loops due to the periodic boundary conditions, hence they finally disappear and the CSL formation is completed.
\begin{sidewaysfigure}[h]
%\begin{figure}[htbp]
    \centering
    \includegraphics[width=0.99\linewidth]{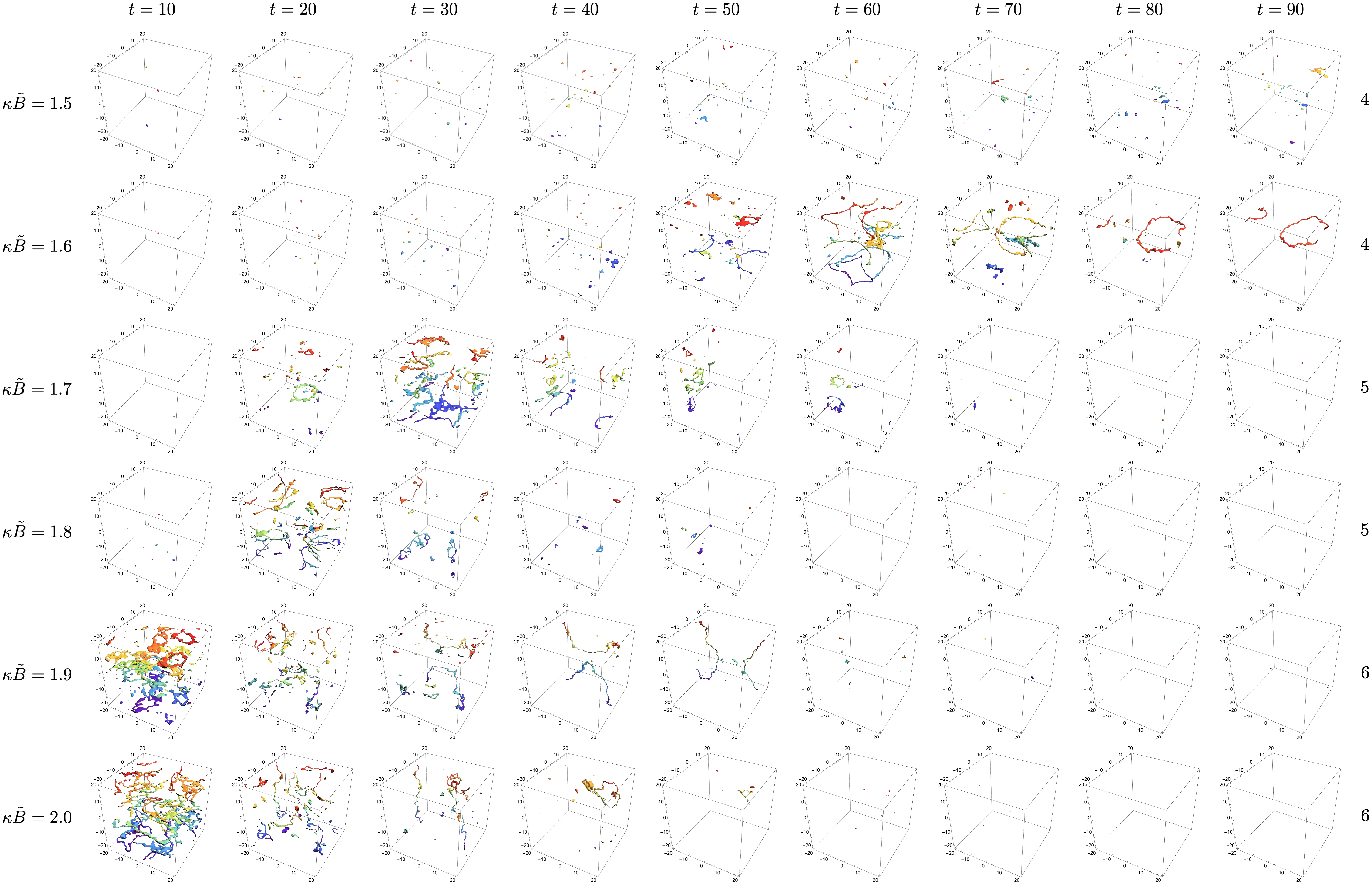}
  \caption{The real time dynamics of the CSL formation. We show the surfaces of $|\tilde\phi| = 0.3$ for the same simulations in Fig.~\ref{fig:seed8_soliton_various_S_longtime}. The horizontally expanded strings correspond to the edge dislocations, and the vertical strings correspond to the screw dislocations. The diagonally stretched ones should be understood as mixed dislocations. }
  \label{fig:seed8_dislocation_various_S_longtime}
%\end{figure}
\end{sidewaysfigure}

\begin{figure}[htbp]
    \centering
    \includegraphics[width=0.85\linewidth]{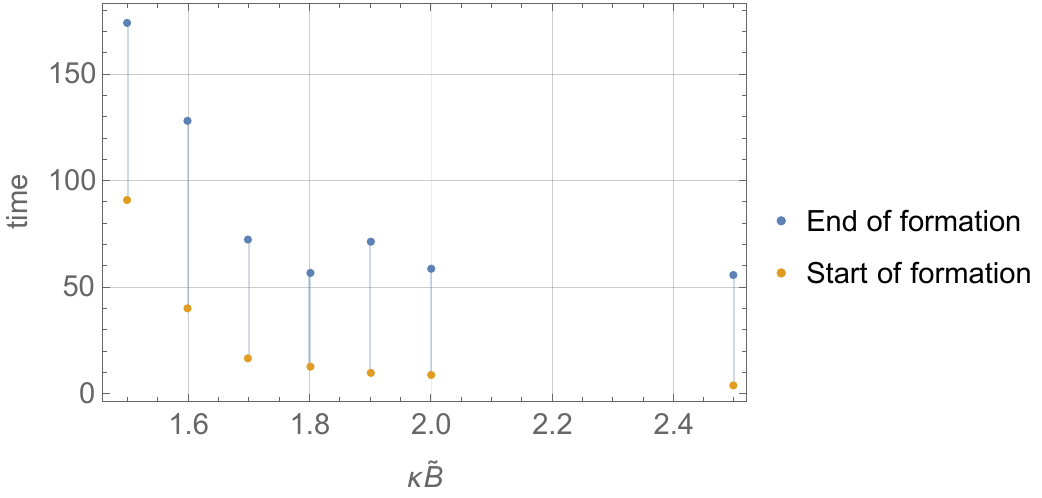}
  \caption{The start and end of the formation CSL for various $\kappa \tilde B$.}
  \label{fig:seed8_formation_time_data}
\end{figure}

We summarize in Fig.~\ref{fig:seed8_formation_time_data} the start time of formation that the first large seed is created and the end time of formation that the CSL is completed for the simulations in Figs.~\ref{fig:seed8_soliton_various_S_longtime} and \ref{fig:seed8_dislocation_various_S_longtime}. This well explains the above three properties 1) -- 3). Of course, our numerical analysis are highly sensitive on many factors like initial configurations, boundary conditions, size of numerical domains, etc, so we should treat our results explain qualitative aspect of dynamics of the formation of the CSL.
However, final configurations that we have reached are robust. As an example, we have obtained the same configurations starting from initial configurations different from the vacuum as shown in Appendix \ref{sec:appendixC}.

%%%%%%%%%%%%%%%%%%%%%
\section{Summary and discussion}
\label{sec:summary}

In this paper, we have investigated 
dynamical crystalization of the CSL from the vacuum.
The previously studied chiral sine-Gordon model (the model I) admits 
the CSLs as the ground state, 
and it is suitable for determining the phase diagram 
of ground states under an external background field. 
It is, however, not appropriate for the dynamical formation of the CSLs. 
We have pointed out that there are two main obstructions in the model I (the chiral sine-Gordon model): one is that the vortices terminating the solitons are singular and cannot be dealt withing continuum theories. 
The other is that the EOM does not depend on the external field and the ground state cannot be reached by solving the EOM. 
The first problem has been resolved by the axion model with a topological term (the model II), 
but the second point has remained unsolved. We then have arrived at the model III, which is the axion model with a modified topological term so that the presence of an external field affects the EOM.
We have investigated in detail both the kinematics and dynamics of the CSLs in the model III. 
As a result of the kinematical aspect, we have numerically constructed 
in Sec.~\ref{sec:chiral_soliton_modelIII} 
a single flat soliton under the presence of the external field, and reveal the critical point between two phases, the homogeneous vacuum and the single soliton.
Then, we have proceeded to figure out various dynamical aspects of the CSL. In Sec.~\ref{sec:modelIII_single_finite_soliton}, 
we have studied the motion of a single soliton suspended by a vortex and anti-vortex and have found that the soliton expands or shrinks depending on the magnitude of the external field.
In Sec.~\ref{sec:2d}, we have shown the simulations about the formation of CSLs in the model III in two spatial dimensions. We have first shown that the CSLs are indeed formulated from the soliton-free homogeneous vacuum with fluctuations. We then have focused on a single soliton sandwiched between vortices inserted in a CSL background and have found that it can be understood as an edge dislocation in a soliton lattice, pointing out a striking similarity to edge dislocations in atomic crystals both in topological and dynamical aspects.
Finally, we have studied the CSLs in model III in three spatial dimensions in Sec.~\ref{sec:3D}.
We have first studied the motion of an edge dislocation, which is a soliton disk surrounded by a vortex string, both in a homogeneous vacuum and in a CSL background in Sec.~\ref{sec:edge_3d}. 
Next, we have investigated screw dislocations that are specific to three-dimensional space in Sec.~\ref{sec:screw}. We have found that the screw dislocations in the CSL have a helicity associated with right- or left-handed spiral structure. The helicity of the screw dislocations 
is related to
%, but different from, 
the chirality of the soliton lattice. It is locked with the vortex charge, and what combination of helicity and vortex charge is realized depends on the direction of the external background field.  We have shown 
in Sec.~\ref{sec:CSL_formation_3d}
the results of three dimensional numerical simulations of the formation of the CSL from soliton-free vacuum with fluctuations. Our numerical simulations have revealed that the formation of CSL in three dimensions goes as follows: In the early stage of the evolution, small disk-shaped chiral solitons are generated, and then the disks expand and grow along the direction perpendicular to the background field, until finally a parallel array of chiral solitons occupies the entire space, completing a solid CSL. We have observed reconnection of the dislocations and collapse of holes on the chiral solitons repeatedly occurs during the formation process of CSL. 
We have observed mixed dislocations
also in intermediate processes. 
We have also given a relation between magnitude of external field and the number density of CSL and the formation speed.

\bigskip
More quantitative analysis of crystallization of CSL such as the phase ordering dynamics 
\cite{Bray01032002} 
should be discussed elsewhere.
Before closing this paper, here we address several other discussions.
In this paper, we have studied CSLs at zero temperature in which case we observed pair annihilation of dislocation and anti-dislocation. 
On the other hand, 
at finite temperature, 
a crystal melting may be caused by creation of dislocation-anti-dislocation pairs in two spatial dimensions as 
the BKT transition for atomic crystals
\cite{
Berezinsky:1970fr,
Berezinsky:1970-2,
Kosterlitz:1972,Kosterlitz:1973xp,
PhysRevLett.41.121,
PhysRevB.19.2457,
PhysRevB.19.1855}. 
In three spatial dimensions, dislocation loops may play a key  role for crystal melting as
 one in smectic liquid crystals  
in three dimensions \cite{Helfrich:1978,PhysRevB.24.363,Moreau_2006,1520312} 
and an Abrikosov vortex lattice in type-II superconductors   \cite{PhysRevB.34.494,PhysRevLett.57.1347,PhysRevB.34.6514,PhysRevB.41.1910,PhysRevB.51.11887}.

Dynamics of dislocations should be clarified microscopically. 
For instance,
when an edge dislocation moves 
in the direction of CSL (the $z$-axis), 
it feels the 
Peierls-Nabarro (PN) potential barrier
\cite{Peierls_1940,nabarro1947dislocations}. 
This can be calculated from the configurations in 
Fig.~\ref{fig:hopping_edge_dislocation}.
The leftmost and rightmost configurations in 
Fig.~\ref{fig:hopping_edge_dislocation}
are close to the ground states of the dislocation and the middle two configurations 
have higher energy, giving rise to the PN potential.

As mentioned in introduction, there are various physical systems admitting CSLs such as chiral magnets, chiral nematic liquid crystals, and QCD.
Thus, our analysis in this paper can be applied to these systems.  
In the CSL (spiral) phase of chiral magnets, dislocations behave as merons with half magnetic skyrmion charges of $\pi_2(S^2)$
\cite{Schoenherr:2018,PhysRevB.106.224428}. They may be useful for instance for a phase transition between the magnetic skyrmion crystal and CSL phases. 

In QCD in strong magnetic field, 
the ground state is a CSL in which 
a chiral soliton is made of the neutral pion. 
The chiral soliton 
can end on a charged pion string \cite{Son:2007ny,Amari:2024mip}, and  thus, charged pions are confined inside dislocations. 
Due to this, the dislocations
 are electrically charged 
and are superconducting. 
In this system, there are also 
domain-wall Skyrmions 
as two-dimensional Skyrmions on the soliton surface 
\cite{Eto:2023lyo,Eto:2023wul,Amari:2024fbo,Amari:2024mip} 
(similar objects exist under rapid rotation
\cite{Eto:2023tuu}). 
Interactions and relations between dislocations and Skyrmions are 
important directions to explore.
For instance, a closed dislocation line is a Skyrmion carrying a baryon number if the phase of charged pion is wound along it \cite{Amari:2024mip,Amari:2025twm}.

Extensions of our model to
the chiral double sine-Gordon model 
\cite{Ross:2020orc,Eto:2021gyy,Amari:2023bmx,Amari:2024jxx} and 
 to the axion model with the higher domain-wall number 
 will be interesting. 
 For the latter, 
 a potential term takes a form of 
$V  
=v m^2(2v -\phi^N - \phi^{*N}) \sim
 v m^2 \cos N \eta$.
In this case, 
CSL is an ordered array of $N$ different domain walls. 
A single dislocation can appears by joining $N$ different domain walls.

\begin{acknowledgments}
This work is supported in part by 
 JSPS KAKENHI [Grants No. 25K17392 (KN), 
 and No. JP22H01221 (ME and MN)], JSPS Fellows Grant 25KJ0148 (KN) and the WPI program ``Sustainability with Knotted Chiral Meta Matter (WPI-SKCM$^2$)'' at Hiroshima University (KN and MN).
\end{acknowledgments}

\clearpage 
%%%%%%%%%%%%%%%%%

\begin{appendix}

%%%%%%%%%%%%%%%%%%%%%%%%%%%%%%%%%%%%%%%%%
\section{The model II: Anomaly matching for chiral separation effect}
\label{sec:appendix}

Let us consider a single massless fermion $\psi$.
The axial $U(1)_{\rm A}$ symmetry is only classically preserved as $\p_\mu j_5^\mu = 0$ with 
$j^\mu_5 = \bar\psi\gamma^\mu\gamma_5\psi$, 
but it is not a symmetry at quantum level due to the quantum anomaly as
\be
\p_\mu j_5^\mu 
= -\frac{e^2}{16\pi^2}\epsilon_{\mu\nu\rho\lambda}F^{\mu\nu}F^{\rho\lambda}
= \frac{e^2}{2\pi^2}{\bm E}\cdot{\bm B}.
\label{eq:chiral_anom}
\ee
In addition, if a chemical potential $\mu$ is turned on and the system is put under an external magnetic field ${\bm B}$,
this system exhibits the so-called the chiral separation effect (CSE): 
an anomalous axial current is induced along the direction of the magnetic field:
\be
{\bm j}_5 = \frac{\mu_{\rm B}}{2\pi^2} {\bm B} \label{CSE_1flavor} \,.
\ee
The CSE is related by the chiral anomaly which protects the $1/2\pi^2$ factor against quantum collections.
Hence, one expects that the CSE should appear independently of the energy scale involved,
and the absence/presence of the spontaneous symmetry breaking of $U(1)_{\rm A}$.
Under the infinitesimal $U(1)_{\rm A}$ transformation with $|\theta_{\rm A}| \ll 1$, the action of the underlying theory changes as
\be
\delta S_{\rm UV} = \int d ^4x \left(\theta_{\rm A} \p_\mu j^\mu_5 + {\bm \nabla}\theta_{\rm A}\cdot {\bm j}_5\right) \label{transformation_UV} \,.
\ee

This anomaly terms should be reproduced in any low energy effective theories.
Here, we consider a linear sigma model (LSM) as a low energy effective theory in terms of a fermion condensate $\phi = \left<\bar \psi\psi\right>$:
\be
{\cal L}_{\rm IR} = \left|\p_\mu \phi\right|^2 - \frac{\lambda}{4}\left(|\phi|^2 - v^2\right)^2 + v m^2(\phi + \phi^*) + {\cal L}^{\rm (anom)}_{\rm IR}.
\ee
The chiral rotation,
$\psi_{\rm R} \to e ^{-i \theta_{\rm R}}\psi_{\rm R}$ and 
$\psi_{\rm L} \to e ^{-i \theta_{\rm L}}\psi_{\rm L}$,
is represented by $\phi$ as
\be
\phi \to e ^{-i  \theta_{\rm L} + i  \theta_{\rm R}} \phi \,.
\ee
Hence, the axial $U(1)_{\rm A}$ rotation is expressed as 
\be
\phi \to e^{i2\theta_{\rm A}}\phi,\qquad
\theta_{\rm A} \equiv \theta_{\rm R} = - \theta_{\rm L}.
\ee
The third term of ${\cal L}_{\rm IR}$ explicitly breaks $U(1)_{\rm A}$ symmetry but we turn it off for a while.
Then the classically preserving Noether current of  $U(1)_{\rm A}$  is given by
\be
j_5^\mu = -2i(\phi^*\p^\mu\phi - \phi\p^\mu\phi^*)  = 4v^2 \p^\mu\eta\,.
\ee
where we have expressed $\phi = \rho\, e^{i\eta}$ and 
we fixed the VEV by $\left<\rho\right> = v$.
The $U(1)_{\rm A}$ transformation is 
\be
\delta \eta = 2\theta_{\rm A}.
\ee
The last term ${\cal L}^{\rm (anom)}_{\rm IR}$ is added as manifestation of the chiral anomaly.
Since responses to the anomalous transformation must be identical in both the UV and IR actions, we should determine
${\cal L}^{\rm (anom)}_{\rm IR}$ in such a way taht
the changes of the IR action to be
\be
\delta S_{\rm IR} = 
\int d^4x\, \left[\theta_{\rm A} \frac{e^2}{2\pi^2}\bm{E}\cdot\bm{B}
+ \left({\bm \nabla}\theta_{\rm A}\right)\cdot \left(\frac{\mu_{\rm B}}{2\pi^2}{\bm B}\right)
\right].
\ee
where we have used 
Eqs.~(\ref{eq:chiral_anom}) and (\ref{eq:CSE}).
These can be realized by 
\be
{\mathcal L}_{\rm IR}^{\rm (anom)}
= \frac{e^2}{4\pi^2} \eta \bm{E}\cdot\bm{B}
+\frac{\mu_{\rm B}}{4\pi^2} {\bm B} \cdot {\bm \nabla}\eta
\ee
Below we will consider the case with $\bm{E} = 0$ and $\bm{B} = (0,0,B_z)$, then the anomalous terms reduce to
\be
{\mathcal L}_{\rm IR}^{\rm (anom)}
= \frac{\mu_{\rm B}  B_z}{4\pi^2} \frac{d\eta}{dz}.
\ee

%%%%%%%%%%%%%%
\section{Comparison with dislocations in atomic crystals}
\label{sec:atomic_crystal}

We compare the edge dislocation in the CSL and an edge dislocation of an atomic crystal.\footnote{Dislocations in an atomic crystal can be described as a soliton in
a discrete sine-Gordon model 
\cite{frenkel1939theory,Peierls_1940,
nabarro1947dislocations}. 
Relations between such a discrete model and our continuous sine-Gordon model are yet unclear.}
The blue and red solid lines in Fig.~\ref{fig:dislocation_CSL_vs_crystal}(a,c) stand for the trajectories of $\arg \phi = (2n+1)\pi$, and 
The black and white diamonds represent the vortex and the anti-vortex, respectively.
Fig.~\ref{fig:dislocation_CSL_vs_crystal}(b,d) is a schematic picture of an edge dislocation of an atomic crystal. 
\begin{figure}[htbp]
    \centering
    \includegraphics[width=0.99\linewidth]{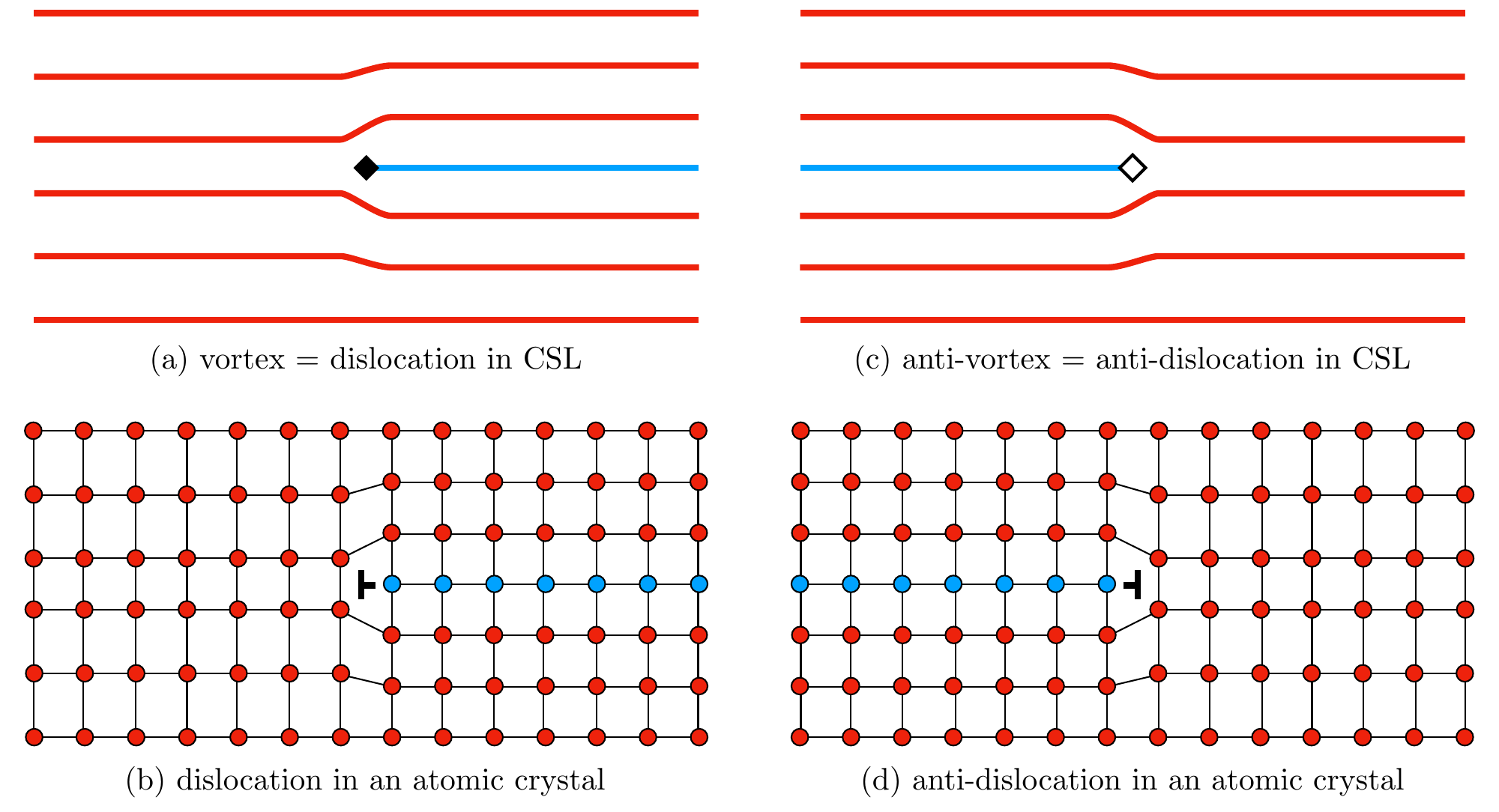}
  \caption{(a) Schematic picture of a finite soliton (blue solid half-line) in CSL background (red solid kinky curves. The black diamond corresponds to the vortex terminating the blue soliton. (b) Schematic picture of an atomic crystal with an edge dislocation. The symbol $\perp$ stands for the edge dislocation.}
  \label{fig:dislocation_CSL_vs_crystal}
\end{figure}
The edge dislocation is expressed by $\perp$ that can be thought of as the termination of an array of atoms in the middle of the  crystal. The similarity between the CSL in Fig.~\ref{fig:dislocation_CSL_vs_crystal}(a)[(c)] and the atomic crystal in Fig.~\ref{fig:dislocation_CSL_vs_crystal}(b)[(d)] should be clear now.
Each (red) soliton in the background CSL corresponds to each line of horizontally aligned (red) atoms in the crystal.  Furthermore, the (blue) finite soliton corresponds to the (blue) half line of atoms pushed in the middle of the crystal.
Finally, the vortex and anti-vortex terminating the finite soliton (black and white diamond) should be identified with the edge dislocation ($\perp$) in the crystal.

The similarities are not only seen in the static properties but also in the dynamic aspects.
In Fig.~\ref{fig:hopping_edge_dislocation}, we observe dislocation hopping in the CSL. Similar motions are known in atomic crystals, see Fig.~\ref{fig:hopping_edge_dislocation_atom}.
%%%%%%%%
\begin{figure}[htbp]
    \centering
    \includegraphics[width=0.99\linewidth]{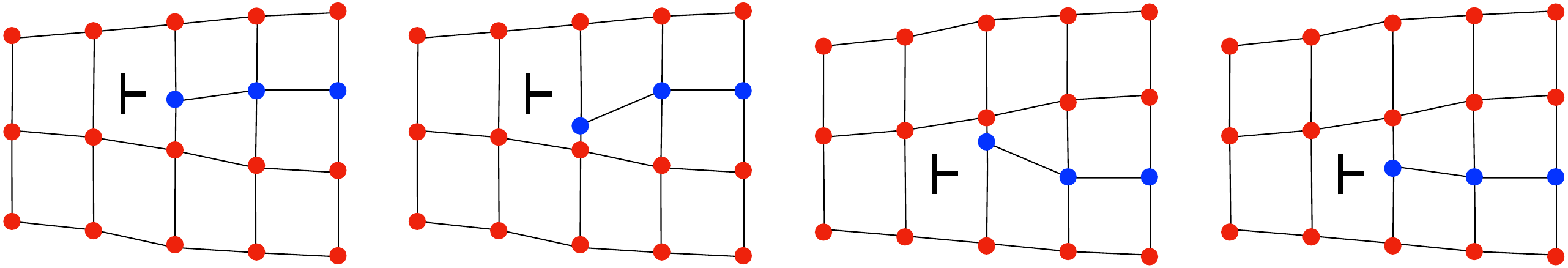}
  \caption{This shows how the atomic bonds at the center of the dislocation break and reform to allow the dislocation to move in an atomic crystal.}
  \label{fig:hopping_edge_dislocation_atom}
\end{figure}

%%%%%%%%%%%%%%%%%%%%%%%%%%%%%%%%%%%%%%%%%%%
\section{Chiral soliton lattice from a twiston}\label{sec:appendixC}

\begin{figure}[htbp]
    \centering
    \includegraphics[width=0.99\linewidth]{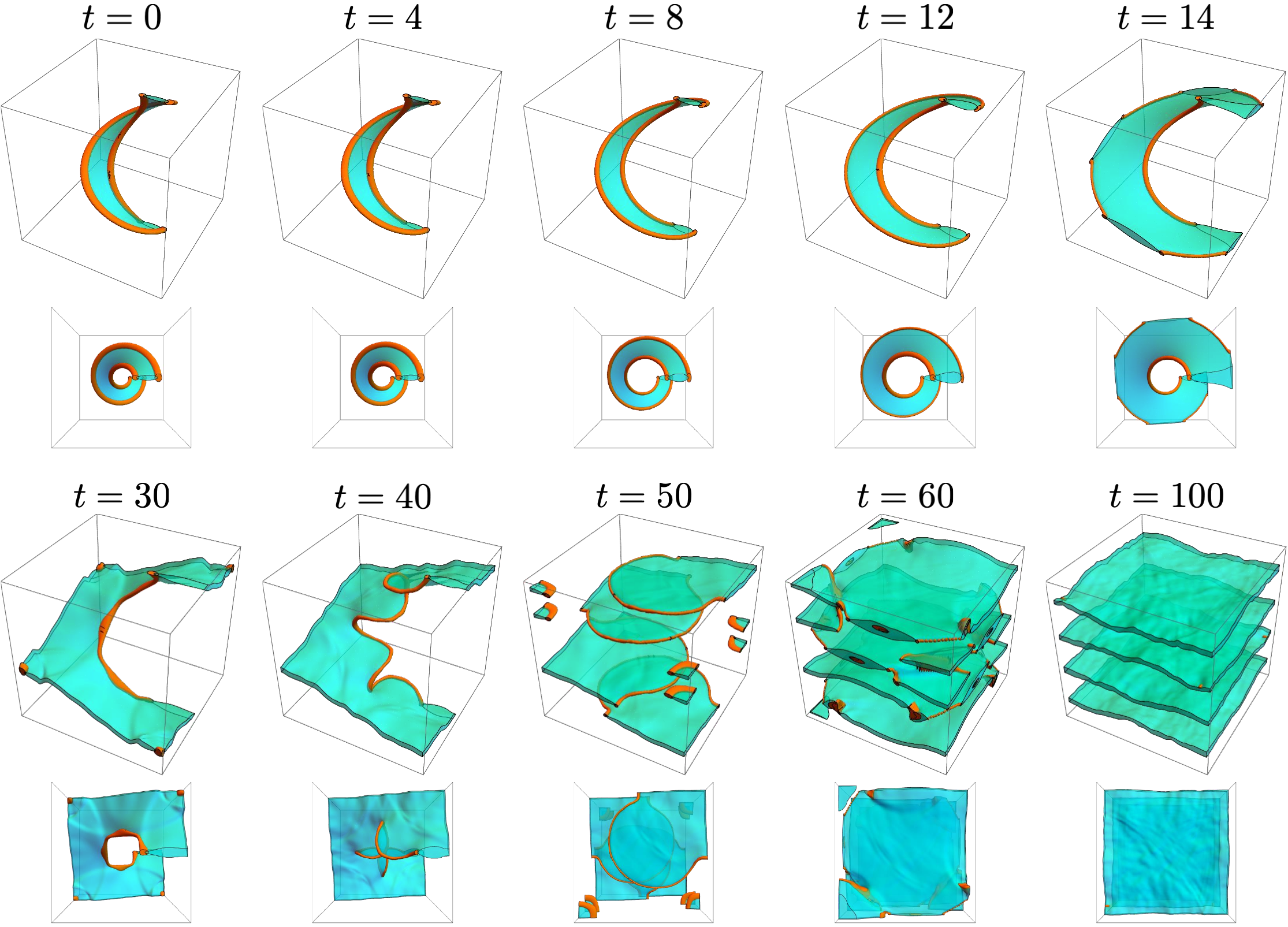}
  \caption{Real time evolution from twiston to CSL with four solitons. The snapshots of evolution of the twiston with zero initial velocity are shown. We take $\tilde m = 0.5$ and $\kappa\tilde B = 1.5$, and the boundary conditions for the $x$ and $y$ axes are the Neumann type while that for the $z$ axis is periodic one.}
  \label{fig:twiston_to_CSL}
\end{figure}
While in the main text, we have studied 
formation of CSL from the vacuum with fluctuations, a CSL can be reached from other configurations. In this appendix, as an example, we study the real time dynamics of a twiston.

First, let us introduce a twiston that is a sort of screw dislocation. It is constructed as follows. Take a finite size soliton stretching a vortex and anti-vortex in the $xy$ plane like the one shown in Fig.~\ref{fig:single_soliton_model_III_S0p256}, then extend it along the $z$ axis with rotating by $2\pi$ in the $xy$ plane.  We show an example of a twiston in the left-top panel of Fig.~\ref{fig:twiston_to_CSL} where the rotation is performed on an off-center axis. When we slice the twiston by a constant $z$ plane, the inner helical string has vortex winding $1$ while the outer helical string has $-1$. We set $\kappa \tilde B = 1.5$ which is sufficiently large  so that the twiston grows radially outward around the $z$ axis until the outer helical string reaches the computation boundaries, see the top row of Fig.~\ref{fig:twiston_to_CSL}. This evolution is similar to the ring-shape edge dislocation shown in Fig.~\ref{fig:single_ring}.
The area of the soliton continues to increase after the outer string touches the boundary. Although the boundary strongly affects the dynamics, we can pursue the dynamical evolution. The outer string is eventually pushed out of the computational domain, leaving behind a left-handed screw dislocation as shown in the figure at $t=30$. The boundary effect can be seen in the top view at $t=30$, where we can see that the inner helical string becomes a square-ring shape, due to the square boundary. The boundary effect continues to propagate, and it largely deforms the screw dislocation as shown in the panels with $t=40, 50$, and $60$. During the evolution, the four edge dislocations are created from the screw dislocation, and finally pure CSL containing no dislocations is formed as shown in the panel with $t=100$.

\end{appendix}

\bibliographystyle{jhep}
\bibliography{reference-dislocation}

\providecommand{\href}[2]{#2}\begingroup\raggedright\begin{thebibliography}{100}

\bibitem{Volterra:1907}
V.~Volterra, \emph{Sur l'equilibre des corps elastiques multiplement connexes}, {\emph{Annales Scientifiques de l'Ecole Normale Superieure} {\bfseries 24} (1907) 401–517}.

\bibitem{Orowan:1934}
E.~Orowan, \emph{Zur kristallplastizitat. iii}, \href{https://doi.org/10.1007/BF01341480}{\emph{Zeitschrift fur Physik} {\bfseries 89} (1934) 634–659}.

\bibitem{Polanyi:1934}
M.~Polanyi, \emph{Über eine art gitterstorung, die einen kristall plastisch machen könnte}, \href{https://doi.org/10.1007/BF01341481}{\emph{Zeitschrift fur Physik} {\bfseries 89} (1934) 660–664}.

\bibitem{Taylor:1934}
G.~I. Taylor, \emph{The mechanism of plastic deformation of crystals. part i. theoretical}, \href{https://doi.org/10.1098/rspa.1934.0106}{\emph{Proceedings of the Royal Society of London} {\bfseries A. 145} (1934) 362–87}.

\bibitem{Hull:2001}
D.~Hull and D.~J. Bacon, \emph{Introduction to dislocations (4th ed.)}. Butterworth-Heinemann, 2001.

\bibitem{Anderson:2017}
P.~M. Anderson, \emph{Theory of dislocations}. Hirth, John Price, 1930-, Lothe, Jens (Third ed.), 2017.

\bibitem{frenkel1939theory}
J.~Frenkel, \emph{On the theory of plastic deformation and twinning}, {\emph{J. Phys.} {\bfseries 1} (1939) 137}.

\bibitem{Peierls_1940}
R.~Peierls, \emph{The size of a dislocation}, \href{https://doi.org/10.1088/0959-5309/52/1/305}{\emph{Proceedings of the Physical Society} {\bfseries 52} (1940) 34}.

\bibitem{nabarro1947dislocations}
F.~Nabarro, \emph{Dislocations in a simple cubic lattice}, \href{https://doi.org/10.1088/0959-5309/59/2/309}{\emph{Proceedings of the Physical Society} {\bfseries 59} (1947) 256}.

\bibitem{Braun:1998}
O.~M. Braun and Y.~S. Kivshar, \emph{Nonlinear dynamics of the frenkel-kontorova model}, {\emph{Physics Reports} {\bfseries 306} (1998) }.

\bibitem{Braun2004}
O.~Braun and Y.~Kivshar, \emph{The Frenkel-Kontorova Model. Concepts, Methods and Applications}. Springer, Berlin, 2004.

\bibitem{Kosterlitz:1972}
J.~M. Kosterlitz and D.~J. Thouless, \emph{{Ordering, metastability and phase transitions in two-dimensional systems}}, \href{https://doi.org/10.1088/0022-3719/6/7/010}{\emph{J. Phys. C} {\bfseries 5} (1972) L124}.

\bibitem{Kosterlitz:1973xp}
J.~M. Kosterlitz and D.~J. Thouless, \emph{{Ordering, metastability and phase transitions in two-dimensional systems}}, \href{https://doi.org/10.1088/0022-3719/6/7/010}{\emph{J. Phys. C} {\bfseries 6} (1973) 1181}.

\bibitem{PhysRevLett.41.121}
B.~I. Halperin and D.~R. Nelson, \emph{Theory of two-dimensional melting}, \href{https://doi.org/10.1103/PhysRevLett.41.121}{\emph{Phys. Rev. Lett.} {\bfseries 41} (1978) 121}.

\bibitem{PhysRevB.19.2457}
D.~R. Nelson and B.~I. Halperin, \emph{Dislocation-mediated melting in two dimensions}, \href{https://doi.org/10.1103/PhysRevB.19.2457}{\emph{Phys. Rev. B} {\bfseries 19} (1979) 2457}.

\bibitem{PhysRevB.19.1855}
A.~P. Young, \emph{Melting and the vector coulomb gas in two dimensions}, \href{https://doi.org/10.1103/PhysRevB.19.1855}{\emph{Phys. Rev. B} {\bfseries 19} (1979) 1855}.

\bibitem{1520312}
D.~Nelson, \emph{Defects and geometry in condensed matter physics}. Cambridge University Press, 2002.

\bibitem{Berezinsky:1970fr}
V.~L. Berezinsky, \emph{{Destruction of long range order in one-dimensional and two-dimensional systems having a continuous symmetry group. I. Classical systems}}, {\emph{Sov. Phys. JETP} {\bfseries 32} (1971) 493}.

\bibitem{Berezinsky:1970-2}
V.~L. Berezinsky, \emph{{Destruction of long range order in one-dimensional and two-dimensional systems having a continuous symmetry group. II. Quantum systems}}, {\emph{Sov. Phys. JETP} {\bfseries 34} (1972) 610}.

\bibitem{PhysRev.79.722}
F.~C. Frank and W.~T. Read, \emph{Multiplication processes for slow moving dislocations}, \href{https://doi.org/10.1103/PhysRev.79.722}{\emph{Phys. Rev.} {\bfseries 79} (1950) 722}.

\bibitem{Helfrich:1978}
W.~Helfrich, \emph{Defect model of the smectic a-nematic phase transition}, \href{https://doi.org/10.1051/jphys:0197800390110119900}{\emph{J. Phys. France} {\bfseries 39} (1978) 1199}.

\bibitem{PhysRevB.24.363}
D.~R. Nelson and J.~Toner, \emph{Bond-orientational order, dislocation loops, and melting of solids and smectic-$a$ liquid crystals}, \href{https://doi.org/10.1103/PhysRevB.24.363}{\emph{Phys. Rev. B} {\bfseries 24} (1981) 363}.

\bibitem{Moreau_2006}
P.~Moreau, L.~Navailles, J.~Giermanska-Kahn, O.~Mondain-Monval, F.~Nallet and D.~Roux, \emph{Dislocation-loop-mediated smectic melting}, \href{https://doi.org/10.1209/epl/i2005-10348-y}{\emph{Europhysics Letters} {\bfseries 73} (2005) 49}.

\bibitem{PhysRevLett.131.128101}
P.~Pieranski, M.~Zeghal, M.~H. Godinho, P.~Judeinstein, R.~Bouffet-Klein, B.~Liagre et~al., \emph{Topological metadefects: Tangles of dislocations}, \href{https://doi.org/10.1103/PhysRevLett.131.128101}{\emph{Phys. Rev. Lett.} {\bfseries 131} (2023) 128101}.

\bibitem{Pieranski13112024}
P.~Pieranski and M.~H. Godinho, \emph{Unknots, knots, links and necklaces made of dislocations in cholesterics}, \href{https://doi.org/10.1080/02678292.2024.2374948}{\emph{Liquid Crystals} {\bfseries 51} (2024) 2144} [\href{https://arxiv.org/abs/https://doi.org/10.1080/02678292.2024.2374948}{{\ttfamily https://doi.org/10.1080/02678292.2024.2374948}}].

\bibitem{Kamien_2016}
R.~D. Kamien and R.~A. Mosna, \emph{The topology of dislocations in smectic liquid crystals}, \href{https://doi.org/10.1088/1367-2630/18/5/053012}{\emph{New Journal of Physics} {\bfseries 18} (2016) 053012}.

\bibitem{severino2024dislocationsfibrationstopologicalstructure}
P.~G. Severino, R.~D. Kamien and B.~Bode, \emph{Dislocations and fibrations: The topological structure of knotted smectic defects},  2024.

\bibitem{Gromov:2022cxa}
A.~Gromov and L.~Radzihovsky, \emph{{Colloquium: Fracton matter}}, \href{https://doi.org/10.1103/RevModPhys.96.011001}{\emph{Rev. Mod. Phys.} {\bfseries 96} (2024) 011001} [\href{https://arxiv.org/abs/2211.05130}{{\ttfamily 2211.05130}}].

\bibitem{Fulde:1964zz}
P.~Fulde and R.~A. Ferrell, \emph{{Superconductivity in a Strong Spin-Exchange Field}}, \href{https://doi.org/10.1103/PhysRev.135.A550}{\emph{Phys. Rev.} {\bfseries 135} (1964) A550}.

\bibitem{Larkin:1964wok}
A.~I. Larkin and Y.~N. Ovchinnikov, \emph{{Nonuniform state of superconductors}}, {\emph{Zh. Eksp. Teor. Fiz.} {\bfseries 47} (1964) 1136}.

\bibitem{Nitta:2017mgk}
M.~Nitta, S.~Sasaki and R.~Yokokura, \emph{{Spatially Modulated Vacua in a Lorentz-invariant Scalar Field Theory}}, \href{https://doi.org/10.1140/epjc/s10052-018-6235-9}{\emph{Eur. Phys. J. C} {\bfseries 78} (2018) 754} [\href{https://arxiv.org/abs/1706.02938}{{\ttfamily 1706.02938}}].

\bibitem{Nitta:2017yuf}
M.~Nitta, S.~Sasaki and R.~Yokokura, \emph{{Supersymmetry Breaking in Spatially Modulated Vacua}}, \href{https://doi.org/10.1103/PhysRevD.96.105022}{\emph{Phys. Rev. D} {\bfseries 96} (2017) 105022} [\href{https://arxiv.org/abs/1706.05232}{{\ttfamily 1706.05232}}].

\bibitem{Gudnason:2018bqb}
S.~B. Gudnason, M.~Nitta, S.~Sasaki and R.~Yokokura, \emph{{Temporally, spatially, or lightlike modulated vacua in Lorentz invariant theories}}, \href{https://doi.org/10.1103/PhysRevD.99.045011}{\emph{Phys. Rev. D} {\bfseries 99} (2019) 045011} [\href{https://arxiv.org/abs/1810.11361}{{\ttfamily 1810.11361}}].

\bibitem{BjarkeGudnason:2018aij}
S.~Bjarke~Gudnason, M.~Nitta, S.~Sasaki and R.~Yokokura, \emph{{Supersymmetry breaking and ghost Goldstino in modulated vacua}}, \href{https://doi.org/10.1103/PhysRevD.99.045012}{\emph{Phys. Rev. D} {\bfseries 99} (2019) 045012} [\href{https://arxiv.org/abs/1812.09078}{{\ttfamily 1812.09078}}].

\bibitem{Machida:1984zz}
K.~Machida and H.~Nakanishi, \emph{{Superconductivity under a ferromagnetic molecular field}}, \href{https://doi.org/10.1103/PhysRevB.30.122}{\emph{Phys. Rev. B} {\bfseries 30} (1984) 122}.

\bibitem{Basar:2008im}
G.~Basar and G.~V. Dunne, \emph{{Self-consistent crystalline condensate in chiral Gross-Neveu and Bogoliubov-de Gennes systems}}, \href{https://doi.org/10.1103/PhysRevLett.100.200404}{\emph{Phys. Rev. Lett.} {\bfseries 100} (2008) 200404} [\href{https://arxiv.org/abs/0803.1501}{{\ttfamily 0803.1501}}].

\bibitem{Basar:2008ki}
G.~Basar and G.~V. Dunne, \emph{{A Twisted Kink Crystal in the Chiral Gross-Neveu model}}, \href{https://doi.org/10.1103/PhysRevD.78.065022}{\emph{Phys. Rev. D} {\bfseries 78} (2008) 065022} [\href{https://arxiv.org/abs/0806.2659}{{\ttfamily 0806.2659}}].

\bibitem{Casalbuoni:2003wh}
R.~Casalbuoni and G.~Nardulli, \emph{{Inhomogeneous superconductivity in condensed matter and QCD}}, \href{https://doi.org/10.1103/RevModPhys.76.263}{\emph{Rev. Mod. Phys.} {\bfseries 76} (2004) 263} [\href{https://arxiv.org/abs/hep-ph/0305069}{{\ttfamily hep-ph/0305069}}].

\bibitem{Anglani:2013gfu}
R.~Anglani, R.~Casalbuoni, M.~Ciminale, N.~Ippolito, R.~Gatto, M.~Mannarelli et~al., \emph{{Crystalline color superconductors}}, \href{https://doi.org/10.1103/RevModPhys.86.509}{\emph{Rev. Mod. Phys.} {\bfseries 86} (2014) 509} [\href{https://arxiv.org/abs/1302.4264}{{\ttfamily 1302.4264}}].

\bibitem{togawa2012chiral}
Y.~Togawa, T.~Koyama, K.~Takayanagi, S.~Mori, Y.~Kousaka, J.~Akimitsu et~al., \emph{Chiral magnetic soliton lattice on a chiral helimagnet}, {\emph{Physical review letters} {\bfseries 108} (2012) 107202}.

\bibitem{togawa2016symmetry}
Y.~Togawa, Y.~Kousaka, K.~Inoue and J.-i. Kishine, \emph{Symmetry, structure, and dynamics of monoaxial chiral magnets}, {\emph{Journal of the Physical Society of Japan} {\bfseries 85} (2016) 112001}.

\bibitem{KISHINE20151}
J.-i. Kishine and A.~S. Ovchinnikov, \emph{Chapter one - theory of monoaxial chiral helimagnet},  vol.~66 of \emph{Solid State Physics}, pp.~1--130, Academic Press, (2015), \href{https://doi.org/https://doi.org/10.1016/bs.ssp.2015.05.001}{DOI}.

\bibitem{PhysRevB.97.184303}
A.~A. Tereshchenko, A.~S. Ovchinnikov, I.~Proskurin, E.~V. Sinitsyn and J.-i. Kishine, \emph{Theory of magnetoelastic resonance in a monoaxial chiral helimagnet}, \href{https://doi.org/10.1103/PhysRevB.97.184303}{\emph{Phys. Rev. B} {\bfseries 97} (2018) 184303}.

\bibitem{PhysRevB.65.064433}
J.~Chovan, N.~Papanicolaou and S.~Komineas, \emph{Intermediate phase in the spiral antiferromagnet ${\mathrm{ba}}_{2}{\mathrm{cuge}}_{2}{\mathrm{o}}_{7}$}, \href{https://doi.org/10.1103/PhysRevB.65.064433}{\emph{Phys. Rev. B} {\bfseries 65} (2002) 064433}.

\bibitem{Ross:2020orc}
C.~Ross, N.~Sakai and M.~Nitta, \emph{{Exact ground states and domain walls in one dimensional chiral magnets}}, \href{https://doi.org/10.1007/JHEP12(2021)163}{\emph{JHEP} {\bfseries 12} (2021) 163} [\href{https://arxiv.org/abs/2012.08800}{{\ttfamily 2012.08800}}].

\bibitem{Amari:2023gqv}
Y.~Amari and M.~Nitta, \emph{{Chiral magnets from string theory}}, \href{https://doi.org/10.1007/JHEP11(2023)212}{\emph{JHEP} {\bfseries 11} (2023) 212} [\href{https://arxiv.org/abs/2307.11113}{{\ttfamily 2307.11113}}].

\bibitem{Amari:2023bmx}
Y.~Amari, C.~Ross and M.~Nitta, \emph{{Domain-wall skyrmion chain and domain-wall bimerons in chiral magnets}}, \href{https://doi.org/10.1103/PhysRevB.109.104426}{\emph{Phys. Rev. B} {\bfseries 109} (2024) 104426} [\href{https://arxiv.org/abs/2311.05174}{{\ttfamily 2311.05174}}].

\bibitem{Amari:2024jxx}
Y.~Amari and M.~Nitta, \emph{{Skyrmion crystal phase on a magnetic domain wall in chiral magnets}}, \href{https://doi.org/10.1103/PhysRevB.111.134441}{\emph{Phys. Rev. B} {\bfseries 111} (2025) 134441} [\href{https://arxiv.org/abs/2409.07943}{{\ttfamily 2409.07943}}].

\bibitem{Dzyaloshinskii}
I.~Dzyaloshinskii, \emph{{A Thermodynamic Theory of `Weak' Ferromagnetism of Antiferromagnetics}}, \href{https://doi.org/10.1016/0022-3697(58)90076-3}{\emph{J.~Phys.~Chem.~Solids} {\bfseries 4} (1958) 241}.

\bibitem{Moriya:1960zz}
T.~Moriya, \emph{{Anisotropic Superexchange Interaction and Weak Ferromagnetism}}, \href{https://doi.org/10.1103/PhysRev.120.91}{\emph{Phys. Rev.} {\bfseries 120} (1960) 91}.

\bibitem{Chandrasekhar:1992}
S.~Chandrasekhar, \emph{Liquid Crystals (2nd ed.)}. Cambridge: Cambridge University Press, 1992.

\bibitem{DEGENNES1968163}
P.~{De Gennes}, \emph{Calcul de la distorsion d'une structure cholesterique par un champ magnetique}, \href{https://doi.org/https://doi.org/10.1016/0038-1098(68)90024-0}{\emph{Solid State Communications} {\bfseries 6} (1968) 163}.

\bibitem{Son:2007ny}
D.~T. Son and M.~A. Stephanov, \emph{{Axial anomaly and magnetism of nuclear and quark matter}}, \href{https://doi.org/10.1103/PhysRevD.77.014021}{\emph{Phys. Rev. D} {\bfseries 77} (2008) 014021} [\href{https://arxiv.org/abs/0710.1084}{{\ttfamily 0710.1084}}].

\bibitem{Eto:2012qd}
M.~Eto, K.~Hashimoto and T.~Hatsuda, \emph{{Ferromagnetic neutron stars: axial anomaly, dense neutron matter, and pionic wall}}, \href{https://doi.org/10.1103/PhysRevD.88.081701}{\emph{Phys. Rev. D} {\bfseries 88} (2013) 081701} [\href{https://arxiv.org/abs/1209.4814}{{\ttfamily 1209.4814}}].

\bibitem{Brauner:2016pko}
T.~Brauner and N.~Yamamoto, \emph{{Chiral Soliton Lattice and Charged Pion Condensation in Strong Magnetic Fields}}, \href{https://doi.org/10.1007/JHEP04(2017)132}{\emph{JHEP} {\bfseries 04} (2017) 132} [\href{https://arxiv.org/abs/1609.05213}{{\ttfamily 1609.05213}}].

\bibitem{Chen:2021vou}
S.~Chen, K.~Fukushima and Z.~Qiu, \emph{{Skyrmions in a magnetic field and \ensuremath{\pi}0 domain wall formation in dense nuclear matter}}, \href{https://doi.org/10.1103/PhysRevD.105.L011502}{\emph{Phys. Rev. D} {\bfseries 105} (2022) L011502} [\href{https://arxiv.org/abs/2104.11482}{{\ttfamily 2104.11482}}].

\bibitem{Gronli:2022cri}
M.~S. Gr\o{}nli and T.~Brauner, \emph{{Competition of chiral soliton lattice and Abrikosov vortex lattice in QCD with isospin chemical potential}}, \href{https://doi.org/10.1140/epjc/s10052-022-10300-5}{\emph{Eur. Phys. J. C} {\bfseries 82} (2022) 354} [\href{https://arxiv.org/abs/2201.07065}{{\ttfamily 2201.07065}}].

\bibitem{Evans:2022hwr}
G.~W. Evans and A.~Schmitt, \emph{{Chiral anomaly induces superconducting baryon crystal}}, \href{https://doi.org/10.1007/JHEP09(2022)192}{\emph{JHEP} {\bfseries 09} (2022) 192} [\href{https://arxiv.org/abs/2206.01227}{{\ttfamily 2206.01227}}].

\bibitem{Evans:2023hms}
G.~W. Evans and A.~Schmitt, \emph{{Chiral Soliton Lattice turns into 3D crystal}}, \href{https://doi.org/10.1007/JHEP02(2024)041}{\emph{JHEP} {\bfseries 2024} (2024) 041} [\href{https://arxiv.org/abs/2311.03880}{{\ttfamily 2311.03880}}].

\bibitem{Qiu:2023guy}
Z.~Qiu and M.~Nitta, \emph{{Quasicrystals in QCD}}, \href{https://doi.org/10.1007/JHEP05(2023)170}{\emph{JHEP} {\bfseries 05} (2023) 170} [\href{https://arxiv.org/abs/2304.05089}{{\ttfamily 2304.05089}}].

\bibitem{Huang:2017pqe}
X.-G. Huang, K.~Nishimura and N.~Yamamoto, \emph{{Anomalous effects of dense matter under rotation}}, \href{https://doi.org/10.1007/JHEP02(2018)069}{\emph{JHEP} {\bfseries 02} (2018) 069} [\href{https://arxiv.org/abs/1711.02190}{{\ttfamily 1711.02190}}].

\bibitem{Nishimura:2020odq}
K.~Nishimura and N.~Yamamoto, \emph{{Topological term, QCD anomaly, and the $\eta^{'}$ chiral soliton lattice in rotating baryonic matter}}, \href{https://doi.org/10.1007/JHEP07(2020)196}{\emph{JHEP} {\bfseries 07} (2020) 196} [\href{https://arxiv.org/abs/2003.13945}{{\ttfamily 2003.13945}}].

\bibitem{Eto:2021gyy}
M.~Eto, K.~Nishimura and M.~Nitta, \emph{{Phases of rotating baryonic matter: non-Abelian chiral soliton lattices, antiferro-isospin chains, and ferri/ferromagnetic magnetization}}, \href{https://doi.org/10.1007/JHEP08(2022)305}{\emph{JHEP} {\bfseries 08} (2022) 305} [\href{https://arxiv.org/abs/2112.01381}{{\ttfamily 2112.01381}}].

\bibitem{Eto:2023rzd}
M.~Eto, K.~Nishimura and M.~Nitta, \emph{{Non-Abelian chiral soliton lattice in rotating QCD matter: Nambu-Goldstone and excited modes}}, \href{https://doi.org/10.1007/JHEP03(2024)035}{\emph{JHEP} {\bfseries 03} (2024) 035} [\href{https://arxiv.org/abs/2312.10927}{{\ttfamily 2312.10927}}].

\bibitem{Son:2004tq}
D.~T. Son and A.~R. Zhitnitsky, \emph{{Quantum anomalies in dense matter}}, \href{https://doi.org/10.1103/PhysRevD.70.074018}{\emph{Phys. Rev. D} {\bfseries 70} (2004) 074018} [\href{https://arxiv.org/abs/hep-ph/0405216}{{\ttfamily hep-ph/0405216}}].

\bibitem{Vilenkin:1979ui}
A.~Vilenkin, \emph{{Macroscopic Parity Violating Effects: Neutrino Fluxes from Rotating Black Holes and in Rotating Thermal Radiation}}, \href{https://doi.org/10.1103/PhysRevD.20.1807}{\emph{Phys. Rev. D} {\bfseries 20} (1979) 1807}.

\bibitem{Vilenkin:1980zv}
A.~Vilenkin, \emph{{Quamtum Field Theory at Finite Temperature in a Rotating System}}, \href{https://doi.org/10.1103/PhysRevD.21.2260}{\emph{Phys. Rev. D} {\bfseries 21} (1980) 2260}.

\bibitem{Son:2009tf}
D.~T. Son and P.~Surowka, \emph{{Hydrodynamics with Triangle Anomalies}}, \href{https://doi.org/10.1103/PhysRevLett.103.191601}{\emph{Phys. Rev. Lett.} {\bfseries 103} (2009) 191601} [\href{https://arxiv.org/abs/0906.5044}{{\ttfamily 0906.5044}}].

\bibitem{Brauner:2017uiu}
T.~Brauner and S.~V. Kadam, \emph{{Anomalous low-temperature thermodynamics of QCD in strong magnetic fields}}, \href{https://doi.org/10.1007/JHEP11(2017)103}{\emph{JHEP} {\bfseries 11} (2017) 103} [\href{https://arxiv.org/abs/1706.04514}{{\ttfamily 1706.04514}}].

\bibitem{Brauner:2017mui}
T.~Brauner and S.~Kadam, \emph{{Anomalous electrodynamics of neutral pion matter in strong magnetic fields}}, \href{https://doi.org/10.1007/JHEP03(2017)015}{\emph{JHEP} {\bfseries 03} (2017) 015} [\href{https://arxiv.org/abs/1701.06793}{{\ttfamily 1701.06793}}].

\bibitem{Brauner:2021sci}
T.~Brauner, H.~Kole\v{s}ov\'a and N.~Yamamoto, \emph{{Chiral soliton lattice phase in warm QCD}}, \href{https://doi.org/10.1016/j.physletb.2021.136767}{\emph{Phys. Lett. B} {\bfseries 823} (2021) 136767} [\href{https://arxiv.org/abs/2108.10044}{{\ttfamily 2108.10044}}].

\bibitem{Yamada:2021jhy}
A.~Yamada and N.~Yamamoto, \emph{{Floquet vacuum engineering: Laser-driven chiral soliton lattice in the QCD vacuum}}, \href{https://doi.org/10.1103/PhysRevD.104.054041}{\emph{Phys. Rev. D} {\bfseries 104} (2021) 054041} [\href{https://arxiv.org/abs/2107.07074}{{\ttfamily 2107.07074}}].

\bibitem{Brauner:2019aid}
T.~Brauner, G.~Filios and H.~Kole\v{s}ov\'a, \emph{{Chiral soliton lattice in QCD-like theories}}, \href{https://doi.org/10.1007/JHEP12(2019)029}{\emph{JHEP} {\bfseries 12} (2019) 029} [\href{https://arxiv.org/abs/1905.11409}{{\ttfamily 1905.11409}}].

\bibitem{Brauner:2019rjg}
T.~Brauner, G.~Filios and H.~Kole\v{s}ov\'a, \emph{{Anomaly-Induced Inhomogeneous Phase in Quark Matter without the Sign Problem}}, \href{https://doi.org/10.1103/PhysRevLett.123.012001}{\emph{Phys. Rev. Lett.} {\bfseries 123} (2019) 012001} [\href{https://arxiv.org/abs/1902.07522}{{\ttfamily 1902.07522}}].

\bibitem{Nitta:2024xcu}
M.~Nitta and S.~Sasaki, \emph{{Solitonic ground state in supersymmetric theory in background}}, \href{https://doi.org/10.1007/JHEP10(2024)178}{\emph{JHEP} {\bfseries 10} (2024) 178} [\href{https://arxiv.org/abs/2404.12066}{{\ttfamily 2404.12066}}].

\bibitem{Eto:2023lyo}
M.~Eto, K.~Nishimura and M.~Nitta, \emph{{Domain-Wall Skyrmion Phase in Dense QCD at Strong Magnetic Fields Using Leading-Order Chiral Perturbation Theory}}, \href{https://doi.org/10.1103/PhysRevLett.134.181902}{\emph{Phys. Rev. Lett.} {\bfseries 134} (2025) 181902} [\href{https://arxiv.org/abs/2304.02940}{{\ttfamily 2304.02940}}].

\bibitem{Eto:2023wul}
M.~Eto, K.~Nishimura and M.~Nitta, \emph{{Phase diagram of QCD matter with magnetic field: domain-wall Skyrmion chain in chiral soliton lattice}}, \href{https://doi.org/10.1007/JHEP12(2023)032}{\emph{JHEP} {\bfseries 12} (2023) 032} [\href{https://arxiv.org/abs/2311.01112}{{\ttfamily 2311.01112}}].

\bibitem{Amari:2024fbo}
Y.~Amari, M.~Eto and M.~Nitta, \emph{{Domain-wall Skyrmion phase of QCD in magnetic field: gauge field dynamics}}, \href{https://doi.org/10.1007/JHEP05(2025)037}{\emph{JHEP} {\bfseries 05} (2025) 037} [\href{https://arxiv.org/abs/2409.08841}{{\ttfamily 2409.08841}}].

\bibitem{Amari:2024mip}
Y.~Amari, M.~Nitta and R.~Yokokura, \emph{{Spin statistics and surgeries of topological solitons in QCD matter in magnetic field}}, \href{https://doi.org/10.1007/JHEP02(2025)171}{\emph{JHEP} {\bfseries 02} (2025) 171} [\href{https://arxiv.org/abs/2406.14419}{{\ttfamily 2406.14419}}].

\bibitem{Amari:2025twm}
Y.~Amari, M.~Nitta and Z.~Qiu, \emph{{Phase Boundary of Nuclear Matter in Magnetic Field}},  \href{https://arxiv.org/abs/2504.08379}{{\ttfamily 2504.08379}}.

\bibitem{Eto:2023tuu}
M.~Eto, K.~Nishimura and M.~Nitta, \emph{{Domain-wall Skyrmion phase in a rapidly rotating QCD matter}}, \href{https://doi.org/10.1007/JHEP03(2024)019}{\emph{JHEP} {\bfseries 03} (2024) 019} [\href{https://arxiv.org/abs/2310.17511}{{\ttfamily 2310.17511}}].

\bibitem{PhysRevB.87.161107}
Z.~Wang and S.-C. Zhang, \emph{Chiral anomaly, charge density waves, and axion strings from weyl semimetals}, \href{https://doi.org/10.1103/PhysRevB.87.161107}{\emph{Phys. Rev. B} {\bfseries 87} (2013) 161107}.

\bibitem{Roy:2014mia}
B.~Roy and J.~D. Sau, \emph{{Magnetic catalysis and axionic charge-density-wave in Weyl semimetals}}, \href{https://doi.org/10.1103/PhysRevB.92.125141}{\emph{Phys. Rev. B} {\bfseries 92} (2015) 125141} [\href{https://arxiv.org/abs/1406.4501}{{\ttfamily 1406.4501}}].

\bibitem{PhysRevB.94.085102}
Y.~You, G.~Y. Cho and T.~L. Hughes, \emph{Response properties of axion insulators and weyl semimetals driven by screw dislocations and dynamical axion strings}, \href{https://doi.org/10.1103/PhysRevB.94.085102}{\emph{Phys. Rev. B} {\bfseries 94} (2016) 085102}.

\bibitem{PhysRevB.102.115159}
D.~Sehayek, M.~Thakurathi and A.~A. Burkov, \emph{Charge density waves in weyl semimetals}, \href{https://doi.org/10.1103/PhysRevB.102.115159}{\emph{Phys. Rev. B} {\bfseries 102} (2020) 115159}.

\bibitem{Gooth:2019}
J.~Gooth, B.~Bradlyn, S.~Honnali, C.~Schindler, N.~Kumar, J.~Noky et~al., \emph{Axionic charge-density wave in the weyl semimetal (tase4)2i}, \href{https://doi.org/10.1038/s41586-019-1630-4}{\emph{Nature} {\bfseries 575} (2019) 315–319}.

\bibitem{Schoenherr:2018}
P.~Schoenherr, J.~Müller and L.~Köhler~et al., \emph{Topological domain walls in helimagnets}, \href{https://doi.org/10.1038/s41567-018-0056-5}{\emph{Nature Phys} {\bfseries 14} (2018) 465–468}.

\bibitem{PhysRevB.106.224428}
N.~Mukai and A.~O. Leonov, \emph{Skyrmion and meron ordering in quasi-two-dimensional chiral magnets}, \href{https://doi.org/10.1103/PhysRevB.106.224428}{\emph{Phys. Rev. B} {\bfseries 106} (2022) 224428}.

\bibitem{PhysRevLett.128.157204}
M.~Azhar, V.~P. Kravchuk and M.~Garst, \emph{Screw dislocations in chiral magnets}, \href{https://doi.org/10.1103/PhysRevLett.128.157204}{\emph{Phys. Rev. Lett.} {\bfseries 128} (2022) 157204}.

\bibitem{Smalyukh:2020zin}
I.~I. Smalyukh, \emph{{Review: knots and other new topological effects in liquid crystals and colloids}}, \href{https://doi.org/10.1088/1361-6633/abaa39}{\emph{Rept. Prog. Phys.} {\bfseries 83} (2020) 106601}.

\bibitem{PhysRevE.109.L012701}
P.~G. Severino and R.~D. Kamien, \emph{Escape from the second dimension: A topological distinction between edge and screw dislocations}, \href{https://doi.org/10.1103/PhysRevE.109.L012701}{\emph{Phys. Rev. E} {\bfseries 109} (2024) L012701}.

\bibitem{Kamien_2006}
R.~D. Kamien and C.~D. Santangelo, \emph{Smectic liquid crystals: Materials with one-dimensional, periodic order}, \href{https://doi.org/10.1007/s10711-006-9075-y}{\emph{Geometriae Dedicata} {\bfseries 120} (2006) 229–240}.

\bibitem{Agterberg_2008}
D.~F. Agterberg and H.~Tsunetsugu, \emph{Dislocations and vortices in pair-density-wave superconductors}, \href{https://doi.org/10.1038/nphys999}{\emph{Nature Physics} {\bfseries 4} (2008) 639–642}.

\bibitem{PhysRevLett.67.1926}
A.-C. Shi and A.~J. Berlinsky, \emph{Pinning and i-v characteristics of a two-dimensional defective flux-line lattice}, \href{https://doi.org/10.1103/PhysRevLett.67.1926}{\emph{Phys. Rev. Lett.} {\bfseries 67} (1991) 1926}.

\bibitem{PhysRevB.34.494}
R.~W\"ordenweber and P.~H. Kes, \emph{Dimensional crossover in collective flux pinning}, \href{https://doi.org/10.1103/PhysRevB.34.494}{\emph{Phys. Rev. B} {\bfseries 34} (1986) 494}.

\bibitem{PhysRevLett.57.1347}
E.~H. Brandt, \emph{Range and strength of pins collectively interacting with the flux-line lattice in type-ii superconductors}, \href{https://doi.org/10.1103/PhysRevLett.57.1347}{\emph{Phys. Rev. Lett.} {\bfseries 57} (1986) 1347}.

\bibitem{PhysRevB.34.6514}
E.~H. Brandt, \emph{Elastic and plastic properties of the flux-line lattice in type-ii superconductors}, \href{https://doi.org/10.1103/PhysRevB.34.6514}{\emph{Phys. Rev. B} {\bfseries 34} (1986) 6514}.

\bibitem{PhysRevB.41.1910}
M.~C. Marchetti and D.~R. Nelson, \emph{Dislocation loops and bond-orientational order in the abrikosov flux-line lattice}, \href{https://doi.org/10.1103/PhysRevB.41.1910}{\emph{Phys. Rev. B} {\bfseries 41} (1990) 1910}.

\bibitem{PhysRevB.51.11887}
M.~J.~W. Dodgson and M.~A. Moore, \emph{Topological defects in the abrikosov lattice of vortices in type-ii superconductors}, \href{https://doi.org/10.1103/PhysRevB.51.11887}{\emph{Phys. Rev. B} {\bfseries 51} (1995) 11887}.

\bibitem{shimizu2023crystallization}
K.~Shimizu and G.-W. Chern, \emph{Crystallization dynamics of magnetic skyrmions in a frustrated itinerant magnet}, {\emph{arXiv preprint arXiv:2305.16182} (2023) }.

\bibitem{Eto:2022lhu}
M.~Eto and M.~Nitta, \emph{{Quantum nucleation of topological solitons}}, \href{https://doi.org/10.1007/JHEP09(2022)077}{\emph{JHEP} {\bfseries 09} (2022) 077} [\href{https://arxiv.org/abs/2207.00211}{{\ttfamily 2207.00211}}].

\bibitem{Higaki:2022gnw}
T.~Higaki, K.~Kamada and K.~Nishimura, \emph{{Formation of a chiral soliton lattice}}, \href{https://doi.org/10.1103/PhysRevD.106.096022}{\emph{Phys. Rev. D} {\bfseries 106} (2022) 096022} [\href{https://arxiv.org/abs/2207.00212}{{\ttfamily 2207.00212}}].

\bibitem{Eto:2023gfn}
M.~Eto, Y.~Hamada and M.~Nitta, \emph{{Composite topological solitons consisting of domain walls, strings, and monopoles in O(N) models}}, \href{https://doi.org/10.1007/JHEP08(2023)150}{\emph{JHEP} {\bfseries 08} (2023) 150} [\href{https://arxiv.org/abs/2304.14143}{{\ttfamily 2304.14143}}].

\bibitem{Vilenkin:2000jqa}
A.~Vilenkin and E.~P.~S. Shellard, \emph{{Cosmic Strings and Other Topological Defects}}. Cambridge University Press, 7, 2000.

\bibitem{YISHI1990689}
D.~Yishi and Z.~Shengli, \emph{Topological structure of dislocation in the gauge field theory of dislocations and disclinations continuum}, \href{https://doi.org/https://doi.org/10.1016/0020-7225(90)90096-2}{\emph{International Journal of Engineering Science} {\bfseries 28} (1990) 689}.

\bibitem{Bray01032002}
A.~J. Bray, \emph{Theory of phase-ordering kinetics}, \href{https://doi.org/10.1080/00018730110117433}{\emph{Advances in Physics} {\bfseries 51} (2002) 481} [\href{https://arxiv.org/abs/https://doi.org/10.1080/00018730110117433}{{\ttfamily https://doi.org/10.1080/00018730110117433}}].

\end{thebibliography}\endgroup

\end{document}